\documentclass[vecphys]{svmult}

\usepackage{makeidx}         
\usepackage{graphicx}        
\usepackage{multicol}        
\usepackage{natbib}
\usepackage[bottom]{footmisc}

\bibpunct{(}{)}{;}{a}{}{,}  
\newcommand{\mnras}{{MNRAS}}
\newcommand{\aap}{{A\&A}}
\newcommand{\aj}{{AJ}}
\newcommand{\apj}{{ApJ}}
\newcommand{\apjl}{{ApJ}}
\newcommand{\araa}{{ARA\&A}}
\newcommand{\nat}{{Nature}}

\newcommand{\oversim}[2]{\protect{\mbox{\lower0.5ex\vbox{%
  \baselineskip=0pt\lineskip=0.2ex
  \ialign{$\mathsurround=0pt #1\hfil##\hfil$\crcr#2\crcr\sim\crcr}}}}}
\newcommand{\simgreat}{\mbox{$\,\mathrel{\mathpalette\oversim>}\,$}} 
\newcommand{\simless} {\mbox{$\,\mathrel{\mathpalette\oversim<}\,$}} 

\begin{document}

\title*{Dense Stellar Populations: Initial Conditions}
\label{chapter:kroupa}
\author{Pavel Kroupa}
\institute{Argelander-Institut f\"ur Astronomie, Auf dem H\"ugel 71,
D-53121 Bonn, Germany\\
                 \texttt{pavel@astro.uni-bonn.de}}
%
%
\maketitle

\vspace{-9mm} {\tt To appear in {\it The Cambridge N-body Lectures},
Sverre Aarseth, Christo\-pher Tout, Rosemary Mardling (eds), Springer
Verlag: {\it Lecture Notes in Physics}; based on four lectures given at the
Cambridge N-body school {\it Cambody}, July 30 -- August 11, 2006}


\vspace{10mm}

\noindent {\bf \large Contents}

\noindent 
1. Introduction     \dotfill        \pageref{sec_pk:intro}

1.1 Embedded clusters \dotfill       \pageref{sec_pk:embcl}

1.2 Some implications for the astrophysics of galaxies 
                 \dotfill        \pageref{sec_pk:impl}

1.3 Long term, or classical, cluster evolution
                  \dotfill        \pageref{sec_pk:longterm}

1.4 What is a galaxy?  \dotfill       \pageref{sec_pk:galaxy}

\noindent 
2. Initial 6D conditions
                   \dotfill       \pageref{sec_pk:initcond}

2.1 6D structure of classical clusters
                   \dotfill       \pageref{sec_pk:classical}

2.2 Comparison: Plummer vs King models
                   \dotfill       \pageref{sec_pk:Plummer_King}

2.3 Discretisation
                   \dotfill       \pageref{sec_pk:discretisation}

2.4 Cluster birth and young clusters
                   \dotfill       \pageref{sec_pk:clbirth}

\noindent
3. The stellar IMF     \dotfill       \pageref{sec_pk:IMF}

3.1 The canonical or standard form of the stellar IMF
                   \dotfill        \pageref{sec_pk:canonIMF}

3.2 Universality of the IMF: resolved populations
                   \dotfill        \pageref{sec_pk:univ}

3.3 Very low-mass stars (VLMSs) and brown dwarfs (BDs)
                   \dotfill        \pageref{sec_pk:bds}

3.4 Composite populations: the IGIMF
                    \dotfill        \pageref{sec_pk:IGIMF}

3.5 Origin of the IMF: theory vs observations
                    \dotfill        \pageref{sec_pk:originIMF}

3.6 Conclusions: IMF
                    \dotfill         \pageref{sec_pk:IMFconcs}

3.7 Discretisation
                    \dotfill         \pageref{sec_pk:IMF_discretisation}

\noindent
4. The initial binary population
                    \dotfill         \pageref{sec_pk:binpop}

4.1 Introduction       \dotfill          \pageref{sec_pk:intro_bin}

4.2 The initial binary population -- late-type stars
                    \dotfill          \pageref{sec_pk:initialpop_lowmass}

4.3 The initial binary population -- massive stars
                    \dotfill          \pageref{sec_pk:initialpop_massive}

\noindent 5. Summary   \dotfill          \pageref{sec_pk:summary}

\noindent Acknowledgements    \dotfill   \pageref{sec_pk:acknowl}

\noindent Bibliography       \dotfill   \pageref{sec_pk:biblio}

\section{Introduction}
\label{sec_pk:intro}

Most stars form in dense star clusters deeply embedded in residual
gas. The populations of these objects range from small groups of stars
with $N\approx\;$dozen binaries within a volume with a typical radius
of $r\approx 0.3\,$pc through to objects formed in extreme star bursts
containing $N\approx 10^8\,$stars within $r\approx\;$a~few~dozen~pc.
Star clusters, or more generally dense stellar systems, must therefore
be seen as the fundamental building blocks of galaxies, whereby a
differentiation of the term {\it star cluster} from a {\it spheroidal
dwarf galaxy} becomes blurred near $N\approx 10^6\,M_\odot$. Both are
mostly pressure-supported, that is, random stellar motions dominate
any bulk streaming motions such as rotation.  The physical processes
that drive the formation, evolution and dissolution of star clusters
have a deep impact on the appearance of galaxies. This impact has many
shades, ranging from the properties of stellar populations such as the
binary fraction and the number of type~Ia and type~II supernovae,
through the velocity structure in galactic disks such as the
age--velocity dispersion relation, through to the existence of stellar
halos about galaxies, tidal streams and the survival and properties of
tidal dwarf galaxies, the existence of which challenge current
cosmological perspectives. Apart from this cosmological relevance,
dense stellar systems provide unique laboratories in which to test
stellar evolution theory, gravitational dynamics, the interplay
between stellar evolution and dynamical processes, and the physics of
stellar birth and stellar feedback processes during formation.

Star clusters and other pressure-supported stellar systems on the sky
merely offer snap shots from which we can glean incomplete
information. Because there is no analytical solution to the equations
of motion for more than two stars, these differential equations need
to be integrated numerically.  Thus, in order to proceed to an
understanding of these objects in terms of the above issues, the
researcher needs to resort to numerical experiments in order to test
various hypotheses as to the possible physical initial conditions (to
test star-formation theory) or the outcome (to quantify stellar
populations in galaxies, for example). Initialising a
pressure-supported stellar system such that the initial object is
relevant for the real physical universe is therefore a problem of some
fundamental importance.

In the following, empirical constrains on the initial conditions of
star clusters are discussed, and some problems star clusters are
relevant for are raised. Section~\ref{sec_pk:initcond} contains
information on setting-up a realistic computer model of a star
cluster, including models of embedded clusters. The initial mass
distribution of stars is discussed in section~\ref{sec_pk:IMF}, and
section~\ref{sec_pk:binpop} delves on the initial distribution functions of
multiple stars. A brief summary is provided in
section~\ref{sec_pk:summary}.

\subsection{Embedded clusters}
\label{sec_pk:embcl}\index{embedded clusters}

In this section an outline of some astrophysical aspects of dense
stellar systems is given, in order to help differentiate probable
evolutionary effects from initial conditions. A simple example
clarifies the meaning of this: An observer may see two young
populations with comparable ages (to within one~Myr, say). They have
similar observed masses but different sizes, a somewhat different
stellar content and different binary fractions. Do they signify two
different initial conditions derived from star formation, or can both
be traced back to a $t=0$ configuration which is the same?

\subsubsection{Preliminaries}
\label{sec_pk:prelim}
Assume we observe a very young population of $N$ stars with an age
$\tau_{\rm age}$, and that we have a rough estimate of its half-mass
radius, $r_{0.5}$, \index{half-mass radius} and embedded stellar mass,
$M_{\rm ecl}$.\footnote{Throughout this text all masses, $m$, $M$,
etc. are in units of $M_\odot$, unless noted otherwise. ``Embedded
stellar mass'' refers to the mass in stars at the theoretical time
before residual gas expulsion but when star formation has ceased.}
The average mass is
\begin{equation}
\overline{m} = {M_{\rm ecl} \over N}.
\end{equation}
Also assume we can estimate the star-formation efficiency
(SFE)\index{star formation efficiency}, $\epsilon$, within a
few~$r_{0.5}$ for this object,
\begin{equation}
\epsilon = {M_{\rm ecl} \over M_{\rm ecl} + M_{\rm gas}},
\label{eq_pk:sfe}
\end{equation}
where $M_{\rm gas}$ is the gas left over from the star-formation
process. The tidal radius \index{tidal radius} of the embedded cluster
can be estimated from the Jacobi limit (eq.~7-84 in \citealt{BT87}) as
determined by the host galaxy and ignoring any contributions by
surrounding molecular clouds,
\begin{equation}
r_{\rm tid} = \left( {M_{\rm ecl} + M_{\rm gas} \over 3\,M_{\rm gal}}
\right)^{1\over 3}\,D,
\label{eq_pk:rtid}
\end{equation}
where $M_{\rm gal}$ is the mass of the Galaxy within the
galacto-centric distance $D$ of the cluster. This radius is a rough
estimate of that distance from the cluster at which stellar motions
begin to be significantly influenced by the host galaxy.

The following quantities that allow us to judge the formal dynamical
state of the system: the formal crossing time \index{crossing time} of
the stars through the object can be defined as
\begin{equation}
t_{\rm cr} \equiv {2\,r_{0.5} \over \sigma},
\label{eq_pk:tcr}
\end{equation}
where\footnote{As an aside, note that $G=0.0045\,{\rm
pc}^3/M_\odot\,{\rm Myr}^2$ and that $1\,{\rm km/s} = 1.02\,{\rm
pc/Myr}$.}
\begin{equation}
\sigma = \sqrt{G\,M_{\rm ecl} \over \epsilon\,r_{0.5}}
\label{eq_pk:sigma}
\end{equation}
is, up to a factor of order unity, the three-dimensional velocity
dispersion \index{velocity dispersion} of the stars in the embedded
cluster. Note that these equations serve to estimate the possible
amount of mixing of the population: If $\tau_{\rm age}<t_{\rm cr}$
then the object cannot be mixed and we are seeing it close to its
initial morphology.  It takes a few $t_{\rm cr}$ for a dynamical
system out of dynamical equilibrium to virialise back to dynamical
equilibrium. This is not to be mistaken for a relaxation process.

Once the stars orbit within the object, they exchange orbital energy
through weak gravitational encounters and rare strong encounters, and
the system evolves towards a state of energy equipartition.  The
energy equipartition time-scale, $t_{\rm ms}$, \index{energy
equipartition time scale}\index{mass segregation time scale} between
massive and average stars (\citealt{Spitzer87}, p.74), which is an
estimate of the time massive stars need to sink to the centre of the
system through dynamical friction on the lighter stars, is
\begin{equation}
t_{\rm ms} = {\overline{m} \over m_{\rm max}}\,t_{\rm relax}.
\label{eq_pk:tms}
\end{equation}
Here, $m_{\rm max}$ is the massive-star mass, and the characteristic
two-body relaxation time \index{relaxation time}(e.g. eq.~4-9 in
\citealt{BT87}) is
\begin{equation}
t_{\rm relax} = 0.1\,{N \over {\rm ln}N} \, t_{\rm cr}.
\label{eq_pk:trelax}
\end{equation}
This formula refers to a pure $N-$body system without embedding gas. A
rough estimate of $t_{\rm relax, emb}$ for an embedded cluster can be
found in eq.~8 of \cite{AM01}. The above eq.~\ref{eq_pk:trelax} is a
measure for the time a star needs to change its orbit significantly
away from the initial trajectory and is often estimated by calculating
the amount of time that is required in order to change the velocity of
a star, $v$, by an amount $\Delta v \approx v$.

Thus, if for example, $\tau_{\rm age} > t_{\rm cr}$ and $\tau_{\rm
age}<t_{\rm relax}$, then the system is probably mixed and close to
dynamical equilibrium, \index{dynamical equilibrium} but it is not
relaxed yet, i.e., did not have sufficient time for the stars to
exchange significant amount of orbital energy. Such a cluster may have
erased its sub-structures. \index{sub clusters}

\subsubsection{Fragmentation and size} 
\index{fragmentation}\index{cluster size}

The very early stages of cluster evolution on a scale of a few~pc are
dominated by gravitational fragmentation of a turbulent magnetized
contracting molecular cloud core
\citep{Clarkeetal2000,MacLowKlessen2004,TilleyPudritz2007}. The
existing gas-dynamical simulations show the formation of contracting
filaments which fragment into denser cloud cores that form
sub-clusters of accreting proto-stars. As soon as the proto-stars emit
radiation and outflows of sufficient energy and momentum to affect the
cloud core these computations become expensive as radiative transport
and deposition of momentum and mechanical energy by non-isotropic
outflows are difficult to handle with given present computational
means \citep{Stamatellosetal2007,Daleetal2007}.\index{sub clusters}

Observations of the very early stages at times $\simless\,{\rm
few}\,10^5\,$yr suggest proto-clusters to have a hierarchical
proto-stellar distribution: a number of sub-clusters with radii
$\simless 0.2\,$pc and separated in velocity space are often seen
embedded within a region less than a~pc across
\citep{Testietal2000}. Most of these sub-clusters may merge to form a
more massive embedded cluster
\citep{ScallyClarke2002,FellhauerKroupa2005}.

It is unclear though if sub-clumps typically merge before residual gas
blow-out or if the residual gas is removed before the sub-clumps can
interact significantly, nor is it clear if there is a systematic mass
dependence of any such possible behaviour.

\subsubsection{Mass segregation} 
\label{sec_pk:ms}\index{mass segregation}

Whether or not star clusters or sub-clusters form mass-segregated
remains an open issue. Mass segregation by birth is a natural
expectation because proto-stars near the density maximum of the
cluster have more material to accrete. For these, the ambient gas is
at a higher pressure allowing proto-stars to accrete longer before
feedback termination stops further substantial gas inflow, and the
coagulation of protostars is more likely there
\citep{ZinneckerYorke2007, Bonnelletal07}. Initially mass-segregated
sub-clusters preserve mass segregation upon merging
\citep{McMillanetal2007}. However, for ${\overline{m}/ m_{\rm
max}} = 0.5/100$ and $N\simless 5\times10^3\,$stars it follows from
eq.~\ref{eq_pk:tms} that
\begin{equation}
t_{\rm ms} \simless  t_{\rm cr},
\end{equation}
i.e., a $100\,M_\odot$ star sinks to the cluster centre within
roughly a crossing time (see table~\ref{tab_pk:youngcl} below for
typical values of $t_{\rm cr}$).

Currently we cannot say conclusively if mass segregation is a birth
phenomenon (e.g. \citealt{Gouliermisetal2004}), or whether the more
massive stars form anywhere throughout the proto-cluster volume.  Star
clusters that have already blown out their gas at ages of one to a
few~Myr are typically mass-segregated (e.g. R136, Orion Nebula
Cluster).\index{R136}\index{Orion Nebula Cluster}

Affirming natal mass segregation would impact positively on the notion
that massive stars ($\simgreat 10\,M_\odot$) only form in rich
clusters, and negatively on the suggestion that they can also form in
isolation (for recent work on this topic see
\citealt{Lietal2003,ParkerGoodwin2007}).

\subsubsection{Feedback termination}
\index{star formation efficiency}

The observationally estimated SFE (eq.~\ref{eq_pk:sfe}) is
\citep{LadaLada2003}
\begin{equation}
0.2 \simless \epsilon \simless 0.4
\label{eq_pk:obs_sfe}
\end{equation}
implying that the physics dominating the star-formation process on
scales less than a few~pc is stellar feedback.  Within this volume,
the pre-cluster cloud core contracts under self gravity thereby
forming stars ever more vigorously, until feedback energy suffices to
halt the process ({\it feedback-termination}).

\subsubsection{Dynamical state at feedback termination} 
\label{sec_pk:dynstate}
Each proto-star needs about $t_{\rm ps}\approx10^5\,$yr to accumulate
about 95~per cent of its mass \citep{WuchterlTscharnuter2003}. The
proto-stars form throughout the pre-cluster volume as the
proto-cluster cloud core contracts. The overall pre-cluster cloud-core
contraction until feedback-termination takes
(eqs~\ref{eq_pk:tcr},~\ref{eq_pk:sigma})
\begin{equation}
t_{\rm cl,form}\approx {\rm few} \times {2\over \sqrt{G}}\,\left({M_{\rm
ecl}\over \epsilon}\right)^{-{1\over2}}\,r_{0.5}^{3\over2},
\end{equation}
(a few times the crossing time), which is about the time over which
the cluster forms.  Once a proto-star condenses out of the
hydro-dynamical flow it becomes a ballistic particle moving in the
time-evolving cluster potential.  Because many generations of
proto-stars can form over the cluster-formation time-scale and if the
crossing time through the cluster is a few times shorter than $t_{\rm
cl,form}$, then the very young cluster is mostly virialised when star
formation stops and at the removal of the residual gas.\footnote{A
brief transition time $t_{\rm tr} \ll t_{\rm cl, form}$ exists during
which the star-formation rate decreases in the cluster while the gas
is being blown out, but for the purpose of the present discussion this
time may be neglected.}  It is noteworthy that for $r_{0.5}=1\,$pc
\begin{equation}
t_{\rm ps}\simgreat t_{\rm cl,form} \quad {\rm for} \quad {M_{\rm
ecl}\over \epsilon} \simgreat 10^{4.9}\,M_\odot
\end{equation}
(the proto-star formation time formally surpasses the cluster
formation time) which is near the turnover mass in the
old-star-cluster mass function (e.g. \citealt{Baumg1998}).

A critical parameter is thus the ratio
\begin{equation}
\tau = {t_{\rm cl,form}\over t_{\rm cr}}.
\end{equation} \index{cluster formation time scale}
If it is less than unity then proto-stars condense from the gas and
cannot virialise in the potential before the residual gas is
removed. Such embedded clusters may be kinematically cold if the
pre-cluster cloud core was contracting, or hot if the pre-cluster
cloud core was pressure confined, because the young stars do not feel
the gas pressure.

In those cases where $\tau > 1$ the embedded cluster is approximately
in virial equilibrium because generations of proto-stars that drop out
of the hydrodynamic flow have time to orbit the potential.  The
pre-gas-expulsion stellar velocity dispersion in the embedded cluster
(eq.~\ref{eq_pk:sigma}) may reach $\sigma=40\,$pc/Myr if $M_{\rm ecl} =
10^{5.5}\,M_\odot$ which is the case for $\epsilon\,r_{0.5} <
1\,$pc. This is easily achieved since the radius of one-Myr old
clusters is $r_{\rm 0.5}\approx 0.8\,$pc with no dependence on
mass. Some observationally explored cases are discussed by
\citet{Kroupa2005}. Notably, using K-band number counts,
\citet{Gutermuthetal2005} appear to find evidence for expansion after
gas removal.

Interestingly, recent Spitzer results suggest a scaling of the
characteristic projected radius $R$ with mass,\footnote{Throughout
this text, projected radii are denoted by $R$, while the 3D radius is
$r$.}
\begin{equation}
M_{\rm ecl}\propto R^2
\label{eq_pk:obsR}
\end{equation} 
\citep{Allenetal2007}, so the question how compact embedded clusters
form and whether there is a mass--radius relation \index{mass--radius
relation} needs further clarification. Note though that such a scaling
is obtained for a stellar population that expands freely with a
velocity given by the velocity dispersion in the embedded cluster
(eq.~\ref{eq_pk:sigma}),
\begin{equation}
r(t) \approx r_{\rm o} + \sigma\,t \quad \Longrightarrow \quad M_{\rm ecl} =
{1\over G} \, \left({r(t) - r_o \over t}\right)^2,
\end{equation}
where $r_{\rm o} \simless 1\,$pc is the birth radius of the cluster.
Is the observed scaling then a result of expansion from a compact
birth configuration after gas expulsion? If so, then it would require
more-massive systems to be dynamically older, which is at least
qualitatively in-line with the dynamical time-scales decreasing with
mass.  Note also that the observed scaling (eq.~\ref{eq_pk:obsR})
cannot carry through to $M_{\rm ecl}\simgreat 10^4\,M_\odot$ because
the resulting objects would not resemble clusters.

There are two broad camps suggesting on the one hand side that
molecular clouds and star clusters form on a free-fall time-scale
\index{free-fall time} \citep{Elmegreen2000,Hartmann2003,Elm07} and on
the other that many free-fall times are needed
\citep{KrumholzTan2007}. The former implies $\tau \approx 1$ while the
latter implies $\tau > 1$.

Thus, currently unclear issues concerning the initialisation of
$N$-body models of embedded clusters is the ratio $\tau$, and whether
a mass--radius relation exists for embedded clusters {\it before} the
development of HII regions. To make progress I assume for now that the
embedded clusters are in virial equilibrium at feedback termination
($\tau > 1$) and that they form highly concentrated with $r\simless
1\,$pc independently of mass.

\subsubsection{The mass of the most massive star} 
\label{sec_pk:mmax}\index{most massive star}

\begin{figure}[t]
\begin{center} 
\rotatebox{0}{\resizebox{0.8
\textwidth}{!}{\includegraphics{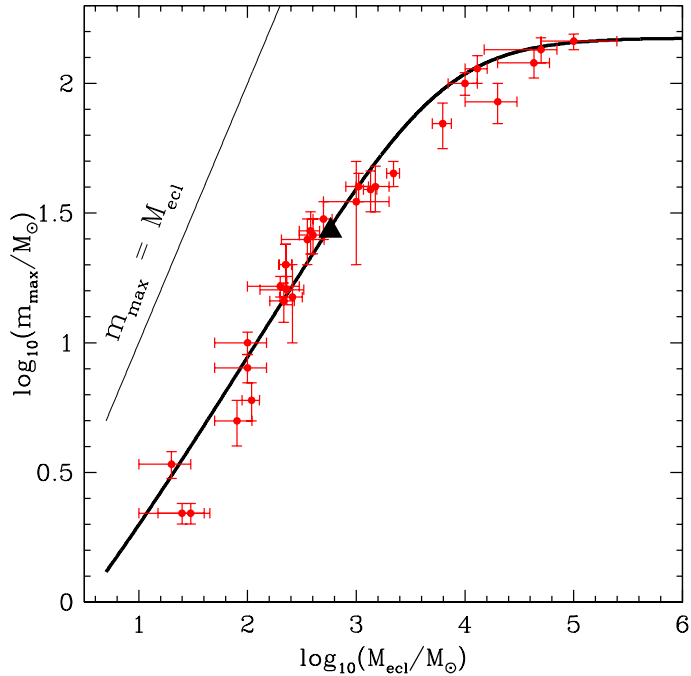}}}
\vspace{-20mm}
  \caption{\small{The maximum stellar mass, $m_{\rm max}$, as a
  function of the stellar mass of the embedded cluster, $M_{\rm ecl}$
  (Weidner, private communication: an updated version of the data
  presented in \citealt{WK06}). The solid triangle is an SPH model of
  star-cluster formation by \cite{Bonnelletal03}, while the solid
  curve stems from stating that there is exactly one most massive star
  in the cluster, $1=\int_{m_{\rm max}}^{150}\xi(m)\,dm$ with the
  condition $M_{\rm ecl} = \int_{0.08}^{m_{\rm max}}m\,\xi(m)\,dm$,
  where $\xi(m)$ is the stellar IMF. The solution can only be obtained
  numerically but an easy to use well-fitting function has been derived
  by \cite{PWK07}.  }}
\label{fig_pk:mmax_Mecl}
\end{center} 
\end{figure}

Young clusters show a well-defined correlation between the mass of the
most massive star, $m_{\rm max}$, in dependence of the stellar mass of
the embedded cluster, $M_{\rm ecl}$, which appears to saturate at
$m_{\rm max*}\approx 150\,M_\odot$ \citep{WK04, WK06}. This is
visualised in fig.~\ref{fig_pk:mmax_Mecl}.
This correlation may indicate feedback termination of star formation
within the proto-cluster volume coupled to the most massive stars
forming latest, or turning-on at the final stage of cluster formation
\citep{Elmegreen1983}. 

The evidence for a universal upper mass cutoff \index{upper stellar
mass cutoff}\index{physical maximum stellar mass} near 
\begin{equation}
m_{\rm max*}\approx 150\,M_\odot
\label{eq_pk:mmax}
\end{equation} 
\citep{WK04, Figer2005, OeyClarke2005, Koen2006, MU07,
ZinneckerYorke2007} seems to be rather well established in populations
with metallicities ranging from the LMC ($Z\approx 0.008$) to the
super-solar Galactic centre ($Z\simgreat 0.02$) such that the stellar
mass function (MF) simply stops at that mass. This mass needs to be
understood theoretically (see discussion in \citealt{KW05} and
\citealt{ZinneckerYorke2007}).  It must be a result of stellar structure
stability, but may be near $80\,M_\odot$ as predicted by theory if the
most massive stars reside in near-equal component-mass binary systems
\citep{KW05}. It may also be that the calculated stellar masses are
significantly overestimated \citep{Martinsetal2005}.

\subsubsection{The cluster core of massive stars} 

Irrespectively of whether the massive stars ($\simgreat 10\,M_\odot$)
form at the cluster centre or whether they segregate there due to
energy equipartition (eq.~\ref{eq_pk:tms}), they ultimately form a
compact sub-population that is dynamically highly unstable.  Massive
stars are ejected from such cores very efficiently on a core-crossing
time-scale, and for example the well-studied Orion Nebula cluster
(ONC) has probably already shot out 70~per cent of its stars more
massive than~$5\,M_\odot$ \citep{PfK2006}. The properties of O and B
runaway stars \index{runaway stars} have been used by
\citet{ClarkePringle1992} to deduce the typical birth configuration of
massive stars, finding them to form in binaries with similar-mass
components in compact small-$N$ groups devoid of low-mass stars. Among
others, the core of the Orion Nebula Cluster (ONC) \index{Orion Nebula
Cluster} is just such a system.

\subsubsection{The star-formation history in a cluster} 
\label{sec_pk:cl_sfh} \index{star formation history of cluster}

The detailed star-formation history in a cluster contains information
about the events that build-up the cluster. Intriguing is the recent
evidence for some clusters that while the bulk of the stars have ages
different by less than a few~$10^5\,$yr, a small fraction of older
stars are often harboured (\citealt{PallaStahler2000} for the ONC,
\citealt{Saccoetal07} for the $\sigma$~Orionis cluster).  This may be
interpreted to mean that clusters form over about 10~Myr with a final
highly accelerated phase, in support of the notion that turbulence of
a magnetized gas determines the early cloud-contraction phase
\citep{KrumholzTan2007}. 

A different interpretation would be that as a pre-cluster cloud core
contracts on a free-fall time-scale, it traps surrounding field stars
which thereby become formal cluster members: Most clusters form in
regions of a galaxy that has seen previous star formation.  The
velocity dispersion of the previous stellar generation, such as an
expanding OB association, is usually rather low, around a few km/s to
10~km/s. The deepening potential of a newly-contracting pre-cluster
cloud core will be able to capture some of the preceding generation of
stars such that these older stars become formal cluster members
although they did not form in this cluster. \citet{PfK2007} study this
problem for the ONC \index{Orion Nebula
Cluster} showing that the reported age spread \index{age
spread} by \citet{Pallaetal2007} can be accounted for in this
way. This suggests that the star-formation history of the ONC may in
fact not have started about 10~Myr ago, supporting the argument by
\citet{Elmegreen2000,Elm07} and \citet{Hartmann2003} that clusters
form on a timescale comparable to the crossing time of the pre-cluster
cloud core. Additionally, the sample of cluster stars may be
contaminated by enhanced fore- and back-ground densities of field
stars due to focussing of stellar orbits during cluster formation
\citep{PfK2007}.\index{field-star capture}\index{capture of field
stars}

For very massive clusters such as $\omega$~Cen,
\citet{Fellhaueretal2006} show that the potential is sufficiently deep
such that the pre-cluster cloud core may capture the field stars of a
previously existing dwarf galaxy. Up to 30~per cent or more of the
stars in $\omega$~Cen may be captured field stars.  This would be able
to explain an age spread of a few~Gyr in the cluster, and is
consistent with the notion that $\omega$~Cen formed in a dwarf galaxy
that was captured by the Milky Way. The attractive aspect of this
scenario is that $\omega$~Cen need not have been located at the center
of the incoming dwarf galaxy as a nucleus, but within its disk,
because it opens a larger range of allowed orbital parameters for the
putative dwarf galaxy moving about the Milky Way.  The currently
preferred scenario in which $\omega$~Cen was the nucleus of the dwarf
galaxy implies that the galaxy was completely stripped while falling
into the Milky Way leaving only its nucleus on its current retrograde
orbit \citep{Zhao04}. The new scenario allows the dwarf galaxy to be
absorbed into the Bulge of the MW with $\omega$~Cen being stripped
from it on its way in.

Another possibility for obtaining an age spread of a few~Gyr in a
massive cluster such as $\omega$~Cen is gas accretion \index{gas
accretion} from a co-moving inter-stellar medium \citep{PfK2008}. This
could only have worked for $\omega$~Cen before it became unbound from
its mother galaxy, though. That is, the cluster must have spent about
2--3$\,$Gyr in its mother galaxy before it was captured by the Milky
Way.

This demonstrates beautifully how an improved understanding of
dynamical processes on scales of a few$\,$pc impinges on problems
related to the formation of galaxies and cosmology (through the
sub-structure problem).

Finally, the increasingly well-documented evidence for stellar
populations in massive clusters with different metallicities and ages,
and in some cases even significant He enrichment, may also suggest
secondary star formation occurring from material that has been
pre-enriched from a previous generation of stars in the cluster.
Different IMFs need to be invoked for the populations of different
ages (see \citealt{Piotto08} for a review).

\subsubsection{Expulsion of residual gas:}
\label{sec_pk:gasblowout}\index{gas expulsion}\index{expulsion of residual gas}

When the most massive stars are~O stars they destroy the proto-cluster
nebula and quench further star formation by first ionising most of it
(feedback termination). The ionised gas, being now at a temperature
near $10^4$~K and in serious over-pressure, pushes out and escapes the
confines of the cluster volume with the sound speed (near $10\,$km/s)
or faster if the winds being blown off O~stars with velocities of
thousands of km/s impart sufficient momentum.

There are two analytically tractable regimes of behaviour: instantaneous gas
removal and slow gas expulsion over many crossing times:

\begin{itemize}

\item Instantaneous gas expulsion, $\tau_{\rm gas}=0$: The binding energy of
the object of mass $M$ is
\begin{equation}
E_{\rm cl,bind} = -{G\,M^2 \over r} + {1\over 2} \, M\, \sigma^2 \; <0.
\end{equation}
Before gas expulsion, $M = M_{\rm init} = M_{\rm gas} + M_{\rm ecl} \;
\longrightarrow {\rm (becomes)} \; M$, and 
\begin{equation}
\sigma_{\rm init}^2 = {G\,M_{\rm init} \over r_{\rm init}} \;
\longrightarrow \; \sigma.
\label{eq_pk:initsigma}
\end{equation}
After instantaneous gas expulsion, $M_{\rm after} = M_{\rm ecl} \;
\longrightarrow M$, but $\sigma_{\rm after} = \sigma_{\rm init}
\longrightarrow \sigma$ and we get the new binding energy 
\begin{equation}
E_{\rm cl, bind, after} = -{G\,M_{\rm after}^2 \over r_{\rm init}} +
{1\over 2} \, M_{\rm after}\, \sigma_{\rm init}^2.
\end{equation}
But the cluster virialises into a new equilibrium, such that, by the
scalar virial theorem\footnote{The scalar virial theorem: $2\,K+W=0 \;
\Longrightarrow \; E=K+W = (1/2)\,W$, where $K, W$ are the kinetic and
potential energy and $E$ is the total energy of the system.}
\begin{equation}
E_{\rm cl, bind, after} = -{1\over2} \, {G\,M_{\rm after} \over r_{\rm after}},
\end{equation}
and on equating these two expressions for the final energy and using
eq.~\ref{eq_pk:initsigma} it follows that
\begin{equation}
{r_{\rm after} \over r_{\rm init}} = {M_{\rm ecl} \over M_{\rm ecl} - M_{\rm gas}}.
\label{eq_pk:r_rat}
\end{equation}
thus, as $M_{\rm gas} \; \longrightarrow \; M_{\rm ecl}$,
i.e. $\epsilon \; \longrightarrow \; 0.5$ from above, $r_{\rm after}
\; \longrightarrow \; \infty$. This means that as the SFE approaches
50~per cent from above, the cluster unbinds. But by
eq.~\ref{eq_pk:obs_sfe}, this result would imply either (see
\citealt{KAH} and references therein)
\begin{itemize}

\item all clusters with OB stars (and thus $\tau_{\rm gas} \ll t_{\rm
cr}$) do not survive gas expulsion, or

\item the clusters expel their gas slowly, $\tau_{\rm gas} \gg t_{\rm
cr}$, which may be the case if surviving clusters such as the Pleiades
\index{Pleiades} or Hyades \index{Hyades} formed without OB stars.

\end{itemize}

\item Slow gas removal, $\tau_{\rm gas} \gg t_{\rm cr}, \; \tau_{\rm
gas} \,\longrightarrow \, \infty$: By eq.~\ref{eq_pk:r_rat} and
assuming that an infinitesimal mass of gas is removed instantaneously,
\begin{equation}
{r_{\rm init} - \delta r \over r_{\rm init}} = {M_{\rm init} - \delta M_{\rm gas}
\over M_{\rm init} - \delta M_{\rm gas} - \delta M_{\rm gas}}.
\end{equation}
For infinitesimal steps and taking for convenience $dM < 0$ but $dr > 0$,
\begin{equation}
{r - dr \over r} = {M + dM \over M + 2\,dM}.
\end{equation}
Re-arranging, 
\begin{equation}
{dr \over r} = {dM \over M}\, \left(1 - 2\,{dM \over M}\, ... \, \right),
\end{equation}
so that
\begin{equation}
{dr \over r} = {dM \over M} \; \Longrightarrow \, {\rm ln}{r_{\rm
after} \over r_{\rm init}} = {\rm ln}{M_{\rm init} \over M_{\rm
after}},
\end{equation}
upon integration of the differential equation. Thus, 
\begin{equation}
{r_{\rm after} \over r_{\rm init}} = {M_{\rm ecl} + M_{\rm gas} \over
M_{\rm ecl}} = {1\over \epsilon},
\label{eq_pk:r_adiabatic}
\end{equation}
and for example for a SFE of 20~per cent, the cluster expands by a
factor of five, $r_{\rm after} = 5\,r_{\rm init}$, without dissolving.
\end{itemize}

Table~\ref{tab_pk:youngcl} gives an overview of the type of behaviour
one might expect for clusters with increasing number of stars, $N$,
and stellar mass, $M_{\rm ecl}$ and for two characteristic radii of
the embedded stellar distribution, $r_{0.5}$. It can be seen that the
gas-evacuation time scale becomes longer than the crossing time
through the cluster for $M_{\rm ecl}\simgreat 10^5\,M_\odot$. Such
clusters would thus undergo adiabatic expansion as a result of gas
blow out. Less-massive clusters are more likely to undergo an
evolution that is highly dynamic and that can be described as an
explosion (the cluster {\it pops}).  For clusters without O~and
massive~B stars, nebula disruption probably occurs on the
cluster-formation time-scale, $\approx 10^6\,$yr, and the evolution is
again adiabatic.
\begin{table}
\begin{tabular}{|p{8mm}p{15mm}p{15mm}p{15mm}p{13mm}p{13mm}p{13mm}p{13mm}|}
\hline
&$M_{\rm ecl}/M_\odot$ &N &O stars? 
&$t_{\rm cr}/$Myr &$\tau_{\rm gas}/t_{\rm cr}$
&$t_{\rm cr}/$Myr &$\tau_{\rm gas}/t_{\rm cr}$\\
&&&($r_{0.5}=$ &$0.5\,$pc &$0.5\,$pc &$1\,$pc &$1\,$pc)\\
\hline
&40 &100                  &N    &0.9   &--    &2.6  &--  \\
&100 &250                 &Y/N  &0.6   &0.08  &1.6  &0.2 \\
&500 &1250                &Y    &0.3   &0.2   &0.7  &0.1 \\
&$10^3$ &$2.5\times10^3$  &Y    &0.2   &0.25  &0.5  &0.2 \\
&$10^4$ &$2.5\times10^4$  &Y    &0.06  &0.8   &0.2  &0.5 \\
&$10^5$ &$2.5\times10^5$  &Y    &0.02  &2.5   &0.05 &2   \\
&$10^6$ &$2.5\times10^6$  &Y    &0.006 &8.3   &0.02 &5   \\
\hline
\end{tabular}
\caption{Notes: O~stars $=$~``Y'' if the maximum stellar mass in the
cluster surpasses $8\,M_\odot$ (fig.~\ref{fig_pk:mmax_Mecl}); the
average stellar mass is taken to be $\overline{m}=0.4\,M_\odot$ in all
clusters; a star-formation efficiency of $\epsilon=0.3$ is assumed;
the crossing time, $t_{\rm cr}$, is eq.~\ref{eq_pk:tcr}; the
pre-supernova gas evacuation time-scale is $\tau_{\rm gas}=r/v_{\rm
th}$, where $v_{\rm th}=10\,$km/s is the approximate sound velocity of
the ionised gas: $\tau_{\rm gas} = 0.05\,$Myr for $r=0.5\,$pc, while
$\tau_{\rm gas} = 0.1\,$Myr for $r=1\,$pc.}
\label{tab_pk:youngcl}
\end{table}
A simple calculation of the amount of energy deposited by an O~star
into its surrounding cluster-nebula suggests it to be larger than the
nebula binding energy \citep{Kroupa2005}. This, however, only gives at
best a rough estimate of the rapidity with which gas can be expelled;
an inhomogeneous distribution of gas leads to the gas removal
occurring preferably along channels and asymmetrically, such that the
overall gas-excavation process is highly non uniform and variable
\citep{Daleetal2005}.

\begin{figure}
\vspace{-30mm}
\hspace{-20mm}
\rotatebox{0}{\resizebox{1.2 \textwidth}{!}{\includegraphics{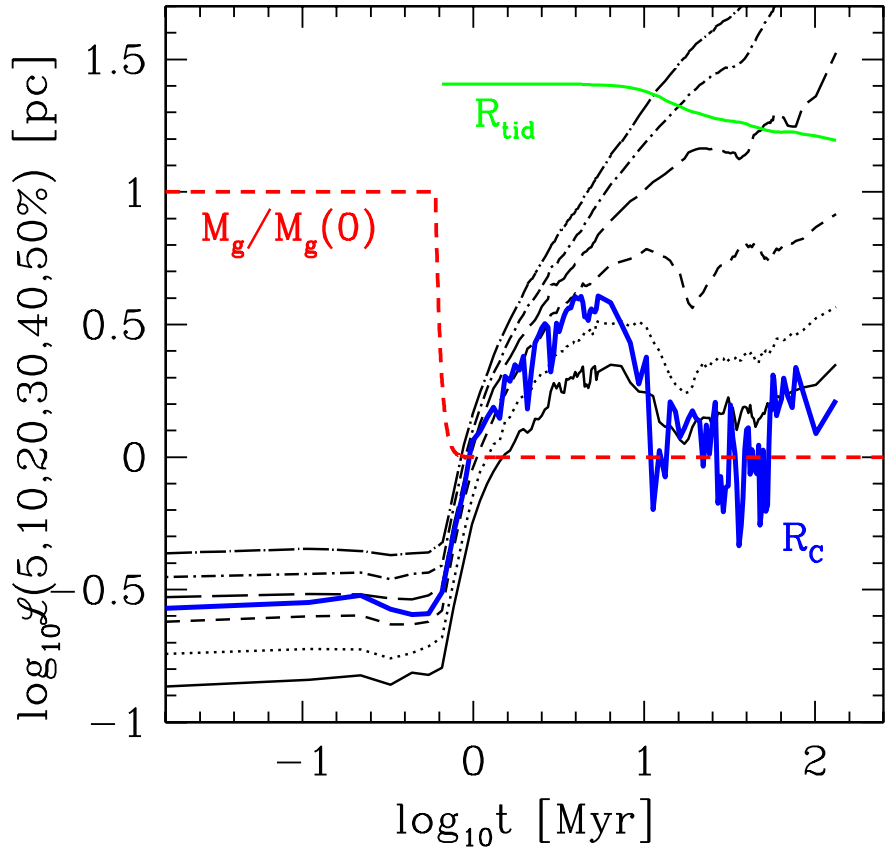}}}
\vskip -40mm
\caption{\small{ The evolution of the 5, 10, 20..., 50~per cent
Lagrange radii and the core radius ($R_c = r_c$, thick lower curve) of
the ONC-type cluster discussed in the text. The gas-mass is shown as
the dashed line: the cluster spends $0.6\,$Myr in an embedded phase
before the gas is blown out on a thermal time-scale. The tidal radius
(eq.~\ref{eq_pk:rtid}) is shown by the upper thick solid curve. From
\cite{KAH}.}}
\label{fig_pk:lagr}
\end{figure}

The reaction of clusters to gas expulsion is best studied numerically
with $N-$body codes. The pioneering such experiments were performed by
\cite{Tutukov78} followed by \cite{LMD84}, and \citep{Goodwin97a,
Goodwin97b, Goodwin98} studied gas expulsion by supernovae from young
globular clusters. Fig.~\ref{fig_pk:lagr} shows the evolution of an
ONC-type initial cluster with a stellar mass $M_{\rm ecl}\approx
4000\,M_\odot$ and a canonical IMF (eq.~\ref{eq_pk:canonIMF}) and
stellar evolution, a 100~per cent initial binary population
(section~\ref{sec_pk:initialpop_lowmass}) in a solar-neighbourhood
tidal field, $\epsilon=1/3$, and spherical gas blow-out on a thermal
time-scale ($v_{\rm th} = 10\,$km/s).
The figure demonstrates that the evolution is far more complex than
the simple analytical estimates above suggest, and in fact a
substantial Pleiades-type \index{Pleiades} cluster emerges after
loosing about $2/3$rd of the initial stellar population (see also
p.~\pageref{sec_pk:assoc}).  Subsequent theoretical work based on an
iterative scheme according to which the mass of unbound stars at each
radius is removed successively shows that the survival of a cluster
depends not only on $\epsilon, \tau_{\rm gas}/t_{\rm cr}$ and $r_{\rm
tid}$, but also on the detailed shape of the stellar distribution
function \citep{BoilyKroupa03}. For instantaneous gas removal,
$\epsilon \approx 0.3$ is a lower limit for the SFE below which
clusters cannot survive rapid gas blow-out. This is significantly
smaller than the critical value of $\epsilon = 0.5$ below which the
stellar system becomes formally unbound (eq.~\ref{eq_pk:r_rat}).
However, if clusters form as complexes of sub-clusters, each of which
pop individually, then overall cluster survival is enhanced to even
smaller values of $\epsilon \approx 0.2$ \citep{FellhauerKroupa2005}.

{\it If} clusters pop, and which fraction of stars remain in a
post-gas expulsion cluster, depends critically on the ratio between
the gas-removal time scale and the cluster crossing time. This ratio
thus mostly defines which clusters succumb to {\it infant
mortality}\index{infant mortality}, and which clusters merely suffer
{\it cluster infant weight loss}\index{infant weight loss}.  The
well-studied observational cases do indicate that the removal of most
of the residual gas does occur within a cluster-dynamical time,
$\tau_{\rm gas}/t_{\rm cr} \simless 1$.  Examples noted
\citep{Kroupa2005} are the ONC \index{Orion Nebula Cluster} and R136
\index{R136} in the LMC both having significant super-virial velocity
dispersions. Other examples are the Treasure-Chest cluster and the
very young star-bursting clusters in the massively-interacting
Antennae galaxy which appear to have HII regions expanding at
velocities such that the cluster volume may be evacuated within a
cluster dynamical time. However, improved empirical constraints are
needed to further develop an understanding of cluster survival.  Such
observations would best be the velocities of stars in very young star
clusters, as they should show a radially expanding stellar population.

Indeed, \citet{BastianGoodwin2006} note that many young clusters have
a radial-density profile signature expected if they are expanding
rapidly, supporting the notion of fast gas blow out.  For example, the
$0.5-2\,$Myr old ONC, which is known to be super-virial with a virial
mass about twice the observed mass \citep{HillenbrandHartmann98}, has
already expelled its residual gas and is expanding rapidly thereby
probably having lost its outer stars \citep{KAH}. The super-virial
state of young clusters makes measurements of their mass-to-light
ratia a bad measure of the stellar mass within them
\citep{GoodwinBastian2006}, and rapid dynamical mass-segregation
likewise makes naive measurements of the $M/L$ ratia wrong
\citep{Boilyetal05, Flecketal06}.  \cite{GoodwinBastian2006} and
\cite{degrijs_parm07} find the dynamical mass-to-light ratia of young
clusters to be too large strongly implying they are in the process of
expanding after gas expulsion.  

\citet{Weidneretal2007} attempted to measure infant weight loss by
using a sample of young but exposed Galactic clusters and applying the
maximal-star-mass vs cluster mass relation from above to estimate the
birth mass of these clusters. The uncertainties are large, but the
data firmly suggest that the typical cluster looses at least about
50~per~cent of its stars.

\subsubsection{Binary stars} 
\index{binary stars}

Most stars form as binaries with, as far as can be stated today,
universal orbital distribution functions
(section~\ref{sec_pk:binpop}). Once a binary system is born in a dense
environment, it is perturbed which changes its eccentricity and
semi-major axis, or it undergoes a relatively strong encounter which
disrupts the binary or hardens it perhaps with exchanged
companions. The initial binary population therefore evolves on a
cluster crossing time scale, and most soft binaries are disrupted. It
has been shown that the properties of the Galactic field binary
population can be explained in terms of the binary properties observed
for very young populations if these go through a dense cluster
enviroment ({\it dynamical population
synthesis},\citealt{K95d}). \index{dynamical population synthesis} A
dense cluster environment hardens existing binaries
(p.\pageref{law_pk:heggie_hills}) which increases the SNIa rate in a
galaxy with many dense clusters \citep{SharaHurley02}.

Binaries have been realised to be significant energy sources (see also
section~\ref{sec_pk:intro_bin}): a hard binary that interacts via a
resonance with a cluster field star ejects one star with a velocity
$v_{\rm ej}\gg \sigma$. The ejected star either leaves the cluster
causing cluster expansion such that $\sigma$ drops, or it shares some
of its kinetic energy with the other cluster field stars through
gravitational encounters causing cluster expansion. Binaries in a
cluster core can thus halt and reverse core collapse
\citep{MeylanHeggie97, HeggieHut2003}.

\subsubsection{Mass loss from evolving stars} 
\index{mass loss}\index{stellar evolution mass loss}

An old globular cluster with a turn-off mass near $0.8\,M_\odot$ will
have lost 30~per cent of the mass that remained in it after gas
expulsion due to stellar evolution \citep{BaumgardtMakino2003}. As the
mass loss is most rapid during the earliest times after
re-virialisation after gas expulsion, the cluster expands further
during this time. This is nicely seen in the Lagrange radii of
realistic cluster-formation models \citep{KAH}.

\subsection{Some implications for the astrophysics of galaxies}
\label{sec_pk:impl}\index{astrophysics of galaxies} 

In general, the above have a multitude of implications 
for galactic and stellar astrophysics:

\begin{enumerate}

\item The heaviest-star---star-cluster-mass correlation constrains
feedback models of star cluster formation \citep{Elmegreen1983}.  It
also implies that by adding up all IMFs in all young clusters in a
galaxy, the {\it integrated galaxial initial mass function} (IGIMF) is
steeper than the invariant stellar IMF observed in star clusters with
important implications for the mass--metallicity relation of galaxies
\citep{KWK07}. Additionally, star-formation rates (SFRs) of dwarf
galaxies can be underestimated by up to three orders of magnitude
because H$\alpha$-dark star formation becomes possible
\citep{PWK07}. This indeed constitutes an important example of how
sub-pc processes influence the physics on cosmological scales.

\item \label{point_pk:ML} The deduction that type~II clusters probably
pop (p.~\pageref{sec_pk:gasblowout}) implies that young clusters will
appear to an observer to be super-virial, i.e. to have a dynamical
mass larger than the luminous mass
\citep{BastianGoodwin2006,degrijs_parm07}.

\item \label{point_pk:thick} It also implies that galactic fields can
be heated, and may also lead to galactic thick-disks and stellar halos
around dwarf galaxies \citep{Kroupa2002}.

\item \label{point_pk:cmf} The variation of the gas expulsion
time-scale among clusters of different type implies that the
star-cluster mass function (CMF) is re-shaped rapidly, on a time-scale
of a few ten~Myr \citep{KroupaBoily2002}.

\item \label{point_pk:popII} Associated with this re-shaping of the
CMF is the natural production of population~II stellar halos during
cosmologically early star-formation bursts
\citep{KroupaBoily2002,ParmGil2007,BKP08}.

\item The properties of the binary-star population observed in
Galactic fields are shaped by dynamical encounters in star clusters
before the stars leave their cluster (section~\ref{sec_pk:binpop}).

\end{enumerate}

Points~\ref{point_pk:ML}--\ref{point_pk:popII} are considered in more
detail in the rest of~section~\ref{sec_pk:intro}.

\subsubsection{Stellar associations, open clusters and moving groups}
\label{sec_pk:assoc}\index{associations}\index{open clusters}\index{moving groups}

As one of the important implications of point~\ref{point_pk:ML}, a
cluster in the age range $1-50\,$Myr will have an unphysical $M/L$
ratio because it is out of dynamical equilibrium rather than having an
abnormal stellar IMF \citep{BastianGoodwin2006,degrijs_parm07}.

Another implication is that a Pleiades-like \index{Pleiades} open
cluster would have been born in a very dense ONC-type configuration
and that, as it evolves, a {\it moving-group-I} is established during
the first few dozen~Myr which comprises roughly 2/3rd of the initial
stellar population and is expanding outwards with a velocity
dispersion which is a function of the pre-gas-expulsion configuration
\citep{KAH}.  These computations were in fact the first to
demonstrate, using high-precision $N$-body modelling, that the
re-distribution of energy within the cluster during the embedded phase
and during the expansion phase leads to the formation of a substantial
remnant cluster despite the inclusion of all physical effects that are
disadvantageous for this to happen (explosive gas expulsion, low SFE
$\epsilon = 0.33$, Galactic tidal field and mass loss from stellar
evolution and an initial binary-star fraction of 100~per cent, see
fig.~\ref{fig_pk:lagr}).  Thus, expanding OB associations may be
related to star-cluster birth, and many OB associations ought to have
remnant star clusters as nuclei (see also \citealt{Clarketal2005}).

As the cluster expands becoming part of an OB association, the
radiation from its massive stars produce expanding HII regions that
may trigger further star formation in the vicinity
(e.g. \citealt{Gouliermisetal2007}).

A {\it moving-group-II} establishes later -- the {\it classical}
moving group made-up of stars which slowly diffuse/evaporate out of
the re-virialised cluster remnant with relative kinetic energy close
to zero. The velocity dispersion of moving group~I is thus comparable
to the pre-gas-expulsion velocity dispersion of the cluster, while
moving group~II has a velocity dispersion close to zero.

\subsubsection{The velocity dispersion of  galactic-field populations and 
galactic thick disks}
\label{sec_pk:heat}\index{thick disk}\index{galactic disk morphology}

Thus, the moving-group-I would be populated by stars that carry the
initial kinematical state of the birth configuration into the field of
a galaxy.  Each generation of star clusters would, according to this
picture, produce overlapping moving-groups-I (and~II), and the overall
velocity dispersion of the new field population can be estimated by
adding in quadrature all expanding populations. This involves an
integral over the embedded-cluster mass function, $\xi_{\rm
ecl}(M_{\rm ecl})$, which describes the distribution of the stellar
mass content of clusters when they are born.  Because the embedded
cluster mass function is known to be a power-law, this integral can be
calculated for a first estimate \citep{Kroupa2002, Kroupa2005}.  The
result is that for reasonable upper cluster mass limits in the
integral, $M_{\rm ecl}\simless10^5\,M_\odot$, the observed
age--velocity dispersion relation of Galactic field stars can be
re-produced.

This theory can thus explain the much debated {\it energy deficit}:
\index{age--velocity dispersion relation for Galactic disk stars}
namely that the observed kinematical heating of field stars with age
could not, until now, be explained by the diffusion of orbits in the
Galactic disk as a result of scattering on molecular clouds, spiral
arms and the bar \citep{Jenkins1992}. Because the velocity-dispersion
for Galactic-field stars increases with stellar age, this notion can
also be used to map the star-formation history of the Milky-Way disk
by resorting to the observed correlation between the star-formation
rate in a galaxy and the maximum star-cluster mass born in the
population of young clusters \citep{WKL2004}.

An interesting possibility emerges concerning the origin of thick
disks. If the star formation rate was sufficiently high about 11~Gyr
ago, then star clusters in the disk with masses up to
$10^{5.5}\,M_\odot$ would have been born.  If they popped a thick disk
with a velocity dispersion near 40~km/s would result naturally
\citep{Kroupa2002}. This notion for the origin of thick disks appears
to be qualitatively supported by the observations of
\citet{Elmegreenetal04} who find galactic disks at a redshift
between~0.5 and~2 to show massive star-forming clumps.

\subsubsection{Structuring the initial cluster mass function}
\label{sec_pk:clmfn}\index{cluster mass function}

Another potentially important implication from this theory of the
evolution of young clusters is that {\it if} the
gas-expulsion-time-to-crossing-time ratio and/or the SFE varies with
initial (embedded) cluster mass, then an initially featureless
power-law mass function of embedded clusters will rapidly evolve to
one with peaks, dips and turnovers at cluster masses that characterize
changes in the broad physics involved.

As an example, \cite{Adams00} and \cite{KroupaBoily2002} assumed that
the function
\begin{equation}
M_{\rm icl} = f_{\rm st}\,M_{\rm ecl}
\end{equation}
exists, where $M_{\rm ecl}$ is as above, $M_{\rm icl}$ is the
{\it classical initial cluster mass} and 
\begin{equation}
f_{\rm st} = f_{\rm st}(M_{\rm
ecl}). 
\end{equation}
According to \cite{KroupaBoily2002}, the classical initial cluster
mass is that mass which is inferred by classical $N$-body computations
without gas expulsion (i.e. in effect assuming $\epsilon=1$, which is
however, unphysical). Thus, for example, for the Pleiades, $M_{\rm
cl}\approx 1000\,M_\odot$ \index{Pleiades} at the present time (age
about 100~Myr). A classical initial model would place the initial
cluster mass near $M_{\rm icl}\approx 1500\,M_\odot$ by using standard
$N$-body calculations to quantify the secular evaporation of stars
from an initially bound and virialised {\it classical} cluster
\citep{Portetal2001}. If, however, the SFE was 33~per cent and the
gas-expulsion time-scale was comparable to or shorter than the cluster
dynamical time, then the Pleiades would have been born in a compact
configuration resembling the ONC \index{Orion Nebula Cluster} and with
a mass of embedded stars of $M_{\rm ecl}\approx 4000\,M_\odot$
\citep{KAH}.  Thus, $f_{\rm st}(4000\,M_\odot) = 0.38 \,
(=1500/4000)$.

By postulating that there exist three basic types of embedded clusters
\citep{KroupaBoily2002}, namely
\begin{description}

\item Type~I: clusters without O~stars ($M_{\rm
ecl}\simless 10^{2.5}\,M_\odot$, e.g.  Taurus-Auriga pre-main sequence
stellar groups, $\rho$~Oph),

\item Type~II: clusters with a few O~stars ($10^{2.5} \simless M_{\rm
ecl}/M_\odot \simless 10^{5.5}$, e.g. the ONC),

\item Type~III: clusters with many O~stars and with a velocity
dispersion comparable to or higher than the sound velocity of ionized
gas ($M_{\rm ecl}\simgreat 10^{5.5}\,M_\odot$), \index{Type I--III clusters}

\end{description}
it can be argued that $f_{\rm st}\approx 0.5$ for type~I, $f_{\rm
st}<0.5$ for type~II and $f_{\rm st}\approx 0.5$ for type~III. The
reason for the high $f_{\rm st}$ values for types~I and~III is that
gas expulsion from these clusters may be longer than the cluster
dynamical time because there is no sufficient ionizing radiation for
type~I clusters, or the potential well is too deep for the ionized gas
to leave (type~III clusters). The evolution is therefore adiabatic
(eq.~\ref{eq_pk:r_adiabatic} above).  Type~II clusters undergo a
disruptive evolution and witness a high {\it infant mortality rate}
\citep{LadaLada2003}, therewith being the pre-cursors of OB
associations and Galactic clusters.  This broad categorisation has
easy-to-understand implications for the star-cluster mass function.
\index{infant weight loss}\index{infant mortality}

Under these conditions and an assumed functional form for $f_{\rm
st}=f_{\rm st}(M_{\rm ecl})$, the power-law embedded cluster mass
function transforms into a cluster mass function with a turnover near
$10^5\,M_\odot$ and a sharp peak near $10^3\,M_\odot$
\citep{KroupaBoily2002}. This form is strongly reminiscent of the
initial globular cluster mass function which is inferred by
e.g. \cite{Vesperini1998,Vesperini2001,ParmGil2005,Baumg1998} to be
required for a match with the evolved cluster mass function that is
seen to have a universal turnover near $10^5\,M_\odot$. By the
reasoning given above, this ``initial'' CMF is, however, unphysical,
being a power-law instead.

This analytical formulation of the problem has been verified nicely
using $N$-body simulations combined with a realistic treatment of
residual gas expulsion by \citet{BKP08}, who show the Milky-Way
globular cluster mass function to emerge from a power-law
embedded-cluster mass function. \citet{Parmetal2008} expand on this by
studying the effect that different assumptions on the physics of gas
removal have on shaping the star-cluster mass function within about
50~Myr.

The general ansatz that residual gas expulsion plays a dominant role
in early cluster evolution may thus bear the solution to the
long-standing problem that the deduced initial cluster mass function
needs to have this turnover, while the observed mass functions of
young clusters are feature-less power-law distributions.

\subsubsection{The origin of population~II stellar halos}
\label{sec_pk:popII}\index{population II halo}

The above theory implies naturally that a major field-star component
is generated whenever a population of star clusters forms. About
$12\,$Gyr ago, the MW began its assembly by an initial burst of star
formation throughout a volume spanning about $10\,$kpc in radius. In this
volume, the star formation rate must have reached~$10\,M_\odot$/yr
such that star clusters with masses up to $\approx10^6\,M_\odot$
formed \citep{WKL2004}, probably in a chaotic, turbulent early
interstellar medium. The vast majority of embedded clusters suffered
infant weight loss or mortality, the surviving long-lived clusters
evolving to globular clusters. The so generated field population is
the spheroidal population~II halo, which has the same chemical
properties as the surviving (globular) star clusters, apart from
enrichment effects evident in the most massive clusters. All of these
characteristics emerge naturally in the above model, as pointed out by
\citet{KroupaBoily2002}, by \citet{ParmGil2007} and most recently by
\citet{BKP08}.

\subsection{Long term, or classical, cluster evolution}
\label{sec_pk:longterm} \index{cluster evolution}

The long-term evolution of star clusters that survive infant weight
loss and the mass loss from evolving stars is characterised by three
physical processes: the drive of the self-gravitating system towards
energy equipartition, stellar evolution processes and the heating or
forcing of the system through external tides.  One emphasis of
star-cluster work in this context is on testing stellar-evolution
theory and on the interrelation of stellar astrophysics with stellar
dynamics given that the stellar-evolution and the dynamical-evolution
time-scales are comparable. The reader is directed to
\cite{MeylanHeggie97} and \cite{HeggieHut2003} for further details.

\subsubsection{Tidal tails} 
\index{tidal tails}

Tidal tails contain the stars evaporating from long-lived star
clusters (the moving~group~II above). The typical S-shaped morphology
of tidal tails close to the cluster are easily understood: Stars that
leave the cluster with a slightly higher galactocentric velocity than
the cluster are on slightly outward directed galactic orbits and
therefore fall behind the cluster as the angular velocity about the
galactic centre decreases with distance. The outward directed trailing
arm develops.  Stars that leave the cluster with slower galactocentric
velocities than the cluster fall towards the galaxy and overtake the
cluster.

Given that energy equipartition leads to a filtering in energy space
\index{energy equipartition} of the stars that escape at a particular
time, one expects a gradient in the stellar mass function progressing
along a tidal tail towards the cluster such that the mass function
becomes flatter, i.e. richer in more massive stars. This effect is
difficult to detect, but for example the long tidal tails found
emanating from Pal~5 \citep{Odenkirchenetal2003} may show evidence for
this.

As emphasised by \citet{Odenkirchenetal2003}, tidal tails have another
very interesting use: they probe the gravitational potential of the
Milky Way if the differential motions along the tidal tail can be
measured. They are thus important future tests of gravitational
physics. \index{Galactic potential}

\subsubsection{Death and hierarchical multiple stellar systems} 
\label{sec_pk:death} \index{cluster death} \index{multiple stellar systems}
\index{hierarchical stellar systems}\index{cluster remnants}

Nothing lasts forever, and star clusters that survive initial
re-virialisation after residual gas expulsion and mass loss from
stellar evolution ultimately cease existing after evaporating all
member stars leaving a binary or a long-lived highly hierarchical
multiple system composed of near-equal mass components
\citep{delaFuenteMarcos1997,delaFuenteMarcos1998}. Note that these
need not be stars.  These cluster remnants are interesting, because
they may account for most of the hierarchical multiple stellar systems
in the Galactic field \citep{GoodwinKroupa2005} with the implication
that they would not be a product of star formation, but rather of
star-cluster dynamics.

\subsection{What is a galaxy?}
\label{sec_pk:galaxy} \index{dwarf spheroidal galaxy}\index{dwarf elliptical galaxy}
\index{galactic bulges}\index{elliptical galaxies}

Star clusters, dwarf-spheroidal (dSph) and dwarf-elliptical (dE)
galaxies as well as galactic bulges and giant elliptical (E) galaxies
are all stellar-dynamical systems that are supported by random stellar
motions, i.e. they are pressure-supported. But why is one class of
these pressure supported systems referred to as star clusters, while
the others are galaxies? Is there some fundamental physical difference
between these two classes of systems?

\begin{figure}
\begin{center} 
\rotatebox{0}{\resizebox{1.0
\textwidth}{!}{\includegraphics{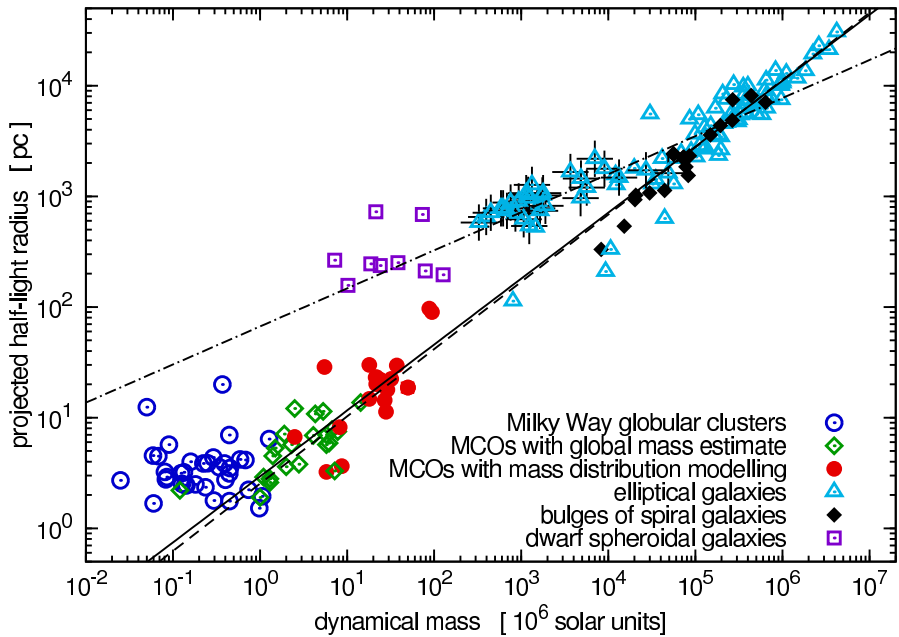}}}
\vspace{-5mm}
  \caption{\small{Mass-radius data in dependence of the dynamical mass
of pressure-supported stellar systems (from \citealt{DHK08}). MCOs are
massive compact objects (also referred to as ultra compact dwarf
galaxies). The solid and dashed lines refer to eq.~\ref{eq_pk:MRR},
while the dash-dotted line is a fit to dSph and dE galaxies.  }}
\label{fig_pk:MRR}
\end{center} 
\end{figure}

Considering the radius as a function of mass, it becomes apparent that
systems with $M\simless 10^6\,M_\odot$ do not show a mass--radius
relation (MRR) and have $r\approx 4\,$pc.  More massive objects,
however, show a well-defined MRR. In fact, \citet{DHK08} find that the
{\it massive compact objects} (MCOs), which have $10^6\simless
M/M_\odot \simless 10^8$, lie on the MRR of giant~E galaxies ($\approx
10^{13}\,M_\odot$) down to normal~E galaxies ($10^{11}\,M_\odot$), as
is evident in fig.~\ref{fig_pk:MRR}: \index{ultracompact dwarf
galaxies} \index{massive compact objects}\index{mass-radius relation}
\begin{equation}
R/{\rm pc} = 10^{-3.15}\,\left({M\over M_\odot}\right)^{0.60\pm0.02}.  
\label{eq_pk:MRR}
\end{equation}

\begin{figure}
\begin{center} 
\rotatebox{0}{\resizebox{1.0
\textwidth}{!}{\includegraphics{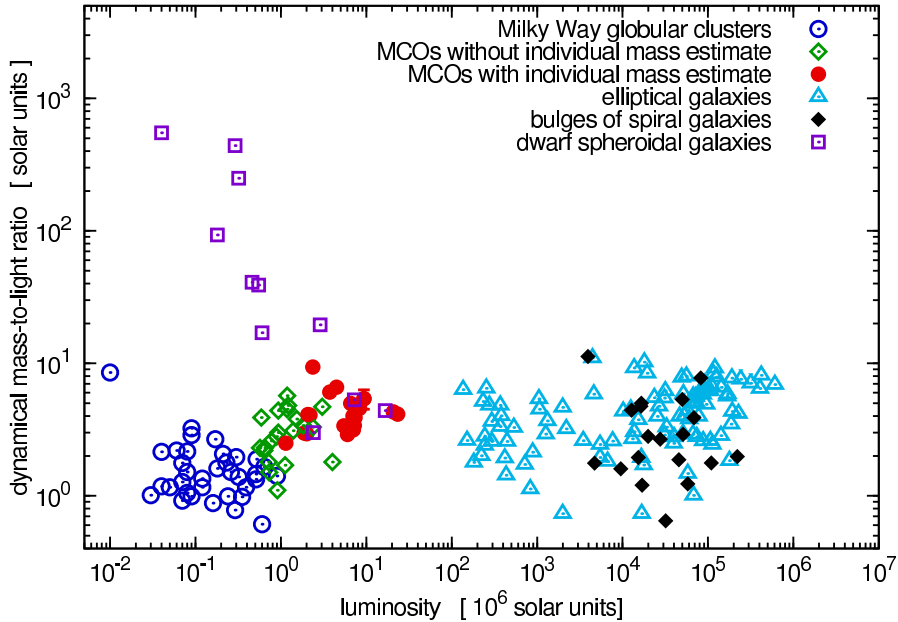}}}
\vspace{-5mm}
  \caption{\small{Dynamical $M/L$ values in dependence of the V-band
luminosity of pressure-supported stellar systems (from
\citealt{DHK08}). MCOs are massive compact objects (also referred to
as ultra compact dwarf galaxies).  }}
\label{fig_pk:ML}
\end{center} 
\end{figure}

Noteworthy is that systems with $M\simgreat 10^6\,M_\odot$ also sport
complex stellar populations, while less massive systems have
single-age, single-metallicity populations.  Remarkably,
\cite{PfK2008} show that a stellar system with $N\simgreat
10^6\,M_\odot$ and a radius as observed for globular clusters can
accrete gas from a co-moving warm inter-stellar medium and may
re-start star formation.  The median two-body relaxation time is
longer than a Hubble time for $M \simgreat 3\times 10^6\,M_\odot$, and
{\it only} for these systems is there evidence for a slight increase
in the dynamical mass-to-light ratio. Intriguingly, $(M/L)_V \approx
2$ for $M<10^6\,M_\odot$, while $(M/L)_V \approx 5$ for
$M>10^6\,M_\odot$ with a possible decrease for $M>10^8\,M_\odot$
(fig.~\ref{fig_pk:ML}).\index{relaxation time}\index{mass-to-light
ratios}
Finally, the average stellar density maximises at $M=10^6\,M_\odot$
with about $3\times10^3\,M_\odot/{\rm pc}^3$ \citep{DHK08}.\index{density}

Thus, 

\begin{itemize}

\item the mass $10^6\,M_\odot$ appears to be special,

\item stellar populations become complex above this mass,

\item evidence for some dark matter {\it only} appears in systems that have
a median two-body relaxation time longer than a Hubble time,

\item dSph galaxies are the {\it only} stellar-dynamical systems with
$10<(M/L)_V<1000$ and as such are {\it total outliers}. 

\item $10^6\,M_\odot$ is a lower accretion limit for massive star
clusters immersed in a warm inter-stellar medium.

\end{itemize}

$M\approx10^6\,M_\odot$ therefore appears to be a characteristic mass
scale such that less-massive objects show characteristics of star
clusters being well-described by Newtonian dynamics, while more
massive objects show behaviour more typical of galaxies. Defining a
galaxy as a stellar-dynamical object which has a median two-body
relaxation time longer than a Hubble time, i.e. essentially a system
with a smooth potential, may be an objective and useful way to define
a {\it galaxy} \citep{Kr98}.

Why {\it only smooth} systems show evidence for dark matter remains at
best a striking coincidence, at worst it may be symptomatic of a
problem in understanding dynamics in such systems.\index{dark matter}

\newpage

\section{Initial 6D conditions}
\label{sec_pk:initcond}

The previous section gave an outline of some of the issues at stake in
the realm of pressure-supported stellar systems. In order to attack
these and other problems, we need to know how to set-up such systems
in the computer. Indeed, as much as analytical solutions may be
preferred, the mathematical and physical complexities of dense stellar
systems leave no alternatives other than to resort to full-scale
numerical integration of the $6N$ coupled first-order differential
equations that describe the motion of the system through
$6N$-dimensional phase-space.  There are three related questions to
ponder: Given a well-developed cluster, how is one to set it up in
order to evolve it forward in time?  How does a cluster form, and how
does the formation process affect its later properties? How do we
describe a realistic stellar population (IMF, binaries)? Each of these
is dealt with in the following sections.

\subsection{6D structure of classical clusters}
\label{sec_pk:classical}\index{structure of star clusters}

Because the state of a star cluster is never known exactly, it is
necessary to perform numerical experiments with conditions that are,
statistically, consistent with the cluster snapshot. To ensure
meaningful statistical results for systems with only a few stars, say
$N<5000$, many numerical renditions of the same object are thus
necessary. For example, systems with $N=100$~stars evolve erratically
and numerical experiments are required to map out the range of
possible states at a particular time: the range of half-mass radii at
an age of 20~Myr in 1000~numerical experiments of a cluster initially
with $N=100$~stars and with an initial half-mass radius $r_{\rm
0.5}=0.5\,$pc can be compared with an actually observed object for
testing consistency with the initial conditions. Excellent recent
examples of this approach can be found in \cite{Hurleyetal05,
Portetal07}, with a recent review available by \cite{Hutetal07}, and
two text books have been written dealing with computational and more
general aspects of the physics of dense stellar systems
\citep{Aarseth2003, HeggieHut2003}.

The six-dimensional structure of a pressure-supported stellar system
at time $t$ is conveniently described by the phase-space distribution
function, $f(\vec{r},\vec{v}; t)$, \index{phase-space distribution
function} where $\vec{r}, \vec{v}$ are the phase-space variables, and
\begin{equation}
dN=f(\vec{r},\vec{v}; t)\,d^3x\,d^3v
\end{equation} 
is the number of stars in 6D phase-space volume element $d^3x\,d^3v$.
In the case of a steady-state, the Jeans theorem (\citealt{BT87},
their chapter~4.4) allows us to express $f$ in terms of the integrals
of motion, i.e. the energy and angular momentum.  The phase-space
distribution function can then be written
\begin{equation}
f = f(\vec{r},\vec{v}) = f(\epsilon_e,l),
\end{equation}
where 
\begin{equation}
\epsilon_e = {1\over2}\,v^2 + \Phi(\vec{r})
\end{equation}
is the specific energy of a star, and 
\begin{equation}
l = |\vec{r} \times \vec{v}|
\end{equation}
is the specific orbital angular momentum of a star. The Poisson equation is
\begin{equation}
\nabla^2\Phi(\vec{r}) = 4\,\pi\,G\,\rho_m(\vec{r}) = 4\,\pi\,G\;\int\,m\,f\,d^3v,
\end{equation}
or in spherical symmetry,
\begin{equation}
{1\over r^2}{d\over dr}\left(r^2\,{d\Phi \over dr}\right) = 4\, \pi \,
G\, \int \, f_m\left({1\over2}v^2 + \Phi, |\vec{r} \times
\vec{v}|\right)\,d^3v,
\end{equation}
where $f_m$ is the phase-space mass-density of all matter and is equal
to $m\,f$ for a system with equal-mass stars.  Most pressure-supported
systems have a near-spherical shape and so in most numerical work it
is convenient to assume spherical symmetry.

For convenience it is useful to introduce the relative potential \index{relative potential}
\footnote{The following discussion is based on \cite{BT87}.},
\begin{equation}
\Psi \equiv -\Phi + \Phi_0
\end{equation}
and the relative energy \index{relative energy}
\begin{equation}
{\cal E} \equiv -\epsilon_e + \Phi_0 = \Psi - {1\over2}v^2, 
\end{equation}
where $\Phi_0$ is a constant such that $f>0$ for ${\cal E}>0$ and
$f=0$ for ${\cal E}\le 0$. The Poisson equation becomes $\nabla^2\Psi
= -4\,\pi\,G\, \rho_m$ subject to the boundary condition $\Psi
\longrightarrow \Phi_0$ as $\vec{r} \longrightarrow \infty$.

One important property of stellar systems is the anisotropy of their
velocity distribution function. Defining the anisotropy parameter
\begin{equation}
\beta(r) \equiv 1 - { \overline{v_\theta^2} \over \overline{v_r^2}},
\end{equation}\index{velocity anisotropy parameter}
where $\overline{v_\theta^2}, \overline{v_r^2}$ are the mean squared
tangential and radial velocities at a particular location $\vec{r}$,
respectively. It follows that systems with $\beta=0$ everywhere have
an isotropic velocity distribution function. 

If $f$ only depends on the energy then the mean squared radial and
tangential velocities are, respectively,
\begin{equation}
\overline{v_r^2} = {1\over\rho} \, \int_{\rm all\,vel.}\, v_r^2 \, 
f\left[\Psi - {1\over2}\left(v_r^2+v_\theta^2+v_\phi^2\right)\right]\;
dv_r\,dv_\theta\,dv_\phi
\label{eq_pk:beta1}
\end{equation}
and
\begin{equation}
\overline{v_\theta^2} = {1\over\rho} \, \int_{\rm all\,vel.}\, v_\theta^2 \, 
f\left[\Psi - {1\over2}\left(v_r^2+v_\theta^2+v_\phi^2\right)\right]\;
dv_r\,dv_\theta\,dv_\phi.
\label{eq_pk:beta2}
\end{equation}
If the labels $\theta$ and $r$ are exchanged in eq.~\ref{eq_pk:beta2},
it can be seen that one arrives at
eq.~\ref{eq_pk:beta1}. Eq.~\ref{eq_pk:beta1} and~\ref{eq_pk:beta2} are
thus identical, apart from the labeling. Thus if $f=f({\cal E})$,
$\beta=0$ and the velocity distribution function is isotropic.

If $f$ depends on the energy and the orbital angular momentum of the
stars ($|\vec{l}|=|\vec{r} \times \vec{v}|$) then the mean squared
radial and tangential velocities are, respectively,
\begin{equation}
\overline{v_r^2} = {1\over\rho} \, \int_{\rm all\,vel.}\, v_r^2 \, 
f\left[\Psi - {1\over}\left(v_r^2+v_\theta^2+v_\phi^2\right), 
r\sqrt{v_\theta^2 + v_\phi^2}\right]\;
dv_r\,dv_\theta\,dv_\phi
\label{eq_pk:beta3}
\end{equation}
and
\begin{equation}
\overline{v_\theta^2} = {1\over\rho} \, \int_{\rm all\,vel.}\, v_\theta^2 \, 
f\left[\Psi - {1\over}\left(v_r^2+v_\theta^2+v_\phi^2\right);
r\sqrt{v_\theta^2 + v_\phi^2}\right]\;
dv_r\,dv_\theta\,dv_\phi.
\label{eq_pk:beta4}
\end{equation}
If the labels $\theta$ and $r$ are exchanged in eq.~\ref{eq_pk:beta4},
it can be seen that this time one does not arrive at
eq.~\ref{eq_pk:beta3}. Thus if $f=f({\cal E},l)$ then $\beta\ne0$ and
the velocity distribution function is not isotropic.

This serves to demonstrate an elementary but useful property of the
phase-space distribution function.

A very useful series of distribution functions can be arrived at from
the following simple form:
\begin{equation}
f_m({\cal E}) = 
\left\{ 
\begin{array}{r@{\quad:\quad}l}
F\,{\cal E}^{n-{3\over2}} & {\cal E}>0,\\
0 & {\cal E} \le 0.
\end{array}
\right.
\label{eq_pk:power_law_df}
\end{equation}
The mass density,
\begin{equation}
\rho_m(r) = 4\,\pi\,F\;\int_0^{\sqrt{2\,\Psi}} \left(\Psi-{1\over2}v^2
\right)^{n-{3\over2}}\,v^2\,dv,
\end{equation}
where the upper integration bound is given by the escape condition,
${\cal E}=\Psi-(1/2)v^2=0$.  Substituting $v^2=2\,\Psi\,{\rm
cos}^2\theta$ for some $\theta$ leads to 
\begin{equation}
\rho_m(r) = \left\{ 
\begin{array}{r@{\quad:\quad}l}
c_n\,\Psi^n & \Psi>0,\\
0 & \Psi \le 0.
\end{array}
\right.
\end{equation}
For $c_n$ to be finite, $n>1/2$, i.e. {\it homogeneous ($n=0$) systems
are excluded}.\index{homogeneous systems}

The {\it Lane-Emden equation} follows from the spherically symmetric
\index{Lane-Emden equation} Poisson equation after introducing
dimensionless variables $s=r/b, \psi = \Psi/\Psi_0$, where
$b=(4\,\pi\,G\,\Psi_0^{n-1}\,c_n)^{-1/2}$ and $\Psi_0=\Psi(0)$,
\begin{equation}
{1\over s^2}{d\over ds}\left(s^2 {d\psi \over ds} \right) =
\left\{ 
\begin{array}{r@{\quad:\quad}l}
-\psi^n & \psi>0,\\
0 & \psi \le 0.
\end{array}
\right.
\label{eq_pk:lane_emden}
\end{equation}
H.~Lane and R.~Emden worked with this equation in the context of
self-gravitating polytropic gas spheres which have an equation of
state
\begin{equation}
p=K\,\rho_m^\gamma, 
\label{eq_pk:polytrope}
\end{equation}
where $K$ is a constant and $p$ the pressure. It can be shown that
$\gamma=1 + 1/n$. i.e. that the density distribution of a stellar
polytrope of index $n$ is the same as that of a polytropic gas sphere
with index $\gamma$.

The natural boundary conditions to be imposed on
eq.~\ref{eq_pk:lane_emden} are at $s=0$,
\begin{enumerate}
\label{eq_pj:bcs}
\item $\psi=1$ since $\Psi(0)=\Psi_0$, and
\item $d\psi/ds=0$ because the gravitational force must vanish at the centre.
\end{enumerate}

Analytical solutions to the Lane-Emden equation are possible only for
a few values of $n$, remembering that a homogeneous ($n=0$) stellar
density distribution has already been excluded as a viable solution of
the general power-law phase-space distribution function.

\subsubsection{The Plummer model}
\label{sec_pk:Plummer}\index{Plummer model}

A particularly useful case is 
\begin{equation}
\psi = {1 \over \sqrt{1+{1\over3}\,s^2}}.
\end{equation}
It follows immediately that this is a solution of the Lane-Emden
equation for $n=5$, and it also satisfies the two boundary conditions
above, and therewith constitutes a physically sensible potential.  By
integrating the Poisson equation it can be shown that the total mass
of this distribution function is finite, 
\begin{equation}
M_\infty = \sqrt{3}\,\Psi_0\,b/G, 
\end{equation}
although the density distribution has no boundary. 
The distribution function is
\begin{equation}
f_m({\cal E}) = 
\left\{ 
\begin{array}{r@{\quad:\quad}l}
F\,\left(\Psi-{1\over2}\,v^2\right)^{7\over2} & v^2<2\Psi,\\
0 & v^2\ge 2\Psi,
\end{array}
\right.
\end{equation}
with the relative potential 
\begin{equation}
\Psi = {\Psi_0 \over \sqrt{ 1+{1\over3} \left( {r \over b} \right)^2}}
\end{equation}
and density law
\begin{equation}
\rho_m = {\rho_{m,0} \over 
\left({1+{1\over3} \left({r\over b}\right)^2}\right)^{5\over2}}
\label{eq_pk:plummer_rho}
\end{equation}
with the above total mass. This density distribution is known as the
{\it Plummer model} named after \cite{Plummer11} where he showed that
the density distribution resulting from this model provides a
reasonable, and in particular very simple analytical description of
globular clusters. The Plummer model is, in fact, a work-horse for
many applications in stellar dynamics because many of its properties
such as the projected velocity dispersion profile can be calculated
analytically. Such formulae are useful for checking numerical codes
used to set-up models of stellar systems.

\subsubsection{Properties of the Plummer model}
\label{sec_pk:plummer_prop}\index{Plummer model: properties}

Some useful analytical results can be derived for the Plummer density
law (see also \citealt{HeggieHut2003}, their p.~73, for another
compilation).

For the Plummer law of mass $M_{\rm ecl}$ the mass-density profile
(eq.~\ref{eq_pk:plummer_rho}) can be written
\begin{equation}
\rho_m(r) = {3\,M_{\rm ecl}\over 4\,\pi\,r_{\rm pl}^3} 
          {1\over \left[1 + \left( {r\over r_{\rm pl}} \right)^2\right]^{5\over2}}.
\label{eq_pk:pldens}
\end{equation}
The central number density is thus
\begin{equation}
\rho_{\rm c} = {3\,N \over 4\,\pi\,r_{\rm pl}^3}.
\label{eq_pk:rhoc}
\end{equation}
The mass within radius $r$ follows from $M(r)=4\,\pi\,\int_0^r
\rho_m(r')\, r'^2\, dr'$,
\begin{equation}
M(r) = M_{\rm ecl}\, 
       {\left({r\over r_{\rm pl}}\right)^3 \over 
\left[1 + \left( {r\over r_{\rm pl}} \right)^2 \right]^{3\over2}}.
\label{eq_pk:plmass}
\end{equation}
Thus, 
\begin{description}
\itemsep=-4mm
\item $r_{\rm pl}$ contains 35.4~per cent of the mass, \\
\item $2\,r_{\rm pl}$ contain 71.6~per cent, \\
\item $5\,r_{\rm pl}$ contain 94.3~per cent and \\
\item $10\,r_{\rm pl}$ contain 98.5~per cent of the total mass.
\end{description}
The half-mass radius contains 50~per cent of the mass, 
\begin{equation}
r_{0.5} = (2^{2\over 3}-1)^{-{1\over 2}}\,r_{\rm pl} \approx
1.305\,r_{\rm pl}.
\label{eq_pk:rhalf_pl}
\end{equation}

The projected surface mass density, $\Sigma_M(R) = 2\,\int_0^\infty
\rho_m(r)\,dz$, where $R$ is the projected radial distance from the
cluster centre and $Z$ is the integration variable along the
line-of-sight ($r^2=R^2+Z^2$), is
\begin{equation}
\Sigma_\rho(R) = {M_{\rm ecl}\over \pi\,r_{\rm pl}^2} 
          {1 \over \left[ 1 + \left({R\over r_{\rm pl}}\right)^2 \right]^2}.
\label{eq_pk:surfdens_pl}
\end{equation}
Assume there is no mass segregation so that the mass-to-light ratio,
$\Upsilon \equiv (M/L)$, measured in some photometric system is
independent of radius. The integrated light within projected radius
$R$ is \index{integrated projected light profile}
\begin{equation}
I(R) = (1/\Upsilon)\, \int_0^R \Sigma_\rho(R')\,2\, \pi\,R'\,dR',
\end{equation}
\begin{equation}
I(R) = {M_{\rm ecl}\,r_{\rm pl}^2 \over \Upsilon} 
          \left[ {1 \over r_{\rm pl}^2} - {1\over R^2+r_{\rm pl}^2}\right].
\label{eq_pk:surfI_pk}
\end{equation}
Thus, $r_{\rm pl}$ is the half-light radius of the projected star
cluster, $I(r_{\rm pl}) = 0.5\,I(\infty)$.

In the above equations $\rho(r)=\rho_m(r)/\overline{m}$,
$N(r)=M(r)/\overline{m}$ and $\Sigma_n=\Sigma_\rho/\overline{m}$ are,
respectively, the stellar number density, the number of stars within
radius $r$ and the projected surface number density profile if there
is no mass segregation within the cluster, the average stellar mass,
$\overline{m}$, therefore not being a function of radius.

The velocity dispersion can be calculated at any radius from
Jeans\index{velocity dispersion profile} eq.~\ref{eq_pk:jeans}.  For
an isotropic velocity distribution ($\sigma_\theta^2 = \sigma_\phi^2 =
\sigma_r^2$), such as the Plummer model, the Jeans equation yields
\begin{equation}
\sigma_r^2(r) = {1\over \rho(r)}\, \int_r^\infty \rho(r') 
              \,\,{G\,M(r')\over r^2}\,\,dr',
\label{eq_pk:sigr_pl}
\end{equation}
since $d\phi(r)/dr = GM(r)/r^2$, and the integration bounds have been
chosen to make use of the vanishing $\rho_m(r)$ as
$r\rightarrow\infty$.  Note that the above equation is also valid if
$M(r)$ consists of more than one spherical component such as a
distinct core plus an extended halo. Combining eqs~\ref{eq_pk:pldens},
\ref{eq_pk:plmass} and~\ref{eq_pk:sigr_pl} leads to
\begin{equation}
\sigma^2(r) = \left({G\,M_{\rm ecl} \over 2\,r_{\rm pl}}\right) {1\over
\left[1+ \left( {r\over r_{\rm pl}} \right)^2\right]^{1\over 2}},
\label{eq_pk:sigpl}
\end{equation}
where $\sigma(r)$ is the three-dimensional velocity dispersion of the
Plummer sphere at radius $r$, $\sigma^2(r) = \sum_{k=r,\theta,\phi}
\sigma_k^3(r)$ or $\sigma^2(r) = 3\,\sigma_{\rm 1D}^2(r)$ 
since isotropy is assumed.

A star with mass $m$ positioned at $r$ and with {\it speed} $v =
\left( \sum_{k=1}^3 v_{k}^2 \right)^{1/2}$ can escape from the cluster if
it has a total energy $e_{\rm bind} = e_{\rm kin} + e_{\rm pot} =
0.5\,m\,v^2 + m\,\phi(r) \ge 0$ such that $v\ge v_{\rm esc}(r)$,
implying for the escape speed at radius $r$, $v_{\rm esc}(r) =
\sqrt{2\,|\phi(r)|}$. The potential at $r$ is given by the mass within
$r$ plus the potential contributed by the surrounding matter which is
calculated by integrating the contributions from each radial mass
shell,
\begin{eqnarray}
\phi(r) &=& - \left[ G\,{M(r)\over r} 
          + \,\int_r^\infty G\, {1\over r'}
            \,\,\rho(r')\,4\,\pi\,r'^2\,dr' \right], \nonumber\\
        &=& - \left( {G\,M_{\rm ecl} \over r_{\rm pl}} \right) 
            {1\over \left[1 + (r/r_{\rm pl})^2\right]^{1/2}}.
\label{eq_pk:phi}
\end{eqnarray}
so that 
\begin{equation}
v_{\rm esc}(r) = \left( {2\,G\,M_{\rm ecl} \over r_{\rm pl}} \right)^{1/2} 
            {1\over \left[1 + (r/r_{\rm pl})^2\right]^{1/4}}.
\label{eq_pk:escv_pl}
\end{equation}

The circular speed, $v_c$, of a star moving on a circular orbit at
a distance $r$ from the cluster centre is obtained from centrifugal
acceleration, $v_c^2/r = d\phi(r)/dr = G\,M(r)/r^2$,
\begin{equation}
v_c^2 = \left( {G\,M_{\rm ecl}\over r_{\rm pl}} \right) 
       {\left(r/r_{\rm pl}\right)^2 \over [1 + (r/r_{\rm pl})^2]^{3/2}}.
\label{eq_pk:vcirc_pl}
\end{equation}

In many but not all instances of interest the initial cluster model is
chosen to be in the state of {\it virial equilibrium}. That is, the
kinetic and potential energies of each star balance such that the
whole cluster is stationary. The scalar virial theorem, \index{scalar
virial theorem} \index{virial equilibrium}
\begin{equation}
2\,K + W = 0,
\end{equation}
where $K$ and $W$ are the total kinetic and
potential energy of the cluster,\footnote{Eq.3.251.4 on p.295 in
\cite{GR80} are useful to solve the integrals for the Plummer sphere}
\begin{eqnarray}
K &=& {1\over2}\,\int_0^\infty \rho(r)\, 
              \sigma^2(r)\, 4\pi r^2dr, \nonumber \\
            &=& {3\pi \over 64}\,{G\,M_{\rm ecl}^2 \over r_{\rm pl}}, 
\,\,\,\,{\rm for\;the\;Plummer\;sphere}, \\
W &=& {1\over2}\,\int_0^\infty\phi(r)\,\rho(r)\,4\pi r^2 dr,
              \nonumber \\
            &=& -{3\pi \over 32}\,{G\,M_{\rm ecl}^2 \over r_{\rm pl}}
\,\,\,\,{\rm for\;the\;Plummer\;sphere}.
\end{eqnarray}
The total, or binding, energy of the cluster, $E_{\rm tot} = W+K$, is
\begin{equation}
E_{\rm tot} = -K = {1\over2}\, W.  
\label{eq_pk:etot_pl}
\end{equation}

The characteristic three-dimensional velocity dispersion of a cluster
can be defined as 
$\sigma_{\rm cl}^2 \equiv 2\,K/M_{\rm ecl}$ so that 
\begin{eqnarray}
\sigma_{\rm cl}^2 
      &=& {3\,\pi \over 32} {G\,M_{\rm ecl}\over r_{\rm pl}},\\
      &\equiv& {G\,M_{\rm ecl} \over r_{\rm grav}},
\label{eq_pk:sigmacl_pk}\\
      &\equiv& s^2\,\left({G\,M_{\rm ecl}\over 2\,r_{0.5}}\right)
\label{chonc_eq:1dvd}
\end{eqnarray}
introducing the {\it gravitational radius} of the cluster, $r_{\rm
grav}\equiv G\,M_{\rm ecl}^2/|W|$.  For the Plummer sphere
$r_{\rm grav}=(32/3\,\pi)r_{\rm pl} = 3.4\,r_{\rm pl}$, and 
the {\it structure factor} 
\begin{eqnarray}
s &=& \left( {6 \times 1.305\,\pi \over 32} \right)^{1\over 2},\nonumber \\
  &\approx& 0.88.
\end{eqnarray}

Defining the {\it virial ratio},\index{virial ratio}
\begin{equation}
Q = {K \over |W|},
\label{eq_pk:virrat}
\end{equation}
a cluster can initially be in three possible states:
\begin{equation}
Q \left\{
    \begin{array}{l@{\quad\quad,\quad}l}
      = {1\over2} & {\rm virial\;equilibrium},\\
      > {1\over2} & {\rm expanding},\\
      < {1\over2} & {\rm collapsing}.
    \end{array}\right.
\end{equation}
Note that if initially $Q<1/2$ the value $Q=1/2$ will be reached
temporarily during collapse, after which $Q$ increases further until
the cluster settles in virial equilibrium after this {\it violent
relaxation phase} (\citealt{BT87}, p.~271).\index{violent relaxation}

The characteristic crossing time through the Plummer cluster,
\begin{eqnarray}
t_{\rm cr} &\equiv& {2\,r_{\rm pl} \over \sigma_{\rm 1D,cl}},\\
              &=& \left({128 \over \pi\,G}\right)^{1\over2}
                  \, M_{\rm ecl}^{-{1\over2}} \, r_{\rm pl}^{3\over2},
\label{chprelim_eq:tcross}
\end{eqnarray}\index{crossing time}
using the characteristic one-dimensional velocity dispersion,
$\sigma_{\rm 1D,cl}=\sigma_{\rm cl}/\sqrt{3}$.

Observationally the {\it core radius} is that radius where the\index{core radius}
projected surface density falls to half its central value. For a real
cluster it is much easier to determine than the other characteristic
radii.  For the Plummer sphere,
\begin{equation}
R_{\rm core} = \left( \sqrt{2} - 1 \right)^{1\over2}\,r_{\rm pl} =
0.64\,r_{\rm pl},
\label{eq_pk:rcorepl_pl}
\end{equation}
from eq.~\ref{eq_pk:surfdens_pl} assuming the mass-to-light ratio,
$\Upsilon$, is independent of radius. For a King model 
\begin{equation}
R_{\rm core}^{king} = \left( {9\over 4\pi\,G}\,
                      {\sigma^2\over \rho_m(0)}\right)^{1\over2},
\label{eq_pk:rcoreking}
\end{equation}
is the {\it King radius}. \index{King radius} From
eq.~\ref{eq_pk:sigpl}, $\sigma^2(0) = G\, M_{\rm ecl}/(2\,r_{\rm pl})$
and from eq.~\ref{eq_pk:pldens}, $\rho_m(0) = 3\,M_{\rm
ecl}/(4\pi\,r_{\rm pl}^3)$ so that
\begin{equation}
r_{\rm pl} = \left( {6\over 4\pi\,G}\,
                      {\sigma(0)^2\over \rho_m(0)}\right)^{1\over2} =
                      0.82\;R_{\rm core}^{\rm king}.
\end{equation}

\subsubsection{The singular isothermal model}
\label{sec_pk:singisothermal}\index{singular isothermal model}

Another useful set of distribution functions can be arrived at by
considering $n=\infty$. The Lane-Emden equation is not well defined in
this limit, but for a polytropic gas sphere eq.~\ref{eq_pk:polytrope}
implies $\gamma \longrightarrow 1$ as $n\longrightarrow \infty$. Thus
$p=K\,\rho_m$, which is the equation of state of an {\it isothermal
ideal gas} with $K=k_{\rm B}\,T/m_{\rm p}$, where $k_{\rm B}$ is
Boltzmann's constant, $T$ the temperature and $m_{\rm P}$ the mass of
a gas particle.  From the equation of hydrostatic support, $dp/dr =
-\rho_m(G\,M(r)/r^2)$, where $M(r)$ is the mass within $r$, the
following equation can be derived
\begin{equation}
{d\over dr} \left( r^2 {d{\rm ln}\rho_m \over dr} \right) = -{G\,m_{\rm p} \over
k_{\rm B}\,T}\,4\,\pi\,r^2\,\rho_m
\label{eq_pk:anequ}
\end{equation}
For a distribution function (our ansatz)
\begin{equation}
f_m({\cal E}) = {\rho_{m,1} \over \left( 2\,\pi\,\sigma^2
\right)^{3\over2}} \, e^{{\cal E}\over \sigma^2},
\label{eq_pk:isothermal_df}
\end{equation}
where $\sigma^2$ is a new quantity related to a velocity dispersion
and remembering ${\cal E} = \Psi-v^2/2$, one obtains from $\rho_m =
\int f_m({\cal E})\,4\,\pi\,v^2\,dv$
\begin{equation}
\Psi(r) = {\rm ln}\left(\rho_m(r) \over \rho_{m,1} \right)\,\sigma^2.
\end{equation}
From the Poisson equation it then follows that 
\begin{equation}
\sigma={\rm const} = {k_{\rm B} \, T \over m_p}
\end{equation}
for consistency with eq.~\ref{eq_pk:anequ}.

Therefore, the structure of an isothermal, self-gravitating ideal
sphere of gas is identical to the structure of a collisionless system
of stars whose phase-space mass-density distribution function is given
by eq.~\ref{eq_pk:isothermal_df}. Note that $f({\cal E})$ is non-zero
at all ${\cal E}$ (cf to King's models below).

The number-distribution function of velocities is $F(v) = \int_{{\rm
all}\,\vec{x}}\, f({\cal E})\,d^3x$, i.e.
\begin{equation}
F(v) = F_0\,e^{-{v^2\over 2\,\sigma^2}}.
\end{equation}
This is the Maxwell-Boltzmann distribution which results from the
kinetic theory of atoms in a gas at temperature $T$ that are allowed
to bounce off each other elastically. This exact correspondence
between a stellar-dynamical system and a gaseous polytrope holds only
for an isothermal case ($n=\infty$). 

The total number of stars in the system is $N_{\rm tot}=N_{\rm tot}\,
\int_0^\infty\, F(v)\; 4\,\pi\, v^2\,dv$ and the number of stars in
the speed interval $v$ to $v+dv$ is
\begin{equation}
dN = F(v)\,4\,\pi\,v^2\,dv = N_{\rm tot} {1\over
\left(2\,\pi\sigma^2\right)^{3\over2}} e^{-{v^2\over 2\, \sigma^2}}\,
4\, \pi\, v^2\, dv,
\end{equation}
which is the Maxwell-Boltzmann distribution of speeds. The
mean-squared speed of stars at any point in the isothermal sphere is
\[
\overline{v^2} = {4\,\pi\,\int_0^\infty \, v^2 \, F(v) \, v^2\,dv \over 
4\,\pi\,\int_0^\infty \, \, F(v) \, v^2\,dv = 3\,\sigma^2},
\]
and the one-dimensional velocity dispersion is $\sigma_{1{\rm D}} =
\sigma_\alpha = \sigma$, where $\alpha=r,\theta,\phi, x, y, z, ...$.

To obtain the radial mass-density of this model, the ansatz
$\rho_m=C\,r^{-b}$ together with Poisson's equation
(eq.~\ref{eq_pk:anequ}) implies
\begin{equation}
\rho_m(r) = {\sigma^2 \over 2\,\pi\,G}\,{1\over r^2},
\label{eq_pk:isothermal_density}
\end{equation}
i.e. a {\it singular isothermal sphere}.

\subsubsection{The isothermal model}
\label{sec_pk:isothermal}\index{isothermal model}

The above model has a singularity at the origin, which is
unphysical. In order to remove this problem, it is possible to force
the central density to be finite. To this end new dimensionless
variables are introduced, $\tilde{\rho_m}\equiv \rho_m / \rho_{m,0}$,
$\tilde{r} \equiv r/r_0$. $\tilde{\rho_m}$ is the finite central
density, while $r_0 = R_{\rm core}^{\rm King}$ is the {\it King
radius} (eq.~\ref{eq_pk:rcoreking}) \index{King radius}\index{core
radius} such that the projected density falls to 0.5013 (i.e. about
half) its central value. $r_0$ is also sometimes called the {\it core
radius} (but see further below for King models on
p.~\pageref{coreradius}).  Poisson's equation (eq.~\ref{eq_pk:anequ})
therewith becomes
\begin{equation}
{d \over d\tilde{r}}\,\left( \tilde{r}^2\,{d{\rm ln}\tilde{\rho_m} \over d\tilde{r}}
\right) = -9\,\tilde{\rho_m}\, \tilde{r}^2.
\label{eq_pk:anequ2}
\end{equation}
This differential equation must be solved numerically for
$\tilde{\rho_m}(\tilde{r})$ subject to the boundary conditions (as before),
\begin{equation}
\tilde{\rho_m}(\tilde{r}=0) = 1, \quad {d\tilde{\rho_m} \over
d\tilde{r}}|_{\tilde{r}=0} = 0.
\label{eq_pk:bc2}
\end{equation}
The solution is the {\it isothermal sphere}. 

By imposing physical reality (central non-singularity) onto our
mathematical ansatz we end-up with a density profile which cannot be
arrived at analytically but only numerically. The isothermal density
sphere must be tabulated in the computer with entries such as 
\begin{equation}
r/r_0, \quad 
{\rm log}_{10}\left({\rho \over \rho_0}\right) \quad
{\rm log}_{10}\left({\Sigma \over r_0\,\rho_0}\right)
\end{equation}
where $\Sigma$ is the projected density (for example see table~4-1 and
fig.~4-7 in \citealt{BT87}). The circular velocity, $v_c(r) =
G\,M(r)/r$ of the isothermal sphere obtains by integrating Poisson's
equation (eq.~\ref{eq_pk:anequ}) from $r=0$ to $r=r'$ such that
$r^2(d{\rm ln}\rho_m /dr) = -(G/\sigma^2)\,M(r)$ and
\begin{equation}
v_c^2(r) = - \sigma^2\,{d{\rm ln}\rho_m(r) \over d{\rm ln}r}.
\end{equation}
Numerical solution of differential eq.~\ref{eq_pk:anequ2} shows that
$v_c \longrightarrow \sqrt{2}\,\sigma$~(constant) for large $r$.

The isothermal sphere is a useful model for describing elliptical
galaxies within a few core radii and disk galaxies because of the
constant rotation curve. However, combining the two equations for
$v_c^2$ above, one finds that $M(r) \approx (2\,\sigma^2/G)\,r$ for
large $r$, i.e. the isothermal sphere has an infinite mass as it is
not bounded.

\subsubsection{The lowered isothermal or King model}
\label{sec_pk:King}\index{lowered isothermal model}\index{King model}

\begin{figure}
\begin{center}
\rotatebox{0}{\resizebox{0.8 \textwidth}{!}{\includegraphics{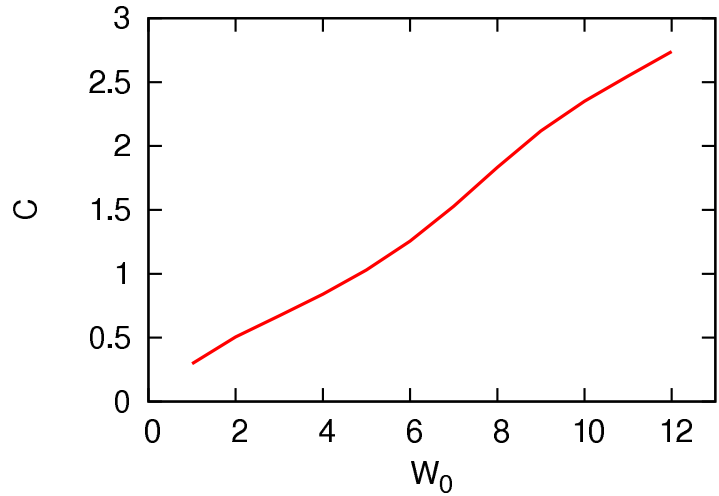}}}
\vskip -0mm
\caption{\small{The King-concentration parameter $W_o$ as a function
of $c$ (cf. with fig.~4-10 in \citealt{BT87}). This figure has been
produced by Andreas K\"upper.}}
\label{fig_pk:w0c}
\end{center}
\end{figure}

We have thus seen that the class of models with $n=\infty$ contains as
the simplest case the singular isothermal sphere. By forcing the
central density to be finite we are led to the isothermal sphere,
which however, has an infinite mass. The final model considered here
within this class is the {\it lowered isothermal model}, or the {\it
King model}\footnote{Note that \cite{King62} suggested three-parameter
(mass, core radius and cutoff/tidal radius) empirical projected (2D)
density laws that fit globular clusters very well. These do not have
information on the velocity structure of the clusters. The {\it King}
(non-analytical) 6D models that are solutions of the Jeans equation
(eq.~\ref{eq_pk:jeans} below) and discussed here, are published by
\cite{King66}.}, which forces not only a finite central density but
also a cutoff in radius. These have a distribution function similar to
that of the isothermal model, except for a cutoff in energy,
\begin{equation}
f_m({\cal E}) = 
\left\{ 
\begin{array}{r@{\quad:\quad}l}
{\rho_{m,1} \over \left( 2\,\pi\,\sigma^2
\right)^{3\over2}} \, \left( e^{{\cal E}\over \sigma^2} - 1\right) & {\cal E} > 0,\\
0 & {\cal E} \le 0.
\end{array}
\right.
\label{eq_pk:loweredisothermal_df}
\end{equation}
The density distribution becomes
\begin{equation}
\rho_m = \rho_{m,1}\,\left[ e^{\Psi \over \sigma^2} \, {\rm
erf}\left({\sqrt{\Psi} \over \sigma}\right) - \sqrt{4\,\Psi \over \pi\,\sigma^2}\,
\left(1 + {2\,\Psi \over 3\sigma^2}\right)\right]
\end{equation}
integrating only to ${\cal E}=0$ as before.
The Poisson eq.~\ref{eq_pk:anequ} becomes 
\begin{equation}
{d \over d\tilde{r}}\,\left( \tilde{r}^2\,{d{\rm ln}\tilde{\rho_m} \over d\tilde{r}}
\right) 
= -4\,\pi\,G\,\rho_{m,1}\,r^2\,\left[
e^{\Psi \over \sigma^2} \, {\rm
erf}\left({\sqrt{\Psi} \over \sigma}\right) - \sqrt{4\,\Psi \over \pi\,\sigma^2}\,
\left(1 + {2\,\Psi \over 3\sigma^2}\right)\right].
\label{eq_pk:anequ3}
\end{equation}
Again, this differential equation must be solved numerically for $\Psi(r)$
subject to the boundary conditions,
\begin{equation}
\Psi(0), \quad {d\Psi \over dr}\left.\right|_{r=0} = 0.
\end{equation}
The density vanishes at $r=r_{\rm tid}$ (the tidal radius), where
\index{tidal radius} $\Psi(r=r_{\rm tid})=0$ also. A King model is
thus limited in mass, has a finite central density but the parameter
$\sigma$ is not the velocity dispersion but is rather related to the
depth of the potential via the {\it concentration parameter}
\begin{equation}
W_o \equiv {\Psi(0) \over \sigma^2}. 
\end{equation}
The {\it concentration} is defined as \index{concentration}
\begin{equation}
c\equiv {\rm log}_{10}\left({r_{\rm tid} \over r_o}\right).
\end{equation}
For globular clusters, $3 < W_o < 9,\; 0.75 < c < 1.75$, and the
relation between $W_o$ and $c$ is plotted in fig.~\ref{fig_pk:w0c}.
Note also that the true {\it core radius} defined as $\Sigma(R_c) =
(1/2)\,\Sigma(0)$, where $\Sigma(R)$ is the projected density profile
and $R$ is the projected radius, is unequal in general to the {\it
King radius}, $r_0$ (eq.~\ref{eq_pk:rcoreking}). \index{King
radius}\index{core radius}\label{coreradius}

Finally, it should be emphasised that it is not physical to use an
arbitrary $r_{\rm tid}$: the tidal radius must always match the value
dictated by the cluster mass and the host galaxy
(e.g. eq.~\ref{eq_pk:rtid}).\index{tidal radius}

\subsection{Comparison: Plummer vs King models}
\label{sec_pk:Plummer_King}\index{Plummer model}\index{King model}
\index{Plummer model, King model: comparison}\index{King model,
Plummer model: comparison}

The above discussion has served to show how various popular models can
be followed through from a power-law distribution function
(eq.~\ref{eq_pk:power_law_df}) with different indices $n$. The Plummer
model (p.~\pageref{sec_pk:Plummer}) and the King model
(p.~\pageref{sec_pk:King}) are particularly useful for describing star
clusters. The Plummer model is determined by two parameters, the mass,
$M$, and the scale radius, $r_{0.5}\approx 1.305\,r_{\rm pl}$. The
King model requires three parameters, $M$, a scale radius, $r_{0.5}$,
and a concentration parameter, $W_o$ or $c$. Which sub-set of
parameters yield models that are similar in terms of the overall
density profile?

\begin{figure}
\begin{center}
\rotatebox{0}{\resizebox{0.8 \textwidth}{!}{\includegraphics{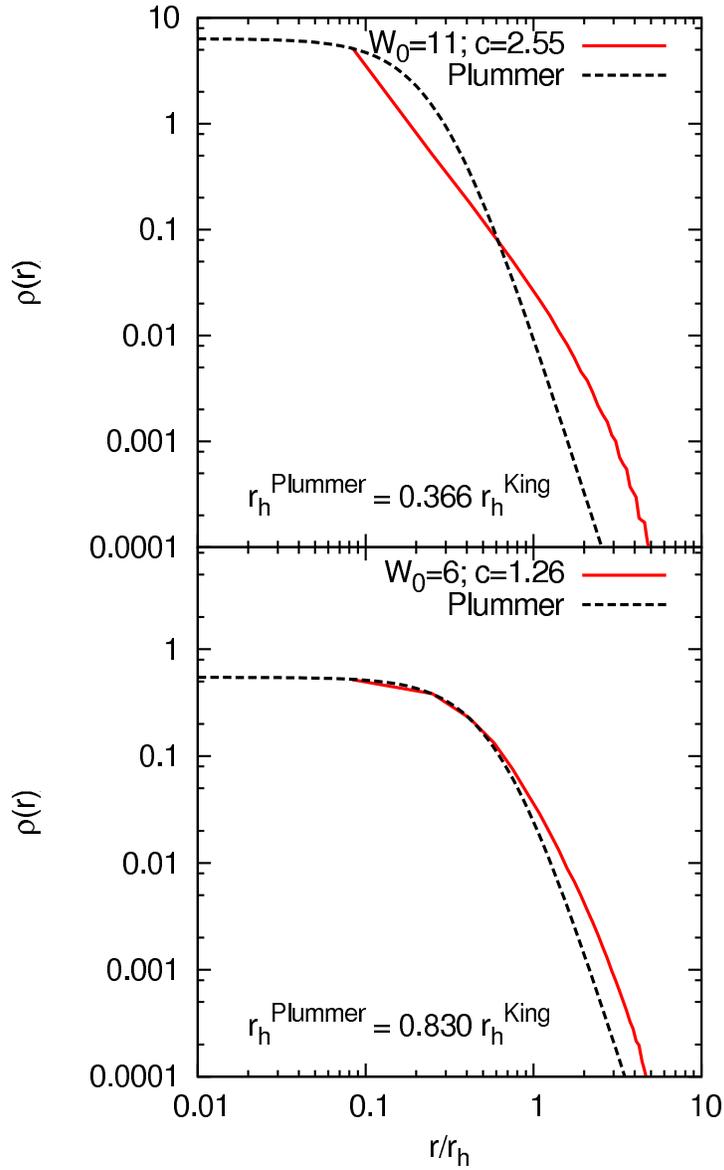}}}
\vskip -0mm
\caption{\small{Comparison of a King model (solid curve) with a
Plummer model (dashed curve). Both have the same mass, and that
Plummer model is sought which minimizes the unweighted reduced
chi-squared between the two models. The upper panel shows a
high-concentration King model with $c=2.55, W_o=11$, and the best-fit
Plummer model has $r_{0.5}^{\rm Plummer} = 0.366\,r_{0.5}^{King}\;
(r_h\equiv r_{0.5})$, as stated in the panel. The lower panel compares
the two best matching models for the case of an
intermediate-concentration King model. This figure has been produced
by Andreas K\"upper.}}
\label{fig_pk:w0bw}
\end{center}
\end{figure}

\begin{figure}
\begin{center}
\rotatebox{0}{\resizebox{0.8 \textwidth}{!}{\includegraphics{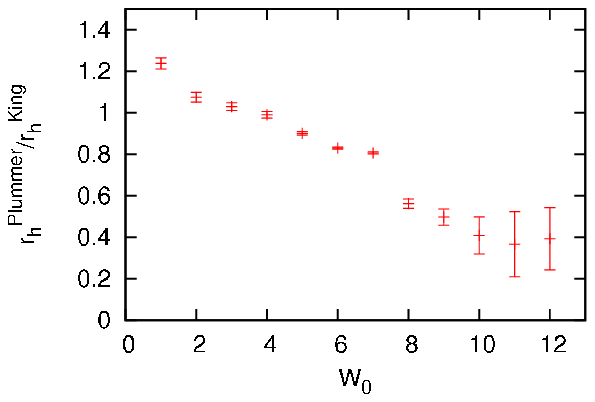}}}
\vskip -0mm
\caption{\small{The ratio $r_{0.5}^{\rm Plummer}/r_{0.5}^{\rm King}\;
(r_h\equiv r_{0.5})$ for the best-matching Plummer and King models
(fig.~\ref{fig_pk:w0bw}) are plotted as a function of the King
concentration parameter $W_o$. The uncertainties are unweighted
reduced chi-squared values between the two density profiles. It is
evident that there are no well-matching Plummer models for low-
($c<2.5$) and high-concentration ($c>7.5$) King models. This figure
has been produced by Andreas K\"upper. }}
\label{fig_pk:w0rh}
\end{center}
\end{figure}

To answer this, the mass is set to be constant. King models with
different $W_o$ and $r_{0.5}$ are computed and Plummer models are
sought that minimise the reduced chi-square value between the two
density profiles. Fig.~\ref{fig_pk:w0bw} shows two examples of
best-matching density profiles, and fig.~\ref{fig_pk:w0rh} uncovers
the family of Plummer profiles that best match King models with
different concentration. Note that a good match between the two is
only obtained for intermediate concentration King models ($2.5
\simless W_o \simless 7.5$).

\subsection{Discretisation}
\label{sec_pk:discretisation}\index{discretisation}

To set up a computer model of a stellar system with $N$ particles
(e.g. stars) the distribution functions need to be sampled $N$ times.
The relevant distribution functions are the {\it phase-space
distribution function}, the {\it stellar initial mass function} and
the {\it three distribution functions governing the properties of
binary stars} (periods, mass-ratios, eccentricities).

Assume the distribution function depends on the variable $\zeta_{\rm
min} \le \zeta \le \zeta_{\rm max}$ (e.g. stellar mass, $m$).  There
are various ways of sampling from a distribution function
\citep{NumRecipes}, but the most efficient way is to use a {\it
generating function} if one exists.  Consider the probability,
$X(\zeta)$, of encountering a value for the variable in the range
$\zeta_{\rm min}$ to $\zeta$,
\begin{equation}
X(\zeta) = \int_{\zeta_{\rm min}}^\zeta\, p(\zeta')\,d\zeta',
\label{eq_pk:Xint}
\end{equation}
with $X(\zeta_{\rm min}) = 0 \le X(\zeta) \le X(\zeta_{\rm max}) = 1$,
and $p(\zeta)$ is the distribution function normalised such that the
latter equal sign holds ($X=1$). $p(\zeta)$ is the probability
density. The inverse of eq.~\ref{eq_pk:Xint}, $\zeta(X)$, is the {\it
generating function}. \index{generating function} It is a one-to-one
map of the uniform distribution $X \in [0,1]$ to $\zeta \in
[\zeta_{\rm min}, \zeta_{\rm max}]$.  If an analytical inverse does
not exist, then it can be found numerically in a straight-forward
manner for example by constructing a table of $X,\; \zeta$ and then
interpolating this table to obtain a $\zeta$ for a given $X$.

\subsubsection{Example: The power-law stellar mass function}
\label{sec_pk:IMF_example}\index{IMF: discretisation}

As an example, consider the distribution function 
\begin{equation}
\xi(m) = k\,m^{-\alpha}, \quad \alpha=2.35; \quad 0.5 \le {m\over M_\odot} \le 150.
\end{equation}
The corresponding probability density is $p(m)=k_p\,m^{-\alpha}$, and
$\int_{0.5}^{150}\,p(m)\,dm = 1$ $\Longrightarrow \; k_p = 0.53$. Thus 
\begin{equation}
X(m) = \int_{0.5}^m\, p(m)\,dm = k_p\,{150^{1-\alpha} - 0.5^{1-\alpha} \over 1-\alpha}
\end{equation}
and the generating function for stellar masses becomes
\begin{equation}
m(X) = \left[
X {1-\alpha \over k_p} + 0.5^{1-\alpha}\right]^{1\over 1-\alpha}.
\end{equation}
It is easy to programme this into an algorithm: obtain a random
variate $X$ from a random number generator and use the above
generating function to get a corresponding mass, $m$. Repeat $N$
times.

\subsubsection{Generating a Plummer model} 
\label{eq_pk:discr_pl}\index{Plummer model: discretisation}

Perhaps the most useful and simplest model of a bound stellar system
is the Plummer model (p.~\pageref{sec_pk:Plummer}). It is worth
introducing the discretisation of this model in some detail, because
analytical formulae go a long way which is important for testing
codes. A condensed form of this material is available in \cite{AHW74}.

The mass within radius $r$ is ($r_{\rm pl} = b$ here)
\begin{equation}
M(r) = \int_0^r\,\rho_m(r')\,4\,\pi\,r'^2\,dr' = M_{\rm cl}\,
{\left(r/r_{\rm pl}\right)^3 \over \left[1+ \left(r/r_{\rm
pl}\right)^2\right]^{3\over2}}.
\end{equation}
A number uniformly distributed between zero and one can then be defined,
\begin{equation}
X_1(r) = {M(r) \over M_{\rm cl}} = {\zeta^3 \over \left[1+\zeta^2\right]}, 
\end{equation}
where $\zeta \equiv r/r_{\rm pl}$ and $X_1(r=\infty)=1$. This function
can be inverted to yield the generating function for particle
distances distributed according to a Plummer density law,
\begin{equation}
\zeta(X_1) = \left(X_1^{-{2\over3}}-1\right)^{-{1\over2}}.
\end{equation}
The coordinates of the particles, $x, y, z, r^2=(\zeta\,r_{\rm
pl})^2=x^2+y^2+z^2$, can be obtained as follows: For a given particle
we already have $r$. For all possible $x$ and $y$, $z$ has a uniform
distribution, $p(z) = {\rm const} = 1/(2\,r)$ over the range $-r \le
z \le +r$. Thus, for a second random variate between zero and one, 
\begin{equation}
X_2(z) = \int_{-r}^z\,p(z')\,dz' = {1\over 2\,r}\,\left(z+r\right),
\end{equation}
with $X_2(+r) = 1$. The generating function for $z$ becomes
\begin{equation}
z(X_2) = 2\,r\,X_2 - r.
\end{equation}
Having obtained $r$ and $z$, $x$ and $y$ can be arrived at as follows,
noting the equation for a circle, $r^2-z^2 = x^2 + y^2$: Choose a
random angle $\theta$ which is uniformly distributed over the range
$0\le \theta \le 2\,\pi$. Thus $p(\theta)=1/(2\,\pi)$ and the third
random variate becomes
\begin{equation}
X_3(\theta) = \int_0^\theta \, {1\over 2\,\pi}\,d\theta' = {\theta \over 2\,\pi}.
\end{equation}
The corresponding generating function is 
\begin{equation}
\theta(X_3) = 2\,\pi\,X_3.
\end{equation}
Finally, 
\begin{equation}
x=\left(r^2-z^2\right)^{1\over2}\,{\rm cos}\theta; \quad 
y = \left(r^2-z^2\right)^{1\over2}\,{\rm sin}\theta.
\end{equation}

The velocity for each particle cannot be obtained as simply as the
positions. In order for the initial stellar system to be in virial
equilibrium, the potential and kinetic energy need to balance according
to the scalar virial theorem. This is ensured by forcing the velocity
distribution function to be that of the Plummer model,
\begin{equation}
f_m(\epsilon_e) = 
\left\{ 
\begin{array}{c@{\quad:\quad}l}
\left( {24\,\sqrt{2} \over 2\,\pi^3}\,{r_{\rm pl}^2 \over (G\,M_{\rm cl})^5}
\right)\,\left(-\epsilon_e\right)^{7\over2}
& \epsilon_e \le 0,\\
0 & \epsilon_e > 0,
\end{array}
\right.
\end{equation}
where 
\begin{equation}
\epsilon_e(r,v) = \Phi(r) + (1/2)\,v^2 
\end{equation}
is the specific energy per star, and 
\begin{equation}
\Phi(r) = -{G\,M_{\rm cl} \over r_{\rm pl}}\, 
\left( 1+ \left({r\over r_{\rm pl}}\right)^2 \right)^{-{1\over2}}
\end{equation}
is the potential. Now, the Plummer distribution function can be
expressed in terms of $r$ and $v$,
\begin{equation}
f(r,v) = f_o\,\left(-\Phi(r) - {1\over2}\,v^2\right)^{7\over2},
\end{equation}
for a normalisation constant $f_o$ and dropping the mass sub-script
because we assume the positions and velocities do not depend on
particle mass.  With the escape speed at distance $r$ from the Plummer
centre, $v_{\rm esc}(r) = \sqrt{-2\,\Phi(r)} \equiv v/\zeta$, it
follows that
\begin{equation}
f(r,v) = f_o\,\left({1\over2}\, v_{\rm
esc}\right)^7\,\left(1-\zeta^2\right)^{7\over2}.
\end{equation}
The number of particles with speeds in the interval $v, v+dv$ is
\begin{equation}
dN = f(r,v)\,4\,\pi\,v^2\,dv \equiv g(v)\,dv.
\end{equation}
Thus
\begin{equation}
g(v) = 16\,\pi\,f_o\,\left({1\over2}\,v_{\rm
esc}(r)\right)^9\,\left(1-\zeta^2(r)\right)^{7\over2}\,\zeta^2(r),
\end{equation}
that is,
\begin{equation}
g(\zeta) = g_o\,\zeta^2(r)\,\left(1-\zeta^2(r)\right)^{7\over2},
\end{equation}
for a normalisation constant $g_o$ determined by demanding that 
\begin{equation}
X_4(\zeta=1) = 1 = \int_0^1\, g(\zeta')\,d\zeta'
\end{equation}
for a fourth random number deviate $X_4(\zeta) =
\int_0^\zeta\,g(\zeta')\,d\zeta'$. It follows that
\begin{equation}
X_4(\zeta) = {1\over2}\,\left(5\,\zeta^3 - 3\,\zeta^5\right).
\label{eq_pk:plummer_speeds}
\end{equation}
This cannot be inverted to obtain an analytical generation function
for $\zeta=\zeta(X_4)$. Therefore, numerical methods need to be used
to solve eq.~\ref{eq_pk:plummer_speeds}. For example, one way to obtain
$\zeta$ for a given random variate $X_4$ is to find the root of the
equation $0=(1/2)\,(5\,\zeta^3 - 3\,\zeta^5) - X_4$, or one can use
the Neumann rejection method \citep{NumRecipes}.

The following procedure can be implemented to calculate the velocity
vector of a particle for which $r$ and $\zeta$ are already known from
above: Compute $v_{\rm esc}(r)$ so that $v=\zeta\,v_{\rm esc}$. Each
speed $v$ is then split into its components $v_x, v_y, v_z$ assuming
velocity isotropy using the same algorithm as above for $x, y, z$:
\begin{equation}
v_z(X_5) = (2\,X_5 -1)\,v; \quad \theta(X_6)=2\,\pi\,X_6;
\end{equation}
\begin{equation}
v_x = \sqrt{v^2-v_z^2}\,{\rm cos}\theta; \quad 
v_y = \sqrt{v^2-v_z^2}\,{\rm sin}\theta.
\end{equation}

Note that a rotating Plummer model can be generated by simply
switching the signs of $v_x$ and $v_y$ such that all particles have
the same direction of motion in the $x-y$ plane.

As an aside, an efficient numerical method to set-up triaxial
spheroids with or without an embedded rotating disk is described by
\cite{BKP01}.

\subsubsection{Generating an arbitrary spherical, non-rotating model} 
\index{arbitrary density distributions: discretisation}

In most cases an analytical density distribution is not known
(e.g. the King models above). Such numerical models can nevertheless
be discretised straightforwardly as follows.  Assume that the density
distribution, $\rho(r)$, is known. Compute $M(r)$ and $M_{\rm
cl}$. Define $X(r) = M(r) / M_{\rm cl}$, as above. We thus have a
numerical grid of numbers $r,\;M(r),\;X(r)$. For a given random
deviate $X\in [0,1]$, interpolate $r$ from this grid. Compute $x, y,
z$ as above.

If the distribution function of speeds is too complex to yield an
analytical generating function $X(\zeta)$ for the speeds $\zeta$, then
one can resort to the following procedure: One of the Jeans equations
for a spherical system is
\begin{equation}
{d\over dr}\left(\rho(r)\,\sigma_r(r)^2\right) + {\rho(r) \over r}\,
\left[
2\,\sigma_r^2(r) - \left(\sigma_\theta(r)^2 + \sigma_\phi(r)^2\right)
\right] =
-\rho(r)\,{d\Phi(r) \over dr}.
\label{eq_pk:jeans}
\end{equation}
For velocity isotropy, $\sigma_r^2=\sigma_\theta^2=\sigma_\phi^2$, this reduces to
\begin{equation}
{d\left( \rho\, \sigma_r^2 \right) \over dr} = -\rho\, {d\Phi \over dr}.
\end{equation}
Integrating this by making use of $\rho \longrightarrow 0$ as $r
\longrightarrow \infty$, and remembering that $d\Phi/dr = -G\,M/r^2$,
\begin{equation}
\sigma_r^2(r) = {1\over \rho(r)} \, \int_r^\infty\,\rho(r')\,
{G\,M(r') \over r'^2}\, dr'.
\end{equation}
For each particle at distance $r$ a one-dimensional velocity
dispersion, $\sigma_r(r)$, is thus obtained. Choosing randomly from a
Gaussian distribution with dispersion $\sigma_i, i=r, \theta, \phi, x,
y, z$ then gives the velocity components (e.g. $v_x, v_y, v_z$) for
this particle.

\subsubsection{Rotating models} 
\index{rotating cluster models}

Star clusters are probably born with some rotation because the
pre-cluster cloud core is likely to have contracted from a cloud
region with differential motions that do not cancel out. How large
this initial angular momentum content of an embedded cluster is
remains uncertain, as the dominant motions are random chaotic ones
owing to the turbulent velocity field of the gas. Once the star
formation process is quenched as a result of gas blow-out
(section~\ref{sec_pk:gasblowout}) the cluster expands which must imply
substantial reduction in the rotational velocity. A case in point is
$\omega$~Cen, which has been found to rotate with a peak velocity of
about~$7\,$km/s (\citealt{Pancinoetal07} and references
therein). 

Setting-up rotating cluster models is easily done, e.g. by increasing
the tangential velocities of stars by a certain factor. A systematic
study of relaxation-driven angular momentum re-distribution within
star clusters has become available through the work of the group of
Rainer Spurzem and Hyung-Mok Lee, and the interested reader is
directed to that body of work (\citealt{Kimetal08} and references
therein). One important outcome of this work is that core collapse is
substantially sped-up in rotating models. The primary reason for this
is that increased rotational support reduces the role of support
through random velocities for the same cluster dimension. Thus, the
relative stellar velocities decrease and the stars exchange momentum
and energy more efficiently, enhancing two-body relaxation and
therewith the dive towards energy equipartition.

\subsection{Cluster birth and young clusters}
\label{sec_pk:clbirth} \index{young clusters: discretisation}

Some astro-physical issues related to the initial conditions of star
clusters have been raised in section~\ref{sec_pk:embcl}. In order to
address most of these issues numerical experiments are required.  The
very initial phase, the first $0.5\,$Myr, can only be treated through
gas-dynamical computations that, however, lack the numerical
resolution for the high-precision stellar-dynamical integrations which
are the essence of collisional dynamics during the gas-free phase of a
cluster's life. This gas-free stage sets-in with the blow-out of
residual gas at an age of about 0.5--$1.5\,$Myr. This time is
dominated by the physics of stellar feedback and radiation transport
in the residual gas as well as energy and momentum transfer to it
through stellar outflows. The gas-dynamical computations cannot treat
all the physical details of the processes acting during this critical
time, which also includes early stellar-dynamical processes such as
mass segregation and binary--binary encounters.

One successful procedure to investigate the dominant macroscopic
physical processes of these stellar-dynamical processes, gas blow-out
and the ensuing cluster expansion through to the long-term evolution
of the remnant cluster, is to approximate the residual gas component
as a time-varying potential in which the young stellar population is
trapped. The pioneering work using this approach has been performed by
\cite{LMD84}, whereby the earlier numerical work by \cite{Tutukov78}
on open clusters and later $N-$body computations by \cite{Goodwin97a,
Goodwin97b, Goodwin98} on globular clusters must also be mentioned in
this context.

The physical key quantities that govern the emergence of embedded
clusters from their clouds and their subsequent appearance are
(section~\ref{sec_pk:embcl} and \citealt{BKP08}):

\begin{itemize}

\item sub-structuring; \index{sub-clustering}

\item initial mass segregation; \index{mass segregation}

\item the dynamical state at feedback termination (dynamical
equilibrium?, collapsing? or expanding?);\index{dynamical state}

\item the star-formation efficiency, $\epsilon$; \index{star formation efficiency}

\item the ratio of the gas-expulsion time-scale to the stellar
crossing time through the embedded cluster, $\tau_{\rm gas}/t_{\rm cross}$;

\item the ratio of the embedded-cluster half-mass radius to its tidal
radius, $r_{\rm h}/r_{\rm t}$.

\end{itemize}

It becomes rather apparent that the physical processes governing the
emergence of star clusters from their natal clouds is terribly messy,
and the research-field is clearly observationally driven. Observations
have shown that star clusters suffer substantial infant weight loss
and probably about 90~per cent of all clusters disperse altogether
(infant mortality). This result is consistent with the observational
insight that clusters form in a compact configuration with a low star
formation efficiency ($0.2\simless \epsilon \simless 0.4$) and that
residual-gas blow-out occurs on a time-scale comparable or even faster
than an embedded-cluster crossing time-scale \citep{Kroupa2005}.
Theoretical work can give a reasonable description of these empirical
findings by compactifying some of the above parameters, such as
working with an {\it effective star-formation efficiency} as being a
measure of the amount of gas removed for a cluster of a given stellar
mass {\it assuming} this cluster was in dynamical equilibrium at
feedback termination, and that the gas and stars were distributed
according to the same radial density function with the same scaling
radius.\\

\noindent
{\bf Embedded clusters:}\index{embedded clusters}
One way of parametrising an embedded cluster is to set-up a Plummer
model in which the stellar positions follow a density law with the
parameters $M_{\rm ecl}$ and $r_{\rm pl}$, and the residual gas is a
time-varying Plummer potential initially with the parameters $M_{\rm
gas}$ and $r_{\rm pl}$ (i.e. same radial density law). The effective
star-formation efficiency is then given by
eq.~\ref{eq_pk:sfe}. Stellar velocities must then be calculated from
a Plummer law with total mass $M_{\rm ecl} + M_{\rm gas}$ following
the recipes of section~\ref{sec_pk:discretisation}. The gas can be removed by
evolving $M_{\rm gas}$ or $r_{\rm pl}$. For example, \cite{KAH} and
\cite{BKP08} assumed the gas-mass decreases on an exponential
time-scale after an embedded phase lasting about $0.5\,$Myr during which
the cluster is allowed to evolve in dynamical
equilibrium. \cite{BastianGoodwin2006}, as another example, do not
include a gas potential but take the initial velocities of stars to be
$1/\sqrt{\epsilon}$ times larger, $v_{\rm
embedded}=(1/\sqrt{\epsilon})\,v_{\rm no\;gas}$, to model the effect
of instantaneous gas removal.

Many variations of these assumptions are possible, and \cite{Adams00},
for example, investigated the fraction of stars left in a cluster
remnant if the radial scale length of the gas is different to that of
the stars, i.e. for a radially-dependent star-formation efficiency,
$\epsilon(r)$.\\

\noindent {\bf Sub-clustering:} \index{sub-clustering} Initial
sub-clustering has been barely studied: \cite{ScallyClarke2002}
considered the degree of sub-structuring of the ONC \index{Orion
Nebula Cluster} allowed by its current morphology, while
\cite{FellhauerKroupa2005} computed the evolution of massive
star-cluster complexes assuming each member cluster in the complex
undergoes its own individual gas-expulsion
process. \cite{McMillanetal2007} showed that initially mass-segregated
sub-clusters retain mass-segregation upon merging, this being an
interesting mechanism for speeding-up dynamical mass segregation as it
occurs faster in smaller-$N$ systems which have a shorter relaxation
time.

The simplest initial conditions for such numerical experiments are to
set-up the star-cluster complex (or proto-ONC-type cluster for
example) as a Plummer model, where each particle is a smaller
sub-cluster. Each sub-cluster is also a Plummer model, embedded in a
gas potential given as a Plummer model. The gas-expulsion process from
each sub-cluster can be treated as above.\\

\noindent
{\bf Mass segregation and gas blow-out:} \index{mass
segregation}\index{gas blow out}The problem of how initially
mass-segregated clusters react to gas blow-out has not been studied at
all in the past owing partially to the lack of convenient algorithms
for setting-up mass-segregated clusters that are in dynamical
equilibrium and which do not go into core-collapse as soon as the
$N-$body integration begins. An interesting aspect here is that gas
blow-out will unbind mostly the low-mass stars, while the massive
stars are retained. These, however, evolve rapidly such that the
mass-lost from the remnant cluster owing to the evolution of the
massive stars can become destructive, enhancing infant mortality.

Ladislav Subr has developed a numerically efficient method for
setting-up initially mass-segregated clusters close to core-collapse
based on a novel concept using the potentials of sub-sets of stars
ordered by their mass \citep{SKB08}\footnote{The C-language software
package {\it plumix} can be down-loaded at
{\tt http://www.astro.uni-bonn.de/$\sim$webaiub/english/downloads.php}~.}.
An alternative algorithm based on ordering the stars by increasing
mass and increasing total energy leading to total mass segregation,
but also to a model that is not in core collapse and which therefore
evolves towards core collapse, has been developed by \cite{BKdeM08}.

An application concerning the effect on the observed stellar mass
function in globular clusters shows that gas-expulsion leads to
bottom-light stellar mass functions in clusters with a low
concentration, consistent with observational data \citep{MKB08}.

\newpage

\section{The stellar IMF}
\label{sec_pk:IMF}\index{stellar initial mass function}\index{IMF}

The stellar initial mass function (IMF), $\xi(m)\,dm$, where $m$ is
the stellar mass, is the parent distribution function of the masses of
stars formed in {\it one} event. Here, the number of stars in the mass
interval $m,m+dm$ is
\begin{equation}
dN = \xi(m)\,dm.
\end{equation}  
The IMF is, strictly speaking, an abstract theoretical construct
because any {\it observed} system of $N$ stars merely constitutes a
particular {\it representation} of this universal distribution
function, if such a function exists \citep{Elm97,MU05}. The probable
existence of a {\it unique} $\xi(m)$ can be inferred from observations
of an ensemble of systems each consisting of $N$ stars
(e.g. \citealt{Massey03}). If, after corrections for (a) stellar
evolution, (b) unknown multiple stellar systems, and (c)
stellar-dynamical biases, the individual distributions of stellar
masses are similar {\it within the expected statistical scatter}, then
we (the community) deduce that the hypothesis that the stellar mass
distributions are not the same can be excluded. That is, we make the
case for a {\it universal, standard} or {\it canonical} {\it stellar
IMF} within the physical conditions probed by the relevant physical
parameters (metallicity, density, mass) of the populations at hand.

Related overviews of the IMF can be found in
\cite{Kroupa2002Sci,Chrev03, Bonnelletal07, Kroupa2007a}, and a review
with an emphasis on the metal-rich problem is available in
\cite{Kroupa2007b}, while \cite{ZinneckerYorke2007} provide an
in-depth review of the formation and distribution of massive
stars. \cite{Elm07} discusses the possibility that star-formation
occurs in different modes with different IMFs.

\subsection{The canonical or standard form of the stellar IMF}
\label{sec_pk:canonIMF}\index{canonical IMF}\index{standardd IMF}

The {\it canonical stellar IMF}\index{canonical stellar IMF}
\index{stellar IMF} \index{IMF} is a two-part-power law
(eq.~\ref{eq_pk:multiIMF}), the only structure with confidence found
so far being the change of index from the Salpeter/Massey value to a
smaller one near $0.5\,M_\odot$\footnote{The uncertainties in
$\alpha_i$ are estimated from the alpha-plot
(section~\ref{sec_pk:alphapl}), as shown in fig.~5 in
\cite{Kroupa2002}, to be about 95~per cent confidence limits } :
\vfill
\[
\xi(m) \propto m^{-\alpha_i}, \quad i=1,2
\]
\vspace{-7mm}
\begin{equation} 
\label{eq_pk:canonIMF} 
\end{equation}
\vspace{-7mm}
\[
\begin{array}{l@{\quad\quad,\quad}r@{\;}l} 
\alpha_1 = 1.3\pm0.3, &0.08 \simless &m/M_\odot \simless 0.5,\\
\alpha_2 = 2.3\pm0.5, &0.5 \simless &m/M_\odot \simless m_{\rm max},
\end{array} 
\]
where $m_{\rm max}\le m_{\rm max*}\approx 150\,M_\odot$ follows from
fig.~\ref{fig_pk:mmax_Mecl}.  Brown dwarfs have been found to form a
separate population with $\alpha_0\approx 0.3\pm0.5$
(eq.~\ref{eq_pk:bdIMF}, \citealt{ThiesKroupa07}).

It has been corrected for bias through unresolved multiple stellar
systems \index{IMF: unresolved binaries} in the low-mass ($m <
1\,M_\odot$) regime \citep{KTG91} using a multi-dimensional
optimisation technique. The general outline of this technique is as
follows \citep{KTG93}: first the correct form of the
stellar--mass-luminosity relation is extracted using observed stellar
binaries {\it and} theoretical constraints on the location, amplitude
and shape of the minimum of its derivative, $dm/dM_V$, near
$m=0.3\,M_\odot, M_V\approx 12, M_I \approx 9$ in combination with the
observed shape of the nearby and deep Galactic-field stellar
luminosity function (LF) \index{luminosity function}
\begin{equation}
\Psi(M_V) = -\left({dm \over dM_V}\right)^{-1}\,\xi(m),
\end{equation}
where $dN=\Psi(M_V)\,dM_V$ is the number of stars in the magnitude
inteval $M_V$ to $M_V+dM_V$.  Having established the semi-empirical
mass--luminosity relation \index{mass--luminosity relation} of stars,
which is an excellent fit to the most recent observational constraints
by \cite{Delfosseetal00}, a model of the Galactic field
\index{Galactic field population: model} is then calculated assuming a
parametrised form for the MF and different values for the scale-height
of the Galactic-disk, and different binary fractions in it.
Measurement uncertainties and age and metallicity spreads must also be
considered in the theoretical stellar population.  Optimisation in
this multi-parameter space (MF parameters, scale-height and binary
population) against observational data leads to the canonical stellar
MF for $m< 1\,M_\odot$.

One important result from this work is the finding that the LF of main
sequence stars has a universal sharp peak near $M_V\approx 12,
M_I\approx 9$. It results from changes in the internal constitution of
stars that drive a non-linearity in the stellar mass--luminosity
relation.

\begin{figure}
\begin{center} 
\rotatebox{0}{\resizebox{0.9
\textwidth}{!}{\includegraphics{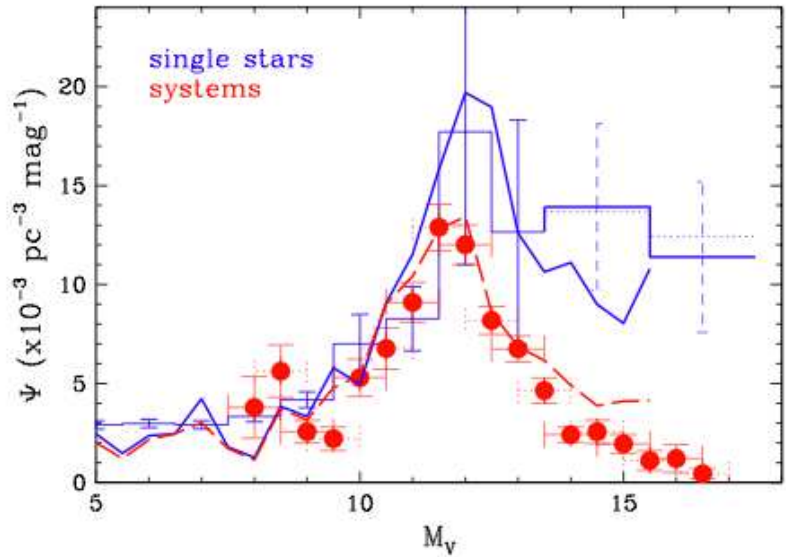}}}
  \caption{\small{The Galactic-field population as resulting from
  disrupted star clusters: Unification of {\it both}, the nearby
  (solid blue histogramme) and deep (filled red circles) LFs with {\it
  one} parent MF (eq.~\ref{eq_pk:canonIMF}). The theoretical nearby LF
  (the solid line) is simply the LF of all individual stars, while the
  dashed line is a theoretical LF with a mixture of about 50~per cent
  unresolved binaries and single stars stemming from a clustered
  star-formation mode. According to this model all stars are formed as
  binaries in modest clusters which disperse to the field, and the
  resulting Galactic-field population has a binary fraction and a
  mass-ratio distribution as observed. After \cite{K95a,K95b}. Note
  the peak in both theoretical LFs -- it stems from the extremum in
  the derivative of the stellar mass--luminosity relation in the mass
  range $0.2-0.4\,M_\odot$ \citep{Kroupa2002}.}}
\label{fig_pk:LFs}
\end{center} 
\end{figure}

A consistency-check is then performed as follows: the above MF is used
in creating young populations of binary systems
(section~\ref{sec_pk:initialpop_lowmass}) that are born in modest star
clusters consisting of a few hundred stars. Their dissolution into the
Galactic field is computed with an $N$-body code, and the resulting
theoretical field is compared to the observed LFs
(fig.~\ref{fig_pk:LFs}).  Further confirmation of the form of the
canonical IMF comes from independent sources, most notably by
\cite{RGH02} and also \cite{Chrev03}.
\index{luminosity function}\index{solar neighbourhood}\index{nearby stars}

In the high-mass regime, \cite{Massey03} reports the same {\it slope
or index} $\alpha_3=2.3\pm0.1$ for $m\simgreat 10\,M_\odot$ in many OB
associations and star clusters in the Milky Way (MW), the Large- and
Small-Magellanic clouds (LMC, SMC, respectively). It is therefore
suggested to refer to $\alpha_2=\alpha_3=2.3$ as the {\it
Salpeter/Massey slope} or {\it index}, given the pioneering work of
\cite{S55} who derived this value for stars with masses
$0.4-10\,M_\odot$.\index{Salpeter IMF}\index{Massey IMF}\index{IMF of
massive stars}

Multiplicity corrections \index{multiplicity corrections} await to be
published once we learn more about how the components are distributed
in massive stars (cf. \citealt{Preibischetal99, Zinnecker03}).
\cite{WK08} are in the process of performing a very detailed study of
the influence of unresolved binary and higher-order multiple stars on
determinations of the high-mass IMF.

Contrary to the Salpeter/Massey index ($\alpha=2.3$) \cite{Sc86} found
$\alpha_{\rm MWdisk} \approx 2.7$ ($m\simgreat 1\,M\odot$) from a very
thorough analysis of OB star counts in the MW disk. Similarly, the
star-count analysis of \cite{RGH02} leads to $2.5\simless \alpha_{\rm
MWdisk}\simless 2.8$, and \cite{Tinsley80}, \cite{Kennicutt83} (his
{\it extended Miller-Scalo IMF}), \cite{Portinarietal04} and
\cite{Romanoetal05} find $2.5 \simless \alpha_{\rm MWdisk} \simless
2.7$. That $\alpha_{\rm MWdisk} > \alpha_2$ follows naturally is shown
in section~\ref{sec_pk:IGIMF}.\index{IGIMF}\index{integrated initial
mass function (IGIMF)} \index{Miller-Scalo IMF}
 
Below the hydrogen-burning limit (see also section~\ref{sec_pk:bds})
there is substantial evidence that the IMF flattens further to
$\alpha_0\approx 0.3\pm0.5$ \citep{Martinetal00, Chrev03,
Morauxetal04}.  Therefore, the canonical IMF most likely has a peak at
$0.08\,M_\odot$. Brown dwarfs, however, comprise only a few~per cent
of the mass of a population and are therefore dynamically irrelevant
(table~\ref{tab_pk:IMFdata}).  The logarithmic form of the canonical
IMF,
\begin{equation}
\xi_{\rm L}(m) = {\rm ln}(10)\;m\;\xi(m),
\end{equation}
which gives the number of stars in log$_{10}m$-intervals, also has a
peak near $0.08\,M_\odot$. However, the {\it system} IMF (of stellar
single and multiple systems combined to system masses) has a maximum
in the mass range $0.4-0.6\,M_\odot$ \citep{Kroupaetal03}.

The above canonical or standard form has been derived from detailed
considerations of Galactic-field star-counts thereby representing an
{\it average} IMF: for low-mass stars it is a mixture of stellar
populations spanning a large range of ages ($0-10\,$Gyr) and
metallicities ([Fe/H]$\simgreat -1$).  For the massive stars it
constitutes a mixture of different metallicities ([Fe/H]$\simgreat
-1.5$) and star-forming conditions (OB associations to very dense
star-burst clusters: R136 \index{R136}in the LMC). Therefore it can be
taken as a canonical form, and the aim is to test the

\vspace{2mm}

\centerline{ \fbox{\parbox{12cm}{{\sc IMF Universality Hypothesis}:
the canonical IMF constitutes the parent distribution of all stellar
populations. \label{hyp_pk:univ}}}}\index{IMF Universality Hypothesis}

\vspace{2mm}

\noindent
Negation of this hypothesis would imply a variable IMF. Note that the
work of \cite{Massey03} has already established the IMF to be
invariable for $m\simgreat 10\,M_\odot$ and for densities
$\rho\simless 10^5\,$stars/pc$^3$ and metallicity $Z\simgreat 0.002$.

Finally, table~\ref{tab_pk:IMFdata} compiles some numbers that are
useful for simple insights into stellar populations.
\begin{table}
{\small
\begin{tabular}{l|ccc|ccc|c|c}

\hline\hline

mass range     
&\multicolumn{3}{c|}{$\eta_N$}
&\multicolumn{3}{c|}{$\eta_M$}    
&$\rho^{\rm st}$
&$\Sigma^{\rm st}$  \\

[$M_\odot$]    
&\multicolumn{3}{c|}{[per cent]}
&\multicolumn{3}{c|}{[per cent]}
&[$M_\odot/{\rm pc}^3$]  
&[$M_\odot/{\rm pc}^2$] \\

&\multicolumn{3}{c|}{$\alpha_3$}
&\multicolumn{3}{c|}{$\alpha_3$}    
&$\alpha_3$
&$\alpha_3$  \\

&2.3 &2.7 &4.5 &2.3 &2.7 &4.5 &4.5 &4.5\\

\hline

0.01--0.08 
&37.2
&37.7 
&38.6

&4.1
&5.4
&7.4

&$3.2\times10^{-3}$
&1.60

\\ 

0.08--0.5
&47.8
&48.5
&49.7

&26.6
&35.2
&48.2

&$2.1\times10^{-2}$
&10.5

\\ 

0.5--1
&8.9
&9.1
&9.3

&16.1
&21.3
&29.2

&$1.3\times10^{-2}$
&6.4

\\ 

1 -- 8
&5.7
&4.6
&2.4

&32.4
&30.3
&15.1

&$6.5\times10^{-3}$
&1.2

\\ 

8 -- 120 
&0.4
&0.1
&0.0

&20.8
&7.8
&0.1

&$3.6\times10^{-5}$
&$6.5\times10^{-3}$

\\ 

\hline

$\overline{m}/M_\odot=$
&$0.38$
&$0.29$
&$0.22$

&
&
&

&$\rho_{\rm tot}^{\rm st}=0.043$
&$\Sigma_{\rm tot}^{\rm st}=19.6$

\\

\hline \hline

&\multicolumn{1}{r|}{}
&\multicolumn{2}{c|}{$\alpha_3=2.3$} 
&\multicolumn{2}{c||}{$\alpha_3=2.7$}
&
&\multicolumn{2}{c}{$\Delta M_{\rm cl}/M_{\rm cl}$}
\\ 

&\multicolumn{1}{r|}{$m_{\rm max}$} 
&$N_{\rm cl}$  &$M_{\rm cl}$  
&$N_{\rm cl}$  &\multicolumn{1}{r||}{$M_{\rm cl}$}
&$m_{\rm to}$  &\multicolumn{2}{c}{[per cent]}
\\

&\multicolumn{1}{r|}{[$M_\odot$]} & &[$M_\odot$] & 
&\multicolumn{1}{r||}{[$M_\odot$]} &[$M_\odot$] 
&$\alpha_3=2.3$ &$\alpha_3=2.7$\\

\cline{2-9}

&\multicolumn{1}{r|}{1}      &16    &2.9     &21                 
&\multicolumn{1}{r||}{3.8}
&80 &3.2 &0.7
\\
								     
&\multicolumn{1}{r|}{8}      &245   &74     &725                
&\multicolumn{1}{r||}{195}
&60 &4.9 &1.1
\\
								      
&\multicolumn{1}{r|}{20}     &806   &269     &3442               
&\multicolumn{1}{r||}{967}
&40 &7.5 &1.8
\\
								      
&\multicolumn{1}{r|}{40}     &1984  &703     &$1.1\times10^4$    
&\multicolumn{1}{r||}{2302}
&20 &13 &4.7
\\

&\multicolumn{1}{r|}{60}     &3361 &1225     &$2.2\times10^4$    
&\multicolumn{1}{r||}{6428}
&8 &22 &8.0
\\

&\multicolumn{1}{r|}{80}     &4885 &1812     &$3.6\times10^4$    
&\multicolumn{1}{r||}{$1.1\times10^4$}
&3 &32 &15
\\
  
&\multicolumn{1}{r|}{100}    &6528 &2451     &$5.3\times10^4$    
&\multicolumn{1}{r||}{$1.5\times10^4$}
&1 &44 &29
\\

&\multicolumn{1}{r|}{120}    &8274 &3136     &$7.2\times10^4$    
&\multicolumn{1}{r||}{$2.1\times10^4$}
&0.7 &47 &33
\\

\hline\hline

\end{tabular}
}
\caption{\small{The number fraction $\eta_N=100\,\int_{m_1}^{m_2}
\xi(m)\,dm/ \int_{m_l}^{m_u}\xi(m)\,dm$, and the mass fraction
$\eta_M=100\,\int_{m_1}^{m_2} m\,\xi(m)\,dm/ M_{\rm cl}$, $M_{\rm cl}=
\int_{m_l}^{m_u} m\,\xi(m)\,dm$, in per cent of BDs or main-sequence
stars in the mass interval $m_1$ to $m_2$, and the stellar
contribution, $\rho^{\rm st}$, to the Oort limit and to the
Galactic-disk surface mass-density, $\Sigma^{\rm st}=2\,h\rho^{\rm
st}$, near to the Sun, taking $m_l=0.01\,M_\odot$, $m_u=120\,M_\odot$
and the Galactic-disk scale-height $h=250\,$pc ($m<1\,M_\odot$
\citealt{KTG93}) and $h=90\,$pc ($m>1\,M_\odot$,
\citealt{Sc86}). Results are shown for the canonical IMF
(eq.~\ref{eq_pk:canonIMF}), for the high-mass-star IMF approximately
corrected for unresolved companions ($\alpha_3=2.7, m>1\,M_\odot$),
and for the present-day mass function (PDMF, $\alpha_3=4.5$,
\citealt{Sc86,KTG93}) which describes the distribution of stellar
masses now populating the Galactic disk. For gas in the disk
$\Sigma^{\rm gas}=13\pm3\,M_\odot$/pc$^2$ and remnants $\Sigma^{\rm
rem}\approx3\,M_\odot$/pc$^2$ \citep{Weidemann90}.  The average
stellar mass is $\overline{m}= \int_{m_l}^{m_u} m\,\xi(m)\,dm/
\int_{m_l}^{m_u}\xi(m)\,dm$.  $N_{\rm cl}$ is the number of stars that
have to form in a star cluster such that the most massive star in the
population has the mass $m_{\rm max}$. The mass of this population is
$M_{\rm cl}$, and the condition is $\int_{m_{\rm
max}}^{\infty}\xi(m)\,dm=1$ with $\int_{0.01}^{m_{\rm max}} \xi(m)\,dm
= N_{\rm cl}-1$.  $\Delta M_{\rm cl}/M_{\rm cl}$ is the fraction of
mass lost from the cluster due to stellar evolution, assuming that for
$m\ge8\,M_\odot$ all neutron stars and black holes are kicked out due
to an asymmetrical supernova explosion, but that white dwarfs are
retained \citep{Weidemannetal92} and have masses $m_{\rm WD} =
0.084\,m_{\rm ini}+0.444\,[M_\odot]$. This is a linear fit to the data
in \cite[][their table~3]{Weidemann00} for progenitor masses $1\le
m/M_\odot \le 7$ and $m_{\rm WD}=0.5\,M_\odot$ for $0.7\le m/M_\odot
<1$. The evolution time for a star of mass $m_{\rm to}$ to reach the
turn-off age is available in fig.~20 in \cite{Kroupa2007a}.}}
\label{tab_pk:IMFdata}\index{IMF: some numbers}
\end{table}

\subsection{Universality of the IMF: resolved populations}
\label{sec_pk:univ} \index{IMF: universality}

The strongest test of the {\sc IMF Universality Hypothesis} is
obtained by studying populations that can be resolved into individual
stars. Since one also seeks co-eval populations with stars at the same
distance and with the same metallicity to minimise uncertainties, star
clusters and stellar associations would seem to be the test objects of
choice.  But before contemplating such work some lessons from stellar
dynamics are useful:

\subsubsection{Star clusters and associations}
\index{star clusters}\index{associations}

To access a pristine population one would consider observing
star-clusters that are younger than a few~Myr. However, such objects
carry rather massive disadvantages: the pre-main sequence stellar
evolution tracks are unreliable \citep{Baraffeetal07,
WuchterlTscharnuter2003} such that the derived masses are uncertain by
at least a factor of about~two. Remaining gas and dust lead to patchy
obscuration.  Very young clusters evolve rapidly: the dynamical
crossing time is given by eq.~\ref{eq_pk:tcr} where the cluster radii
are typically $r_{0.5} < 1\,$pc and for pre-cluster cloud-core masses
$M_{\rm gas+stars} > 10^3\,M_\odot$ the velocity dispersion
$\sigma_{\rm cl} > 2\,$km/s such that $t_{\rm cr}<1\,$Myr.

The inner regions of populous clusters have $t_{\rm cr} \approx
0.1\,$Myr, and thus significant mixing and relaxation occurs there by
the time the residual gas is expelled by the winds and photo-ionising
radiation from massive stars, if they are present, being the case in
clusters with $N\simgreat {\rm few}\times 100$ stars
(table~\ref{tab_pk:youngcl}).

Massive stars \index{massive stars} ($m > 8\,M_\odot$) are either
formed at the cluster centre or get there through dynamical mass
segregation, i.e. energy equipartition \citep{Bonnelletal07}. The
latter process is {\it very rapid} (eq.~\ref{eq_pk:tms},
p.~\pageref{sec_pk:ms}) and can occur within $1\,$Myr. A cluster core
of massive stars is therefore either primordial or forms rapidly
because of energy equipartition in the cluster, and it is dynamically
highly unstable decaying within a few $t_{\rm cr, \;core}$. The ONC,
for example, should not be hosting a Trapezium as it is extremely
volatile. The implication for the IMF is that the ONC \index{Orion
Nebula Cluster} and other similar clusters and the OB associations
which stem from them must be very depleted in their massive-star
content \citep{PfK2006}.

Important for measuring the IMF are corrections for the typically high
multiplicity fraction of the very young population. However, these are
very uncertain because the binary population is in a state of change
(fig.~\ref{fig_pk:ftot} below).

The determination of an IMF relies on the assumption that all stars in
a very young cluster formed together.  However, trapping and focussing
of older field or OB~association stars by the forming cluster has been
found to be possible (section~\ref{sec_pk:cl_sfh}). 

Thus, be it at the low-mass end or the high-mass end, the stellar mass
function estimated from very young clusters cannot be the true IMF.
Statistical corrections for the above effects need to be applied and
comprehensive $N$-body modelling is required.

Old open clusters in which most stars are on or near the main sequence
are no better stellar samples: They are dynamically highly evolved,
since they have left their previous concentrated and gas-rich state
and so they contain only a small fraction of the stars originally born
in the cluster \citep{KroupaBoily2002, Weidneretal2007, BK07}. The
binary fraction is typically high and comparable to the Galactic
field, but does depend on the initial density and the age of the
cluster, the mass-ratio distribution of companions also. So, simple
corrections cannot be applied equally to all old clusters.  The
massive stars have died, and secular evolution begins to affect the
remaining stellar population (after gas expulsion) through energy
equipartition.  \cite{BaumgardtMakino2003} have quantified the changes
of the MF for clusters of various masses and on different Galactic
orbits. Near the half-mass radius the local MF resembles the global MF
in the cluster, but the global MF becomes significantly depleted of
its low-mass stars already by about 20~per cent of the cluster
disruption time.

Given that we are never likely to learn the exact dynamical history of
a particular cluster, it follows that we can {\it never} ascertain the
IMF for any individual cluster. This can be summarised concisely with
the following theorem:

\vspace{2mm}

\centerline{
\fbox{\parbox{12cm}{
{\sc Cluster IMF Theorem}: The IMF cannot be extracted for any individual star
cluster. \label{theorem_pk:IMFtheorem}}}}\index{Cluster IMF Theorem}

\vspace{2mm}

\noindent
{\sc Proof:} For clusters younger than about $0.5\,$Myr star formation
has not ceased and the IMF is therefore not assembled yet and the
cluster cores consisting of massive stars have already dynamically
ejected members \citep{PfK2006}. For clusters with an age between~0.5
and a few~Myr the expulsion of residual gas has lead to loss of stars
\citep{KAH}.  Older clusters are either still loosing stars due to
residual gas expulsion or are evolving secularly through evaporation
driven by energy equipartition
\citep{BaumgardtMakino2003}. Furthermore, the birth sample is likely
to be contaminated by captured stars \citep{Fellhaueretal2006, PfK2007}.
There exists thus no time when all stars are assembled in an
observationally accessible volume (i.e. a star cluster).

Note that the {\sc Cluster IMF Theorem} implies that individual
clusters cannot be used to make deductions on the similarity or not of
their IMFs, unless a complete dynamical history of each cluster is
available. Notwithstanding this pessimistic theorem, it is
nevertheless necessary to observe and study star clusters of any
age. Combined with thorough and realistic $N$-body modelling the data
do lead to essential {\it statistical} constraints on the {\sc IMF
Universality Hypothesis} (p.~\pageref{hyp_pk:univ}). Such an approach
is discussed in the next section.

\subsubsection{The alpha plot} 
\label{sec_pk:alphapl}\index{alpha plot}

\begin{figure} 
\hspace{-2mm}
\rotatebox{0}{\resizebox{0.8\textwidth}{!}
{\includegraphics{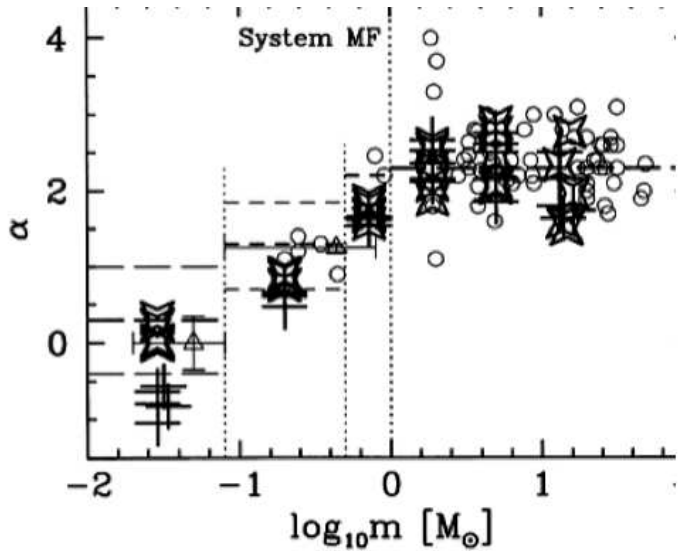}}} 
\caption{\small{The alpha plot. The power-law index, $\alpha$, is
measured over stellar mass-ranges and plotted at the mid-point of the
respective mass range. The power-law indices are measured on the mass
function of system masses, where stars not in binaries are counted
individually.  Open circles are the observational constraints for open
clusters and associations for the MW, Large and Small Magellanic
clouds collated mostly by \cite{Sc98}. The open stars (crosses) are
theoretical star clusters ``observed'' in the computer at an age of 3
(0)~Myr and within a radius of~$3.2\,$pc from the cluster centre. The
5~clusters have 3000~stars in 1500~binaries initially and the assumed
IMF is the canonical one. The theoretical data nicely show a similar
spread as the observational ones; note the binary-star-induced
depression of $\alpha_1$ in the mass range $0.1-0.5\,M_\odot$.  The
{\sc IMF Universality Hypothesis} can therefore not be discarded given
the observational data.  Models from \cite{K01}.  }}
\label{fig_pk:alpha} 
\end{figure}

\cite{Sc98} conveniently summarised a large part of the available
observational constraints on the IMF of resolved stellar populations
with the {\it alpha plot}, as used by \cite{K01, Kroupa2002} for
explicit tests of the {\sc IMF Universality Hypothesis}
(p.~\pageref{hyp_pk:univ}) given the {\sc Cluster IMF Theorem}.  One
example is presented in fig.~\ref{fig_pk:alpha}, which demonstrates
that the observed scatter in $\alpha(m)$ can be readily understood as
being due to Poisson uncertainties (see also \citealt{Elm97, Elm99})
and dynamical effects, as well as arising from biases through
unresolved multiple stars. Furthermore, there is no evident systematic
change of $\alpha$ at a given $m$ with metallicity or density of the
star-forming cloud. More exotic populations such as the Galactic bulge
have also been found to have a low-mass MF indistinguishable from the
canonical form (e.g. \citealt{Zoccalietal00}).  Thus the {\sc IMF
Universality Hypothesis} cannot be falsified for known resolved
stellar populations.

\subsubsection{Very ancient and/or metal-poor resolved populations}
\index{IMF: dSph galaxies}\index{IMF: globular clusters}

Witnesses of the early formation phase of the MW are its globular
clusters.  Such $10^{4-6}\,M_\odot$ clusters formed with individual
star-formation rates of $0.1-1\,M_\odot/$yr and densities $\approx
5\times 10^{3-5}\,M_\odot/\,$pc$^3$. These are relatively high values,
when compared with the current star-formation activity in the MW disk.
For example, a $5\times 10^3\,M_\odot$ Galactic cluster forming in
$1\,$Myr corresponds to a star formation rate of $0.005\,M_\odot$/yr.
The alpha plot, however, does not support any significant systematic
difference between the IMF of stars formed in globular clusters and
present-day low-mass star formation. For massive stars, it can be
argued that the mass in stars more massive than $8\,M_\odot$ cannot
have been larger than about half the cluster mass, because otherwise
the globular clusters would not be as compact as they are today.  This
constrains the IMF to have been close to the canonical IMF (Kroupa
2001).

A particularly exotic star-formation mode is thought to have occurred
in dwarf-sphe\-roidal (dSph) satellite galaxies. The MW has about
19~such satellites at distances from 50~to~$250\,$kpc
\citep{MetzKroupa07}. These objects have stellar masses and ages
comparable to those of globular clusters but are $10-100$ times larger
and are thought to have $10-1000$ times more mass in dark matter than
in stars. They also show evidence for complex star-formation activity
and metal-enrichment histories and must have therefore formed under
rather exotic conditions. Nevertheless, the MFs in two of these
satellites are found to be indistinguishable from those of globular
clusters in the mass range $0.5-0.9\,M_\odot$, thus again showing
consistency with the canonical IMF \citep{Grillmairetal98,
Feltzingetal99}.

The work of \cite{Yasuietal06, Yasuietal08} has been pushing studies of
the IMF in young star clusters to the outer, metal-poor regions of the
Galactic disk.  They find the IMF to be indistinguishable, within the
uncertainties, with the canonical IMF. \index{outer-Galaxy young clusters}

\subsubsection{The Galactic bulge and centre}
\label{sec_pk:gal_bulge}\index{IMF: Galactic bulge}\index{IMF: Galactic centre}

For low-mass stars the Galactic bulge has been shown to have a MF
indistinguishable from the canonical form \citep{Zoccalietal00}.
However, abundance patterns of bulge stars suggest the IMF to have
been top heavy \citep{BKM07}, which may be a result of extreme
star-burst conditions valid in the formation of the bulge
\citep{Zoccalietal06}.

Even closer to the Galactic centre, Hertzsprung-Russell-diagram
modelling of the stellar population within $1\,$pc of Sgr~A$^*$
suggests the IMF to have always been top-heavy there
\citep{Manessetal07}. Perhaps this is the long-sought after evidence
for a variation of the IMF under very extreme conditions, in this case
a strong tidal field and higher temperatures (but note
fig.~\ref{fig_pk:arches} below).

\subsubsection{Extreme star bursts}
\label{sec_pk:star_bursts}\index{IMF: star bursts}

As noted on p.~\pageref{sec_pk:galaxy}, objects with a mass
$M\simgreat 10^6\,M_\odot$ have an increased $M/L$ ratio. If such
objects form in 1~to~$5\,$Myr, then their star-formation rates,
SFR$\,\simgreat 10\,M_\odot/$yr, and they contain $\simgreat
10^4$~O~stars packed within a region spanning at most a few~pc, given
their observed present-day mass-radius relation. Such a star-formation
environment is presently outside the reach of theoretical
investigation. However, it is conceivable that the higher $M/L$ ratios
of such objects may be due to a non-canonical IMF.  One possibility is
that the IMF is bottom heavy as a result of intense photo-destruction
of accretion envelopes of intermediate to low-mass stars 
\citep{MieskeKroupa08}. Another possibility is that the IMF becomes
top-heavy leaving many stellar remnants that inflate the $M/L$ ratio
\citep{DabringhausenKroupa08}.  Work is in progress to achieve
observational constraints on these two possibilities.

\subsubsection{Population III: the primordial IMF}
\label{sec_pk:popIII} \index{IMF: population III}

Most theoretical workers agree that the primordial IMF ought to be top
heavy because the ambient temperatures were much higher and the lack
of metals did not allow gas clouds to cool and to fragment into
sufficiently small cores \citep{Larson98}.  The existence of extremely
metal-poor low-mass stars with chemical peculiarities is interpreted
to mean that low-mass stars could form under extremely metal-poor
conditions, but that their formation was suppressed in comparison to
later star-formation \citep{Tumlinson07}. Modelling of the formation of
stellar populations during cosmological structure formation suggests
that low-mass population~III stars should be found within the Galactic
halo {\it if} they formed. Their absence to-date would imply a
primordial IMF depleted in low-mass stars \citep{Brooketal07}. 

Thus, the last three sub-sections hint at physical environments in
which the IMF Universality Hypothesis may be violated.

\subsection{Very low-mass stars (VLMSs) and brown dwarfs (BDs)}
\label{sec_pk:bds}\index{brown dwarfs}\index{very low mass stars}

The origin of BDs and some VLMSs is being debated fiercely. One camp
believes these objects to form as stars do, because the star-formation
process does not know where the hydrogen burning mass limit is
(e.g. \citealt{EisloeffelSteinacker07}). The other camp believes that BDs
cannot form exactly like stars through continued accretion because the
conditions required for this to occur in molecular clouds are far too
rare (e.g. \citealt{ReipurthClarke01, GoodwinWhitworth07}).

If BDs and VLMSs form like stars then they should follow the same
pairing rules.  In particular, BDs and G~dwarfs would pair in the same
manner, i.e. according to the same mathematical rules, as M~dwarfs and
G~dwarfs.  \cite{Kroupaetal03} tested this hypothesis by constructing
$N$-body models of Taurus-Auriga-like groups and Orion-Nebula-like
clusters finding that it leads to far too many star--BD and BD--BD
binaries with the wrong semi-major axis distribution.  Instead,
star--BD binaries are very rare \citep{GretherLineweaver06}, while
BD--BD binaries are rarer than stellar binaries (BDs have a 15~per
cent binary fraction as opposed to 50~per cent for stars), and BDs
have a semi-major axis distribution significantly narrower than that
of star--star binaries. The hypothesis of a star-like origin of BDs
must therefore be discarded. BDs and some VLMSs form a separate
population, which is however linked to that of the stars.

\cite{ThiesKroupa07} re-address this problem with a detailed analysis
of the underlying MF of stars and BDs given observed MFs of four
populations, Taurus, Trapezium, IC348 and the
Pleiades. \index{Pleiades} By correcting for unresolved binaries in
all four populations, by taking into account the different pairing
rules of stellar and VLMS and BD binaries, a significant discontinuity
of the MF emerges. BDs and VLMSs therefore form a truly separate
population from that of the stars and can be described by a single
power-law MF (eq.~\ref{eq_pk:bdIMF}) which implies that about one BD
forms per 5~stars in all four populations.

This strong correlation between the number of stars and BDs, and the
similarity of the BD MF in the four populations implies that the
formation of BDs is closely related to the formation of stars. Indeed,
the truncation of the binary binding energy distribution of BDs at a
high energy suggests that energetic processes must be operating in the
production of BDs, as discussed in \cite{ThiesKroupa07}. Two such
possible mechanisms are embryo ejection \citep{ReipurthClarke01} and
disk fragmentation \citep{GoodwinWhitworth07}.

\subsection{Composite populations: the IGIMF}
\label{sec_pk:IGIMF} \index{IGIMF}\index{integrated initial mass function}
\index{stellar population: composite}\index{stellar population: complex}

The vast majority of all stars form in embedded clusters and so the
correct way to proceed to calculate a galaxy-wide stellar IMF is to
add-up all the IMFs of all star-clusters born in one {\it
star-formation epoch}.  Such {\it epochs} may be identified with the
\cite{Zoccalietal06} star-burst events creating the Galactic bulge. In
disk galaxies they may be related to the time-scale of transforming
the inter-stellar matter to star clusters along spiral arms. Addition
of the clusters born in one epoch gives the {\it integrated galactic
initial mass function}, the IGIMF \citep{KW03}.

\vspace{2mm}

\centerline{ \fbox{\parbox{12cm}{{\sc IGIMF Definition}: The IGIMF is
the IMF of a composite population which is the integral over a
complete ensemble of simple stellar populations.
\label{kroupa_theorem:igimfdef}}}}\index{IGMF definition}

\vspace{2mm}

\noindent 
Note that a {\it simple population}\index{stellar population: simple}
has a mono-metallicity and a mono-age distribution and is therefore a
star cluster.  Age and metallicity distributions emerge for massive
populations with $M_{\rm cl}\simgreat 10^6\,M_\odot$
(e.g. $\omega$~Cen) indicating that such objects, which also have
relaxation times comparabe to or longer than a Hubble time, are not
{\it simple} (section~\ref{sec_pk:galaxy}).  A {\it complete ensemble}
is a statistically complete representation of the initial cluster mass
function (ICMF) in the sense that the actual mass function of $N_{\rm
cl}$ clusters lies within the expected statistical variation of the
ICMF.

\vspace{2mm}

\centerline{ \fbox{\parbox{12cm}{{\sc IGIMF Theorem}: The IGIMF is
steeper than the canonical IMF if the {\sc IMF Universality
Hypothesis} holds.
\label{theorem_pk:igimf}}}}

\vspace{2mm}

\noindent
{\sc Proof:} \cite{WK06} calculate that the IGIMF is steeper than the
canonical IMF for $m\simgreat 1\,M_\odot$ if the {\sc IMF Universality
Hypothesis} holds. The steepening becomes negligible if the power-law
mass function of embedded star clusters,
\begin{equation}
\xi_{\rm ecl}(M_{\rm ecl}) \propto M_{\rm ecl}^{-\beta}
\end{equation}
is flatter than $\beta=1.8$.

It may be argued that IGIMF$=$IMF (e.g. Elmegreen 2006) because when a
star cluster is born, its stars are randomly sampled from the IMF up
to the most massive star possible. On the other hand, the
physically-motivated ansatz by \cite{WK05, WK06} of taking the mass of
a cluster as the constraint and of including the observed correlation
between the maximal star mass and the cluster mass
(fig.~\ref{fig_pk:mmax_Mecl}), yields an IGIMF which is equal to the
canonical IMF for $m\simless 1.5\,M_\odot$ but which is systematically
steeper above this mass. By incorporating the observed
maximal-cluster-mass vs star-formation rate of galaxies, $M_{\rm
ecl,max} = M_{\rm ecl,max}(SFR)$, for the youngest clusters
\citep{WKL2004}, it follows for $m\simgreat 1.5\,M_\odot$ that
low-surface-brightness (LSB) galaxies ought to have very steep IGIMFs,
while normal or L$_*$ galaxies have Scalo-type IGIMFs,
i.e. $\alpha_{\rm IGIMF} = \alpha_{\rm MWdisk} > \alpha_2$
(section~\ref{sec_pk:canonIMF}) follows naturally. This systematic
shift of $\alpha_{\rm IGIMF}$ ($m\simgreat 1.5\,M_\odot$) with galaxy
type implies that less-massive galaxies have a significantly
suppressed supernova~II rate per low-mass star. They also show a
slower chemical enrichment such that the observed
metallicity--galaxy-mass relation can be nicely accounted for
\citep{KWK07}. Another very important implication is that the
SFR--H$\alpha$-luminosity relation for galaxies flattens such that the
SFR becomes higher by up to three orders of magnitude for dwarf
galaxies than the value calculated from the standard (linear)
Kennicutt relation \citep{PWK07}.\index{Kennicutt relation}
\index{SFR--H$\alpha$ relation}

Strikingly, the IGIMF variation has now been directly measured by
\cite{HG08} using galaxies in the Sloan Digital Sky
Survey. \cite{Leeetal04} have indeed found LSBs to have bottom-heavy
IMFs, while \cite{Portinarietal04} and \cite{Romanoetal05} find the MW
disk to have a steeper-than Salpeter IMF for massive stars which is,
in comparison with \cite{Leeetal04}, much flatter than the IMF of
LSBs, as required by the {\sc IGIMF Theorem}.

\subsection{Origin of the IMF: theory vs observations}
\label{sec_pk:originIMF}\index{IMF: origin}

General physical concepts such as coalescence of proto-stellar cores,
mass-depen\-dent focussing of gas accretion onto proto-stars, stellar
feedback, and fragmentation of molecular clouds lead to predictions of
systematic variations of the IMF with changes of the physical
conditions of star formation (\citealt{MurrayLin96, Elm04,
TilleyPudritz05}, but see \citealt{CasusoBeckman07} for a simple
cloud coagulation/dispersal model leading to an invariant mass
distribution). Thus, the thermal Jeans mass of a molecular cloud
decreases with temperature and increasing density, implying that for
higher metallicity ($=$ stronger cooling) and density the IMF should
shift on average to smaller stellar masses (e.g. \citealt{Larson98,
Bonnelletal07}). The entirely different notion that stars regulate
their own masses through a balance between feedback and accretion also
implies smaller stellar masses for higher metallicity due in part to
more dust and thus more efficient radiation pressure on the gas
through the dust grains.  Also, a higher metallicity allows more
efficient cooling and thus a lower gas temperature, a lower sound
speed and therefore a lower accretion rate \citep{AF96, AL96}.  As
discussed above, a systematic IMF variation with physical conditions
has not been detected. Thus, theoretical reasoning, even at its most
elementary level, fails to account for the observations.

\begin{figure}
\begin{center}
\rotatebox{0}{\resizebox{0.8
\textwidth}{!}{\includegraphics{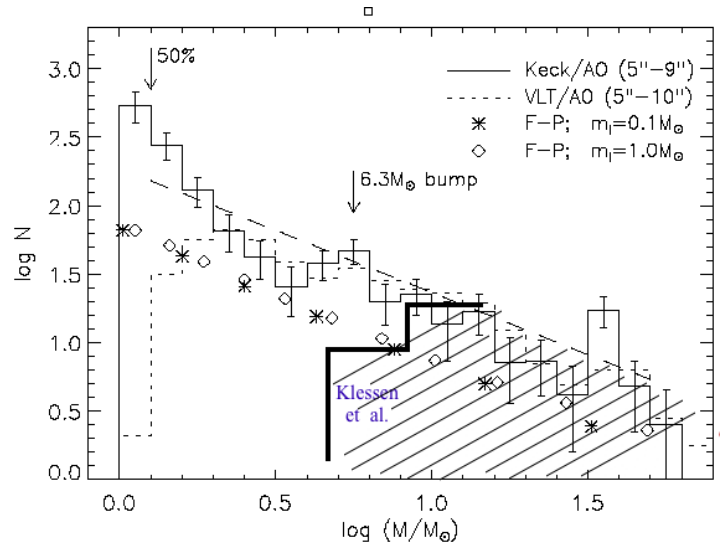}}}
\vskip 0mm
\caption{\small{ The observed mass function of the Arches cluster near
the Galactic centre by \cite{Kimetal06} shown as the thin histogramme is
confronted with the theoretical MF for this object calculated with the
SPH technique by \cite{Klessenetal07}, marked as the hatched
histogramme. The latter has a down-turn (bold steps near $10^{0.7}$)
incompatible to the observations therewith ruling out a theoretical
understanding of the stellar mass spectrum (one counter-example
suffices to bring-down a theory).  One possible reason for the
theoretical failure may be the assumed turbulence driving. For details
on the figure see \cite{Kimetal06}.  }}
\label{fig_pk:arches}
\end{center}
\end{figure}

A dramatic case in point has emerged recently: \cite{Klessenetal07}
report state-of-the art calculations of star-formation under physical
conditions as found in molecular clouds near the Sun and they are able
to reproduce the canonical IMF. Applying the same computational
technology to the conditions near the Galactic centre they obtain a
theoretical IMF in agreement with the previously reported apparent
decline of the stellar MF in the Arches cluster below about
$6\,M_\odot$.  \cite{Kimetal06} published their observations of the
Arches cluster on the astro-physics preprint archive shortly after
\cite{Klessenetal07} and performed the necessary state-of-the art
$N$-body calculations of the dynamical evolution of this young
cluster, revising our knowledge significantly. In contradiction to the
theoretical prediction they find that the MF continues to increase
down to their 50~per cent completeness limit ($1.3\,M_\odot$) with a
power-law exponent only slightly shallower than the canonical
Massey/Salpeter value once mass-segregation is corrected for.  This
situation is demonstrated in fig.~\ref{fig_pk:arches}.

It therefore emerges that there does not seem to exist {\it any solid}
theoretical understanding of the IMF.

Observations of cloud cores appear to suggest that the canonical IMF
is frozen-in already at the pre-stellar cloud-core level
\citep{Motteetal98, Motteetal01}. \cite{NutterWardTh07} and
\cite{Alvesetal07} find, however, the pre-stellar cloud cores to be
distributed according to the same shape as the canonical IMF, but
shifted to larger masses by a factor of about three or more. This is
taken to perhaps mean a star-formation efficiency per star of 30~per
cent or less independently of stellar mass. The interpretation of such
observations in view of multiple star formation in each cloud-core is
being studied by \cite{Goodwinetal08}, while \cite{Krumholz07}
outlines current theoretical understanding of how massive stars form
out of massive pre-stellar cores.

\subsection{Conclusions: IMF} 
\label{sec_pk:IMFconcs}

The {\sc IMF Universality Hypothesis}, the {\sc Cluster IMF Theorem}
and the {\sc IGIMF Theorem} have been stated. Furthermore,

\begin{enumerate}

\item The stellar luminosity function has a pronounced maximum at
$M_V\approx 12, M_I\approx 9$ which is universal and well understood
as a result of stellar physics. Thus by counting stars on the sky we
can look into their interiors.

\item Unresolved multiple systems must be accounted for when the MFs of
different stellar populations are compared.

\item BDs and some VLMSs form a separate population which correlates
with the stellar content; there is a discontinuity in the MF near the
star/BD mass transition.

\item The canonical IMF (eq.~\ref{eq_pk:canonIMF}) fits the
solar-neighbourhood star counts and {\it all} resolved stellar
populations available to-date. Recent data at the Galactic centre
suggest a top-heavy IMF, perhaps hinting at a possible variation with
conditions (tidal shear, temperature).

\item {\it Simple} stellar populations are found in individual
star clusters with $M_{\rm cl}$ $\simless 10^6\,$ $M_\odot$. These
have the canonical IMF.

\item {\it Composite} populations describe entire galaxies. They are a
result of many epochs of star-cluster formation and are described
by the {\sc IGIMF Theorem}.

\item The IGIMF above $\approx 1\,M_\odot$ is steep for LSB galaxies,
flattening to the Scalo slope ($\alpha_{\rm IGIMF} \approx 2.7$) for
$L_*$ disk galaxies. This is nicely consistent with the {\sc IMF
Universality Hypothesis} in the context of the {\sc IGIMF theorem}.

\item Therefore, the {\sc IMF Universality Hypothesis} can not be excluded
despite the {\sc cluster IMF Theorem} for conditions $\rho \simless
10^5$~stars/pc$^3$, $Z\simgreat 0.002$ and non-extreme tidal fields.

\item Modern star-formation computations and elementary theory give
wrong results concerning the variation and shape of the stellar IMF,
as well as the stellar multiplicity \citep{GoodwinKroupa2005}.

\item The stellar IMF appears to be frozen-in at the pre-stellar
cloud-core stage therewith probably being a result of the processes
leading to the formation of self-gravitating molecular clouds.

\end{enumerate}

\subsection{Discretisation}
\label{sec_pk:IMF_discretisation}\index{IMF: discretisation}

As discussed above, a theoretically-motivated form of the IMF which
passes observational tests does not exist. Star-formation theory gets
the rough shape of the IMF right (there are fewer massive stars than
low-mass stars), but other than this, fails to make any reliable
predictions whatsoever as to how the IMF should look like in detail
under different physical conditions. In particular, the overall change
of the IMF with metallicity or density or temperature predicted by
theory is not evident.  An empirical multi-power-law form description
of the IMF is therefore perfectly adequate, and has important
advantages over other formulations. A general formulation of the
stellar IMF in terms of multiple power-law segments is as follows:
\begin{equation}
\xi(m) = k\!  \left\{\!\!\!\!
    \begin{array}{l@{\;,}l}
      \;\left(\frac{m}{m_{\mathrm{H}}}\right)^{-\alpha_0}&
      \;m_{\mathrm{low}}\le m \le m_{\mathrm{H}}\\
      \;\left(\frac{m}{m_{\mathrm{H}}}\right)^{-\alpha_1}&
      \;m_{\mathrm{H}}\le m \le m_0\\
      \;\left(\frac{m_0}{m_\mathrm{H}}\right)^{-\alpha_1}
      \;\left(\frac{m}{m_0}\right)^{-\alpha_2}&
      \;m_0\le m \le m_1\\
      \;\left(\frac{m_0}{m_\mathrm{H}}\right)^{-\alpha_1}
      \;\left(\frac{m_1}{m_0}\right)^{-\alpha_2}
      \;\left(\frac{m}{m_1}\right)^{-\alpha_3}&
      \;m_1\le m \le m_\mathrm{max}\\
    \end{array},
\right.
\label{eq_pk:multiIMF}
\end{equation}
where $m_{\rm max}\le m_{\rm max*}\approx 150\,M_\odot$ depends on the
stellar mass of the embedded cluster
(fig.~\ref{fig_pk:mmax_Mecl}). The empirically determined stellar IMF
is a two-part-form (eq.~\ref{eq_pk:canonIMF}), with a third power-law
for BDs, whereby BDs and VLMSs form a separate population from that of
the stars (p.~\pageref{sec_pk:bds}),
\begin{equation}
\xi_{\rm BD} \propto m^{-\alpha_0}, \quad \alpha_0\approx0.3,
\label{eq_pk:bdIMF}
\end{equation}
\citep{Martinetal00, Chrev03, Morauxetal04}) and
\[
\xi_{\rm BD}(0.075\,M_\odot) \approx 0.25 \;\xi(0.075\,M_\odot), 
\]
\citep{ThiesKroupa07} where $\xi$ is the canonical stellar IMF
(eq.~\ref{eq_pk:canonIMF}).  This implies that about one BD forms per
5~stars.

One advantage of the power-law formulation becomes immediately obvious
by realising that analytical generating functions, and other
quantities, can be derived readily. Another important advantage is
that by using a multi-power-law form, different parts of the IMF can
be varied in numerical experiments without affecting the other parts. A
practical numerical formulation of the IMF is prescribed in
\cite{PfK2006}.  Thus, for example, the canonical two-part power-law IMF
can be changed by adding a third power-law above $1\,M_\odot$ and
making the IMF top-heavy ($\alpha_{m>1\,M_\odot} < \alpha_2$) {\it
without} affecting the shape of the late-type stellar luminosity
function as evident in fig.~\ref{fig_pk:LFs}. The KTG93 \citep{KTG93}
IMF is such a three-part power-law form relevant for describing the
overall young population in the Milky Way disk, which is top-light
($\alpha_{m>1\,M_\odot} > \alpha_2$, \citealt{KW03}).

\begin{figure}
\begin{center}
\resizebox{0.95 \textwidth}{!}{\includegraphics{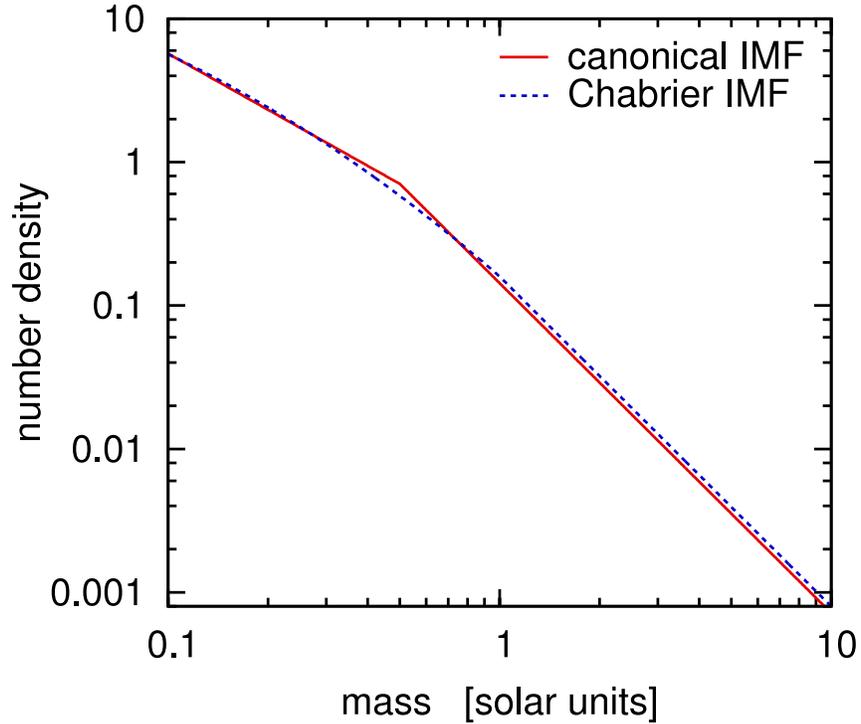}}
\vskip 5mm
\caption{\small{Comparison between the popular Chabrier IMF
(log-normal plus power-law extension above $1\,M_\odot$: dashed curve,
table~1 in \citealt{Chrev03}) with the canonical two-part power-law
IMF (solid line, eq.~\ref{eq_pk:canonIMF}). From \cite{DHK08}.  }}
\label{fig_pk:ch_kr}
\end{center}
\end{figure}

A log-normal formulation does not offer these advantage, and requires
power-law tails above about $1\,M_\odot$, and for brown dwarfs, for
consistency with the observational constraints discussed
above. However, while not as mathematically convenient, the popular
{\it Chabrier log-normal plus power-law IMF} (table~1 in
\citealt{Chrev03}) formulation leads to an indistinguishable stellar
mass distribution to the two-part power-law IMF \index{Chabrier IMF}
(fig.~\ref{fig_pk:ch_kr}).
Various analytical forms for the IMF are compiled in table~3 of
\cite{Kroupa2007a}.

A generating function for the two-part power-law form of the canonical
IMF, eq~\ref{eq_pk:canonIMF}, can be written down by following the
steps taken in section~\ref{sec_pk:discretisation}.  The corresponding
probability density is 
\[
p_1=k_{p,1}\,m^{-\alpha_1}, \quad 0.08 \le m \le 0.5\,M_\odot
\]
\vspace{-9mm}
\begin{equation}
\end{equation}
\vspace{-9mm}
\[
p_2=k_{p,2}\,m^{-\alpha_2}, \quad 0.5 < m \le m_{\rm max},
\]
where $k_{p,i}$ are normalisation constants ensuring continuity at
$0.5\,M_\odot$ and
\begin{equation}
\int_{0.08}^{0.5}\,p_1\,dm + \int_{0.5}^{m_{\rm max}}\,p_2\,dm = 1,
\end{equation}
whereby $m_{\rm max}$ follows from
fig.~\ref{fig_pk:mmax_Mecl}. Defining
\begin{equation}
X_1' = \int_{0.08}^{0.5}\, p_1(m)\,dm,
\end{equation}
it follows that
\begin{equation}
X_1(m) = \int_{0.08}^m\, p_1(m)\,dm, \quad {\rm if}\; m\le 0.5\,M_\odot,
\end{equation}
or 
\begin{equation}
X_2(m) = X_1' + \int_{0.5}^m\, p_2(m)\,dm, \quad {\rm if}\; m>0.5\,M_\odot.  
\end{equation}
The generating function for stellar masses follows by inverting the
above two equations $X_i(m)$. The procedure is then to choose a random
variate $X\in [0,1]$ and to select the generating function $m(X_1=X)$
if $0\le X\le X_1$, or $m(X_2=X)$ if $X_1 < X \le 1$.

This algorithm is readily generalised to any number of power-law
segments (eq.~\ref{eq_pk:multiIMF}), such as including a third segment
for brown dwarfs and allowing the IMF to be discontinuous near
$0.08\,M_\odot$ \citep{ThiesKroupa07}. Such a form has been
incorporated into Aarseth's {\sc Nbody4/6/7} programmes, but hitherto
without the discontinuity.  However, Jan Pflamm-Altenburg developed a
more powerful and general method of generating stellar masses (or any
other quantities) given an arbitrary distribution function
\citep{PfK2006}\footnote{The C-language software package ``libimf'',
can be downloaded at
{\tt http://www.astro.uni-bonn.de/$\sim$webaiub/english/downloads.php}~.}.

\newpage

\section{The initial binary population}
\label{sec_pk:binpop}\index{binary population: initial}\index{binary population: primordial}

\subsection{Introduction}
\label{sec_pk:intro_bin}

It has already been demonstrated that corrections for unresolved
multiple stars are of much importance for correctly deriving the shape
of the stellar MF given an observed LF (fig.~\ref{fig_pk:LFs}). Binary
stars are also of significant importance for the dynamics of star
clusters, because a binary has intrinsic dynamical degrees of freedom
which a single star does not possess.  A binary can therefore exchange
energy and angular momentum with the cluster. Indeed, binaries are
very significant energy sources, as for example, a binary composed of
two $1\,M_\odot$ main sequence stars and with a semi-major axis of
$0.1\,$AU has a binding energy comparable to that of a $1000\,M_\odot$
cluster of size $1\,$pc. Such a binary can interact with cluster-field
stars accelerating them to higher velocities and thereby heating the
cluster.

The dynamical properties describing a multiple system are: \index{dynamical properties}

\begin{itemize}

\item the period $P$ (in days throughout this text) or semi-major axis $a$,

\item the system mass $m_{\rm sys} = m_1 + m_2$,

\item the mass ratio $q\equiv {m_2\over m_1} \le 1$, where $m_1, m_2$
are, respectively, the primary and secondary-star masses,

\item
the eccentricity $e=(r_{\rm apo} - r_{\rm peri})/(r_{\rm apo} +
r_{\rm peri})$, where $r_{\rm apo}, r_{\rm peri}$ are, respectively,
the apo-centric and peri-centric distances.

\end{itemize}

Given a snapshot of a binary, the above quantities can be computed
\index{binary orbital parameters} from the relative position,
$\vec{r}_{\rm rel}$, and velocity, $\vec{v}_{\rm rel}$, vectors and
the masses of the two companion stars by first calculating the binding
energy,
\begin{equation}
E_{\rm b} = {1\over2}\,\mu\,v_{\rm rel}^2 - {G\,m_1\,m_2 \over r_{\rm rel}}
= -{G\,m_1\,m_2 \over 2\,a} \Longrightarrow a,
\label{eq_pk:bin_ebin}
\end{equation}
where $\mu = m_1\,m_2\,/(m_1+m_2)$ is the reduced mass. From Kepler's third law we have
\begin{equation}
m_{\rm sys} = {a_{\rm AU}^3\over P_{\rm yr}^2} \Longrightarrow 
P = P_{\rm yr} \times 365.25\;{\rm days},
\label{eq_pk:kepler}
\end{equation}
where $P_{\rm yr}$ is the period in years and $a_{\rm AU}$ is the
semi-major axis in~AU. Finally, the instantaneous eccentricity can be
calculated using
\begin{equation}
e = \left[\left(1-{r_{\rm rel}\over a}\right)^2 + {\left(\vec{r}_{\rm
rel} \cdot \vec{v}_{\rm rel}\right)^2 \over a\,G\,m_{\rm sys}}
\right]^{1\over2},
\label{eq_pk:bin_ecc}
\end{equation}
which can be derived from the orbital angular momentum, 
\begin{equation}
\vec{L} = \mu\,\vec{v_{\rm rel}} \times \vec{r_{\rm rel}},
\end{equation}
with
\begin{equation}
L = \left[{G\over m_{\rm sys}}\,a\,(1-e^2) \right]^{1\over2}\,m_1\,m_2.
\end{equation}

The relative equation of motion is
\begin{equation}
{d^2\vec{r}_{\rm rel} \over dt^2} = -G {m_{\rm sys} \over r_{\rm
rel}^3}\vec{r}_{\rm rel} + \vec{a}_{\rm pert}(t),
\end{equation}
where $\vec{a}_{\rm pert}(t)$ is the time-dependent perturbation from
other cluster members. It follows that the orbital elements of a
binary in a cluster are functions of time, $P=P(t)$ and
$e=e(t)$. Also, $q=q(t)$ during strong encounters when partners are
exchanged. Since most stars form in embedded clusters it follows that
the binary-star properties of a given population cannot be taken to
represent the initial or primordial values.

The following theorem can therewith be stated:

\vspace{2mm}

\centerline{ \fbox{\parbox{12cm}{{\sc Dynamical Population Synthesis
Theorem}: If initial binary populations are invariant, then a
dynamical birth configuration of a stellar population can be inferred
from its observed binary population. This birth configuration is not
unique, however, but defines a class of dynamically equivalent
solutions.\label{theorem_pk:dynpop}}}} \index{Dynamical Population
Synthesis Theorem} 

\vspace{2mm}

\noindent The proof is simple: Set-up initially identical binary
populations in clusters with different radii and masses, and calculate
the dynamical evolution with an $N-$body programme. For a given
snapshot of a population, there is a scalable starting configuration
in terms of size and mass \citep{K95c,K95d}. 

Binaries can absorb energy and thus cool a cluster. They can also heat
a cluster. There are two extreme regimes that can be understood with a
Gedanken experiment. Define 
\[
E_{\rm bin}\equiv -E_{\rm b} > 0,
\]
\vspace{-9mm}
\begin{equation}
\label{eq_pk:Ebin_Ek}
\end{equation}
\vspace{-9mm}
\[
E_{\rm k} \equiv (1/2)\,\overline{m}\,\sigma^2 \approx (1/N)\times \;
{\rm kinetic \; energy \; of \; cluster}. 
\]
{\it Soft binaries} have $E_{\rm bin} \ll E_{\rm k}$, while {\it hard
binaries} have $E_{\rm bin} \gg E_{\rm k}$. A useful equation in this
context is the relation between the orbital period and circular
velocity of the reduced particle
\footnote{Throughout this text: $lx \equiv {\rm log}_{10}(x)$.}:
\begin{equation}
lP[{\rm days}] = 6.986 + lm_{\rm sys}[M_\odot] - 3\,lv_{\rm orb}[{\rm
km/s}], 
\label{eq_pk:P_v}
\end{equation}

Consider now the case of a soft binary, i.e. a reduced-mass particle
with $v_{\rm orb} \ll \sigma$. By the principle of energy
equipartition, $v_{\rm orb} \rightarrow \sigma$
(eq.~\ref{eq_pk:sigma}) as time progresses. This implies $a \uparrow,
P \uparrow$. A hard binary has $v_{\rm orb} \gg \sigma$. Invoking
energy equipartition, it follows that $v_{\rm orb} \downarrow$,
i.e. $a \downarrow, P \downarrow$. Furthermore, the amount of energy
needed to ``ionise'' a soft binary is negligible compared to the
amount of energy required to ``ionise'' a hard binary. And, the cross
section for suffering an encounter scales with the semi-major axis
implying that a soft binary becomes ever more likely to suffer an
additional encounter as its semi-major axis increases.  Therefore, it
is much more probable for soft binaries to be disrupted rapidly, than
for hard binary to do so.  Thus follows \citep{Heggie75,
Hills75}\index{soft binary}\index{hard binary}

\vspace{2mm}

\centerline{ \fbox{\parbox{12cm}{{\sc Heggie--Hills Law}: Soft binaries
soften and cool a cluster. Hard binaries harden and heat a cluster.
\label{law_pk:heggie_hills}}}}\index{Heggie-Hills Law}

\vspace{2mm}

\noindent Numerical scattering experiments by \cite{Hills75} have
shown that hardening of binaries often involves partner
exchanges. \cite{Heggie75} derived the above law analytically.
Binaries in the energy range $10^{-2}\,E_{\rm k} \simless E_{\rm bin}
\simless 10^2\, E_{\rm k}$, $33^{-1}\,\sigma \simless v_{\rm orb}
\simless 33\,\sigma$ cannot be treated analytically due to the complex
resonances that are created between the binary and incoming star or
binary. It is these binaries that may be important for the early
cluster evolution, depending on its velocity dispersion, $\sigma =
\sigma(M_{\rm ecl})$.  Cooling of a cluster is energetically not
significant but has been seen for the first time by
\cite{Kroupaetal99}.

\begin{figure}
\begin{center}
\rotatebox{0}{\resizebox{0.8
\textwidth}{!}{\includegraphics{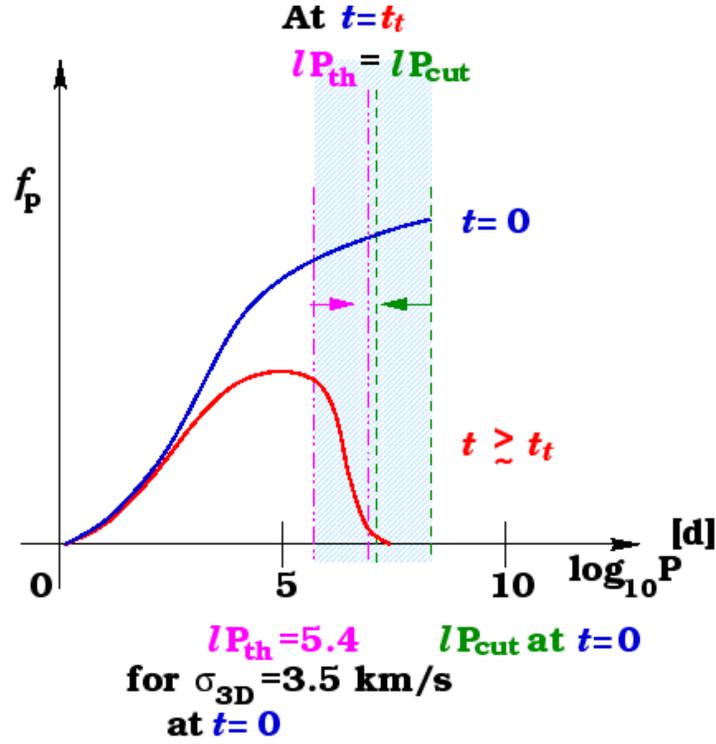}}}
\vskip 0mm
\caption{\small{Illustration of the evolution of the distribution of
binary star periods in a cluster ($l{\rm P}={\rm log}_{10}P$). A
binary has orbital period $P_{\rm th}$ when $\sigma_{\rm
3D}\;(=\sigma)$ equals its circular orbital velocity
(eq.~\ref{eq_pk:P_v}). The initial or birth distribution
(eq.~\ref{eq_pk:fpbirth}) evolves to the form seen at time $t >
t_t$.  }}
\label{fig_pk:fp_thermal}
\end{center}
\end{figure}

Fig.~\ref{fig_pk:fp_thermal} visualises the broad evolution of the
initial period distribution in a star cluster.  At any time, binaries
near the {\it hard/soft} boundary, with energies $E_{\rm bin} \approx
E_{\rm k}$ and periods $P\approx P_{\rm th} (v_{\rm orb}=\sigma)$
(eq.~\ref{eq_pk:sigma}) denoting the {\it thermal
period}\index{thermal period}, are most active in the energy exchange
between the cluster field and the binary population.  The cluster
expands as a result of binary heating \index{binary
heating}\index{binaries as energy source} and mass segregation, and
the hard/soft boundary, $P_{\rm th}$, shifts to longer periods.
Meanwhile, binaries with $P>P_{\rm th}$ continue to be disrupted while
$P_{\rm th}$ keeps shifting to longer periods. This process ends when
\begin{equation}
P_{\rm th}\simgreat P_{\rm cut}, 
\end{equation}
which is the cutoff or maximum period in the surviving period
distribution. At this critical time, $t_t$, further cluster expansion
is slowed because the population of heating sources, i.e. the binaries
with $P\approx P_{\rm th}$, is significantly reduced. The details
strongly depend on the initial value of $P_{\rm th}$ which determines
the amount of binding energy in soft binaries which can cool the
cluster if significant enough.

After the critical time, $t_{\rm t}$, the expanded cluster reaches a
temporary state of {\it thermal equilibrium} \index{thermal
equilibrium}with the remaining binary population. Further evolution of
the binary population occurs with a significantly reduced rate
determined by the velocity dispersion in the cluster, the cross
section given by the semi-major axis of the binaries, their number
density and that of single stars in the cluster.  The evolution of the
binary-star population during this slow phase will usually involve
partner exchanges and unstable but also long-lived {\it hierarchical
systems}. The {\it IMF} is critically important for this stage, as the
initial number of massive stars determines the cluster density at
$t\simgreat5\,$Myr owing to mass loss from evolving stars. Further
binary depletion will occur once the cluster goes into core-collapse
and the kinetic energy in the core rises.

\subsubsection{Frequency of binaries and higher-order multiples}
\label{sec_pk:higherorder}\index{higher-order multiple stars}\label{binary star frequency}

The emphasis here is on late-type binary stars because higher-order
multiples are rare as shown by observation. The information on the
multiplicity of massive stars is very limited \citep{Goodwinetal07}.

Defining, respectively, the number of single stars, binaries, triples,
quadruples, etc., by the numbers
\begin{equation}
(N_{\rm sing}\;:\;N_{\rm bin}\;:\;N_{\rm trip}\;:\;N_{\rm quad}\;:\,...) = 
({\cal S}\;:\;{\cal B}\;:\;{\cal T}\;:\;{\cal Q}\;:\; ...)
\end{equation}
the multiplicity fraction can be defined,
\begin{equation}
f_{\rm mult} = {N_{\rm mult} \over N_{\rm sys}} = {{\cal B} + {\cal T} + {\cal Q} + ...
\over {\cal S} + {\cal B} + {\cal T} + {\cal Q} + ...}
\end{equation}
and the binary fraction is
\begin{equation}
f_{\rm bin} = {{\cal B} \over N_{\rm sys}}.
\end{equation}

In the Galactic field \cite{DM91} derive from a decade-long survey for
G-dwarf primary stars, $^{\rm G}N_{\rm mult} = (\;57\;:\;38\;:\;4\;:\;1\;)$
and for M-dwarfs \cite{FM92} find $^{\rm M}N_{\rm mult} =
(\;58\;:\;33\;:\;7\;:\;1\;)$. Thus,
\begin{equation}
^{\rm G}f_{\rm mult} = 0.43; \quad ^{\rm G}f_{\rm bin} = 0.38
\end{equation} 
\vspace{-9mm}
\begin{equation}
\end{equation}
\vspace{-9mm}
\begin{equation}
^{\rm M}f_{\rm mult} = 0.41; \quad ^{\rm M}f_{\rm bin} = 0.33 
\end{equation}
It follows that most ``stars'' are indeed binaries.

After correcting for incompleteness,
\begin{equation}
^{\rm G}f_{\rm bin} = 0.53\pm0.08,
\end{equation} 
\begin{equation}
^{\rm K}f_{\rm bin} = 0.45\pm0.07, 
\end{equation}
\begin{equation}
^{\rm M}f_{\rm bin} = 0.42\pm0.09,
\end{equation}
where the K-dwarf data have been published by \cite{Mayoretal92}.
It follows that 
\begin{equation}
^{\rm G}f_{\rm bin} \approx ^{\rm K}f_{\rm bin} \approx ^{\rm M}f_{\rm
bin} \approx 0.5 \approx f_{\rm tot},
\end{equation}
in the Galactic field, perhaps with a slight decrease towards lower
masses. In contrast, for brown dwarfs, $^{\rm BD}f_{\rm bin} \approx
0.15 \ll ^{\rm stars}f_{\rm bin}$ (\citealt{ThiesKroupa07}, and
references therein).

An interesting problem arises when one considers that for $1\,$Myr old
stars, $f_{\rm TTauri}\approx 1$ (e.g. \citealt{Duchene99}).  Given
the above information, the following theorem can be stated:

\vspace{2mm}

\centerline{ \fbox{\parbox{12cm}{{\sc Binary-Star Theorem}: The vast majority 
of stars form in binary systems.
\label{theorem_pk:binaries}}}}\index{Binary-Star Theorem}

\vspace{2mm}

\noindent Proof: If a substantial fraction of stars were to form in
higher-order multiple systems, or as small$-N$ systems, then the
typical properties of these at birth imply their decay within
typically $10^4$ to $10^5\,$yr leaving a predominantly single-stellar
population. However, the majority of $10^6\,$Myr old stars are
observed to be in binary systems
\citep{GoodwinKroupa2005}.\index{higher-order multiple stars}

Higher-order multiple systems do exist and can only be hierarchical to
guarantee stability.  Hierarchical systems are multiple stars which
are stable over many orbital times, and are typically tight binaries
orbited by outer tertiary companions, or two tight binaries in orbit
about each other, etc. \cite{EggletonKiseleva1995} discuss stability
issues in more detail, but it suffices here to state that the outer
and inner semi-major axes should typically have a ratio of about four
for stability for comparable system masses. If the stability criterion
is not fulfilled, then higher-order multiple systems typically decay
on a few crossing times by ejecting members until a stable or
long-lived configuration is found. Most often this is a hardened
binary.

Star cluster remnants (or dead star clusters) may be the origin of
most hierarchical, higher-order multiple stellar systems
(p.~\pageref{sec_pk:death}).\index{dead star clusters}\index{star
cluster remnants}

\subsection{The initial binary population -- late-type stars}
\label{sec_pk:initialpop_lowmass} \index{initial binary population} 
\index{primordial binary population}

The initial binary population is described by distribution functions
that are as fundamental for a stellar population as the IMF. There are
four distribution functions defining the initial dynamical state of a
population:
\begin{enumerate}
\item The IMF, $\xi(m)$,

\item the distribution of periods (or semi-major axis), $df=f_P(lP)\,dlP$

\item the distribution of mass-ratios, $df=f_q(q)\,dq$,

\item the distribution of eccentricities, $df=f_e(e)\,de$, 

\end{enumerate}
where $df$ is the fraction of systems with a parameter in the vicinity
of the given value. Thus, for example, $^{\rm G}f_{lP}(lP=4.5) =
0.11$, i.e.  of all G-dwarfs on the sky, 11~per cent have a
companion with a period in the range of 4~to~5$\,$days
(fig.~\ref{fig_pk:pfin}).

\begin{figure}
\vspace{-30mm}
\hspace{-20mm}
\rotatebox{0}{\resizebox{1.2 \textwidth}{!}{\includegraphics{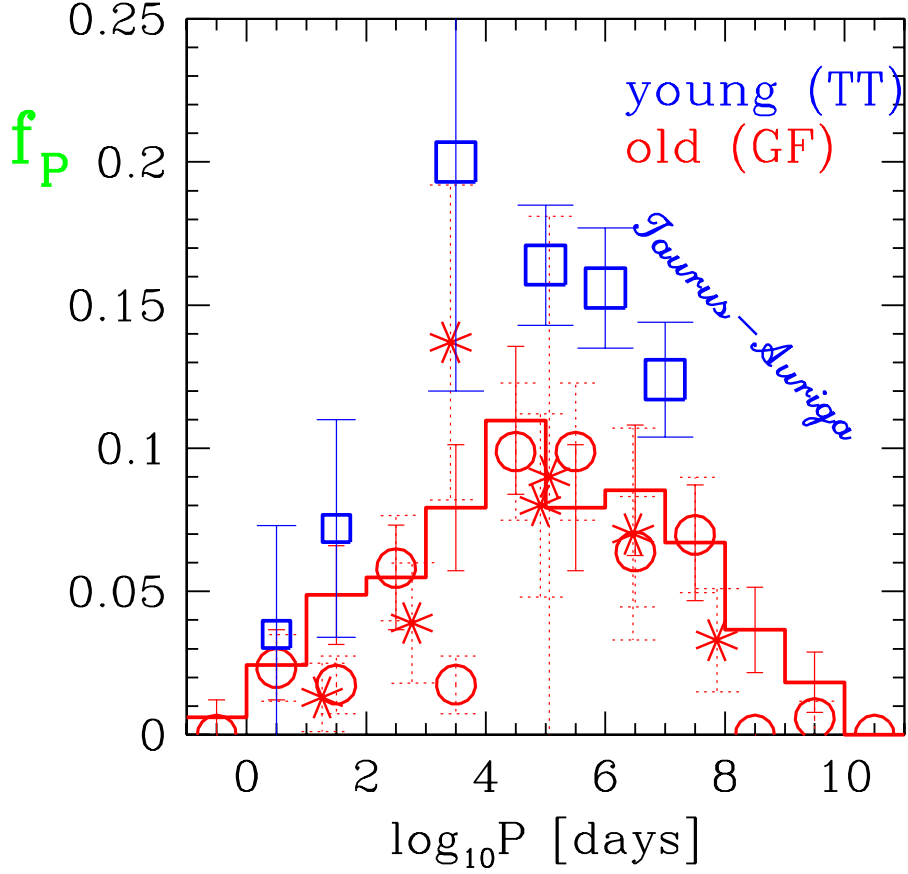}}}
\vskip -40mm
\caption{\small{ Measured period-distribution functions for
Galactic-field G-dwarfs (histogramme, \citealt{DM91}), K-dwarfs (open
circles, \citealt{Mayoretal92}) and M-dwarfs (asterisks,
\citealt{FM92}). The about $1\,$Myr old T~Tauri binary data (open
squares, partially from the Taurus--Auriga stellar groups) are a
compilation from various sources (see fig.~10 in \citealt{KAH}). In
all cases, the areas under the distributions is $f_{\rm tot}$.  }}
\label{fig_pk:fp}
\end{figure}

These distribution functions have been measured for late-type stars in
\index{binary star distribution functions} the Galactic field and in
star-forming regions (fig.~\ref{fig_pk:fp}). According to \cite{DM91}
and \cite{FM92} both G-dwarf and M-dwarf binary systems in the
Galactic field have period distribution functions that are well
described by log-normal functions,\index{log-normal period
distribution function}
\begin{equation}
f_P(lP) = f_{\rm tot}\,\left( {1\over \sigma_{lP}\,\sqrt{2\,\pi}} \right) \,
e^{\left[-{1\over2}\,{(lP - \overline{lP})^2 \over \sigma_{lP}^2} \right]},
\label{eq_pk:dmfp}
\end{equation}
with $\overline{lP} \approx 4.8$ and $\sigma_{lP}\approx 2.3$, and
$\int_{{\rm all}\;lP}f_{lP}(lP)\,dlP = f_{\rm tot} \approx
0.5$. K-dwarfs appear to have an indistinguishable period
distribution.

From fig.~\ref{fig_pk:fp} it follows that the pre-main sequence binary
fraction is larger than that of main-sequence stars (see also
\citealt{Duchene99}). Is this an evolutionary effect? 

Further, \cite{DM91} derived the mass-ratio and eccentricity
distributions for G-dwarfs in the Galactic field. The mass-ratio
distribution of G-dwarf primaries is not consistent with random
sampling from the canonical IMF (eq.~\ref{eq_pk:canonIMF}), as the
number of observed low-mass companions is underrepresented
\citep{K95c}. In contrast, the pre-main sequence mass-ratio
distribution is consistent, within the uncertainties, with random
sampling from the canonical IMF for $q\simgreat 0.2$
\citep{Woitasetal01}. The eccentricity distribution of Galactic-field
G-dwarfs is found to be thermal for $lP\simgreat3$, while it is bell
shaped with a maximum near $e=0.25$ for $lP\simless 3$. Not much is
known about the eccentricity distribution of pre-main sequence
binaries, but numerical experiments show that $f_e$ does not evolve
much in dense clusters, i.e. the thermal distribution must be initial
\citep{K95d}.

The {\it thermal eccentricity distribution},\index{eccentricity distribution function}
\index{thermal eccentricity distribution function}
\begin{equation}
f_e(e) = 2\,e,
\label{eq_pk:thermal}
\end{equation}
follows from a uniform binding-energy distribution (i.e., all energies
are equally populated), as follows. The orbital angular momentum of a
binary is
\begin{equation}
L^2 = {G\over m_{\rm sys}}\,{G\,m_1\,m_2 \over 2 E_{\rm
bin}}\,\left(1-e^2\right)\,\left(m_1\,m_2\right)^2
\end{equation}
from which follows
\begin{equation}
e = \left( 1-2\,E_{\rm bin}\,L^2\,{m_{\rm sys} \over G^2\,(m_1\,m_2)^2}\right)^{1\over2}.
\end{equation}
Differentiation leads to
\begin{equation}
{de \over dE_{\rm bin}} = \left[-L^2 {m_{\rm sys} \over
G^2\,(m_1\,m_2)^2} \right]\,e^{-1} \equiv [\,]\,e^{-1}.
\end{equation}
The number of binaries with eccentricities in the
range $e,e+de$ is the same number of binaries with binding-energy in
the range $E_{\rm bin}, E_{\rm bin} + dE_{\rm bin}$ (the same sample of binaries),
\begin{equation}
f(e)\,de = f(E_{\rm bin})\,dE_{\rm bin} = f(E_{\rm bin})\,[\,]^{-1}\,e\,de,
\end{equation}
where the square brackets are from the previous equation. But 
\begin{equation}
\int_0^1\,f(e)\,de = 1,
\end{equation}
i.e.
\begin{equation}
1 = f(E_{\rm bin})\,[\,]^{-1}\,\int_0^1\,e\,de.
\end{equation}
So
\begin{equation}
f(E_{\rm bin})\,[\,]^{-1} = 2 = {\rm const} \; \Longrightarrow \; f(e)\,de = 2\,e\,de.
\end{equation}
Thus, $f(e) = 2\,e$ is a thermalised distribution: all energies are
equally occupied ($f(E_{\rm bin}) = {\rm const}$).

$N-$body experiments have demonstrated that the period distribution
function must span the observed range of periods at birth, as
dynamical encounters in dense clusters cannot widen an initially
narrow distribution \citep{KroupaBurkert01}.

There are thus three discrepancies between main-sequence and pre-main
sequence late-type stellar binaries: 
\begin{enumerate}
\item the binary fraction is higher
for the latter, 

\item the period distribution function is different, and

\item the mass-ratio distribution is consistent with random paring for
the latter, while it is deficient in low-mass companions in the former,
for G-dwarf primaries. 
\end{enumerate}
Can these be unified, i.e. are there unique initial $f_{lP}, f_q, f_e$
consistent with the pre-main sequence data that can be evolved to the
observed main-sequence distributions?

This question can be solved by framing the following ansatz: Assume
the orbital-parameter distribution function for binaries with
primaries of mass $m_1$ can be separated,
\begin{equation}
{\cal D}(lP, e, q\,:\,m_1) = f_{lP}\,f_e\,f_q.
\end{equation}
The {\it stellar-dynamical operator}, $\Omega^{N,r_{0.5}}$, can now be
introduced such that the initial distribution function is transformed
to the final (Galactic-field) one,
\begin{equation}
D_{\rm fin}(lP, e, q\,:\,m_1) = \Omega^{N,r_{0.5}}\,\left[ {\cal
D}_{\rm in}(lP, e, q\,:\,m_1)\right].
\end{equation}
This operator provides a dynamical environment {\it equivalent} to
\index{stellar-dynamical operator}\index{dynamically equivalent
clusters} that of a star cluster with $N$ stars and a half-mass radius
$r_{0.5}$ (see also the Dynamical Population Synthesis Theorem,
p.~\pageref{theorem_pk:dynpop}). \index{dynamical population
synthesis} \cite{K95c} and \cite{K95d} indeed prove this to be the
case for a cluster $N=200$~binaries and $r_{0.5}=0.77\,$pc and derives
the initial distribution function, ${\cal D}_{\rm in}$, for late-type
binary systems such that it fulfills the above requirement and also
has a simple generating function (see below). Noteworthy is that such
a cluster is very similar to the typical cluster from which most field
stars probably originate. The full solution for $\Omega$, such that the
Galactic-field is re-produced from forming and dissolving star
clusters, requires full-scale {\it inverse dynamical population
synthesis} for the Galactic field.\label{page_pk:inv_dyn_pop_synth}
\index{inverse dynamical population synthesis}

\begin{figure}
\begin{center}
\resizebox{0.95 \textwidth}{!}{\includegraphics{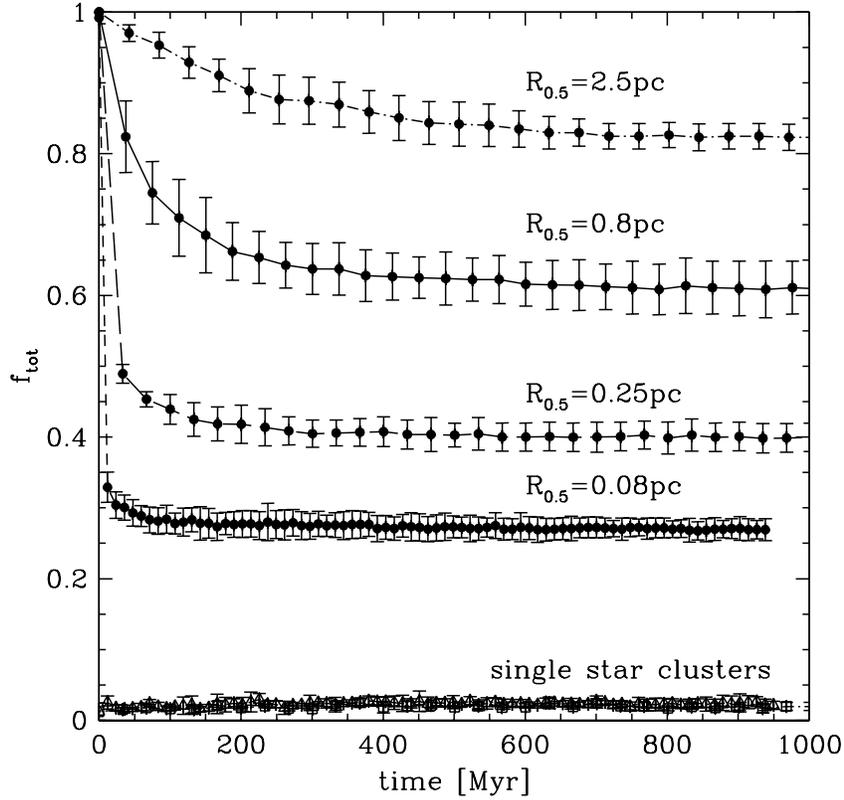}}
\vskip 5mm
\caption[\it Evolution of $f_{\rm tot}$] {\small{ Evolution of the
total binary fraction for stellar mass $0.1\le m_i/M_\odot \le 1.1,
\; i=1,2$ with time for the four star-cluster models initially with
$N=200$~binaries computed by \cite{K95c} in the search of the
existence of an $\Omega^{r_{0.5},N}$ ($R_{0.5}$ is the initial
half-mass radius of the clusters, denoted in this text as $r_{0.5}$).
Note that the $r_{0.5}=0.8\,$pc cluster yields the correct $f_{\rm
tot}\approx 0.5$ for the Galactic field. The period-distribution
function {\it and} the mass-ratio distribution function that emerge
from this cluster also fit to the observed Galactic-field
distribution.  Some binary stars form by three-body encounters in
clusters that initially consist only of single stars, and the
proportion of such binaries is shown for the single star clusters
(with initially $N=400$~stars). Such dynamically formed binaries are
very rare and so $f_{\rm tot}$ remains negligible.}}
\label{fig_pk:ftot}
\end{center}
\end{figure}
Thus, by the {\sc Dynamical Population Synthesis
Theorem}\index{dynamical population synthesis}
(p.~\pageref{theorem_pk:dynpop}), the above ansatz with $\Omega^{N,
r_{0.5}}$ leads to one solution to inverse dynamical population
synthesis (the 200~binary, $r_{0.5}=0.8\,$pc cluster,
fig.~\ref{fig_pk:ftot}; i.e. most stars in the Galactic field stem
from clusters dynamically similar to this one), provided the birth (or
{\it primordial}) distribution functions for $lP, e, q$ are as
follows:
\begin{equation}
f_{lP, {\rm birth}} = \eta \, {lP-lP_{\rm min} \over \delta + (lP-lP_{\rm min})^2}.
\label{eq_pk:fpbirth}
\end{equation}\index{pre-main sequence period distribution function}
This distribution function has a generating function
(section~\ref{sec_pk:discretisation}),
\begin{equation}
lP(X) = \left[\delta\,\left(e^{2\,X \over \eta}
-1\right)\right]^{1\over2} + lP_{\rm min}.
\label{eq_pk:fpbirth_gen}
\end{equation}
The solution obtained by \cite{K95d} has 
\begin{equation}
\eta = 2.5, \quad \delta = 45, \quad lP_{\rm min} = 1,
\label{eq_pk:birth_fp_param}
\end{equation}
such that $lP_{\rm max} = 8.43$ since $\int_{lP_{\rm
min}}^{lP_{\rm max}}\,f_{lP}\,dlP = f_{\rm tot}= 1$ is a requirement
for stars at birth. Intriguingly, similar distributions can be arrived
at semi-empirically assuming an isolated formation of binary stars in
a turbulent molecular cloud \citep{Fisher04}.

The birth-eccentricity distribution is thermal
(eq.~\ref{eq_pk:thermal}) while the birth mass-ratio distribution is
generated from random pairing from the canonical IMF. However, in
order to re-produce (1)~the observed data in the eccentricity--period
diagramme, (2)~the observed eccentricity distribution and (3)~the
observed mass-ratio distribution \index{mass-ratio distribution} for
{\it short-period} ($lP\simless 3$) systems, a correlation of the
parameters needs to be introduced through {\it
eigenevolution}. Eigenevolution is the sum of all dissipative physical
processes that transfer mass, energy and angular momentum between the
companions when they are still very young and accreting.

A formulation which is quite successful in re-producing the overall
observed correlations between $lP, e, q$ for short-period systems has
been derived from tidal-circularisation theory
\citep{K95d}. Most-effective orbital dissipation occurs when the
binary is at its peri-astron, \index{eigenevolution}
\begin{equation}
r_{\rm peri} = (1-e)\,P_{\rm yr}^{2\over3}\,\left(m_1+m_2\right)^{1\over3},
\label{eq_pk:rperi}
\end{equation}
where $P_{\rm yr} = P/365.25$ is the period in years. Let the binary
be born with eccentricity $e_{\rm birth}$, then the system evolves
approximately according to \citep{GoldmanMazeh94}
\begin{equation}
{1\over e}\,{de\over dt} = - \rho' \Longrightarrow {\rm
log}_{10}e_{\rm in} = -\rho + {\rm log}_{10}e_{\rm birth},
\end{equation}
where $1/\rho'$ is the tidal-circularisation time-scale, $e_{\rm in}$
is the initial eccentricity, and
\begin{equation}
\rho = \int_0^{\Delta t} \, \rho'\,dt = \left({\lambda \, R_\odot
\over r_{\rm peri}} \right)^\chi,
\end{equation}
where $R_\odot$ is the Solar radius in AU, $\lambda, \chi$ are
tidal-circularisation parameters and $r_{\rm peri}$ (in AU) is assumed
to be constant because the dissipational force only acts tangentially
at peri-astron.  Note that a large $\lambda$ implies that
tidal-dissipation is effective for large separations of the companions
(e.g. they are puffed-up pre-main sequence structures), and a small
$\chi$ implies the dissipation is soft, i.e. weakly varying with the
separation of the companions.  In this integral, $\Delta t \simless
10^5\,$yr is the time-scale within which pre-main-sequence
eigenevolution completes.  The initial period becomes, from
eq.~\ref{eq_pk:rperi},
\begin{equation}
P_{\rm in} = P_{\rm birth}\,\left({m_{\rm tot, birth} \over m_{\rm
tot, in}}\right)^{1\over2}\,\left({1-e_{\rm birth} \over 1-e_{\rm
in}}\right)^{3\over2},
\end{equation}
\cite{K95d} assumed the companions merge if $a_{\rm in} \le
10\,R_\odot\,:\; m_1+m_2 \rightarrow m$.

In order to re-produce the observed mass-ratio distribution given
random pairing at birth, and to also re-produce the fact that
short-period binaries tend to have similar-mass companions,
\cite{K95d} implemented a {\it feeding algorithm}, \index{feeding
algorithm} \index{binary star mass ratio} according to which the
secondary star accretes high-angular-momentum gas from the
circum-binary accretion material, such that its mass increases while
the primary mass remains constant. Thus, after generating the two
birth-companion masses randomly from the canonical IMF, the initial
mass-ratio becomes
\begin{equation}
q_{\rm in} = q_{\rm birth} + (1-q_{\rm birth})\,\rho^*,
\end{equation}
where
\begin{equation}
\rho^* = 
\left\{ 
\begin{array}{r@{\quad:\quad}l}
\rho & \rho\le 1,\\
1   & \rho > 1.
\end{array}
\right.
\end{equation}
The above is a very simple algorithm which nevertheless re-produces
the essence of orbital dissipation such that the correlations between
the orbital parameters for short-period systems are well accounted
for. The best parameters for the evolution ``birth $\longrightarrow$
initial'' are
\begin{equation}
\lambda=28\, , \; \chi=0.75.
\end{equation}

\begin{figure}
\begin{center}
\rotatebox{0}{\resizebox{0.9 \textwidth}{!}{\includegraphics{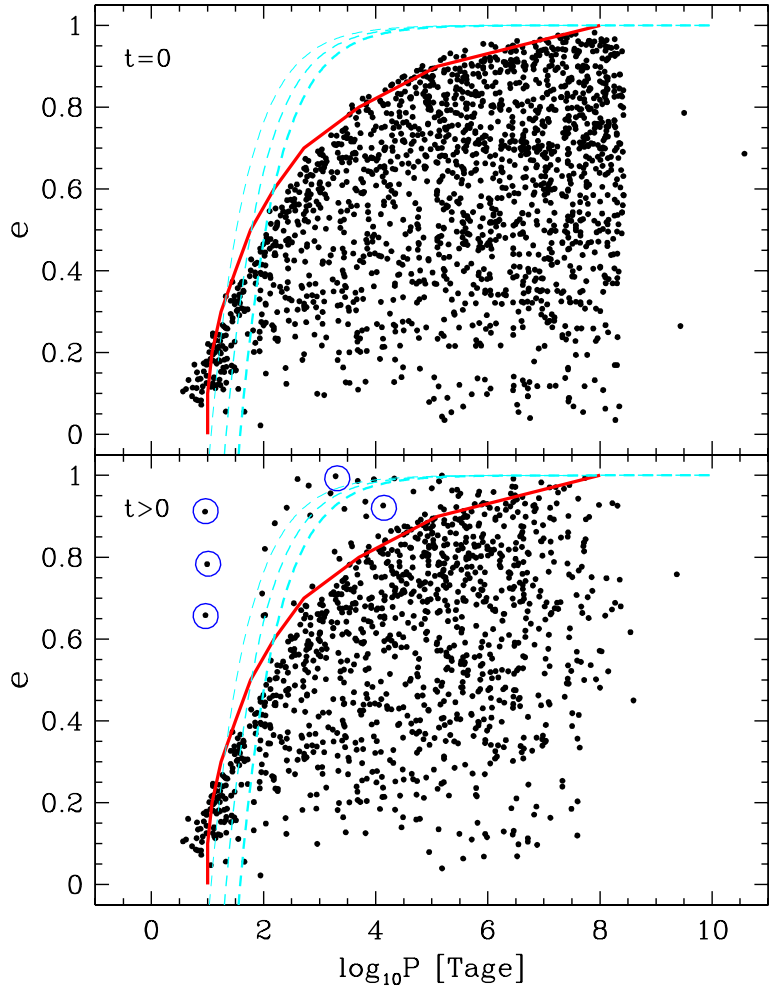}}}
\vskip -1mm
\caption{\small{The eccentricity--period diagram for the model of the
pre-main sequence eigenevolution ($\lambda=28,\chi=0.75$) at $t=0$
(upper panel) for stellar mass $0.1 \le m_i/M_\odot \le 1.1$ and after
cluster disintegration (bottom panel; note: Tage$=$days).  Systems
with semi-major axis after pre-main sequence eigenevolution
$a\le10\,R_\odot$ have been merged. The upper envelope is from
\cite{DM91} -- binaries are only observed to have $e,lP$ below this
envelope, the region above this envelope being {\it forbidden}, as
pre-main sequence dissipational effects are understood to de-populate
this region with $10^5\,$yr.  However, dynamical encounters in the
cluster populate the eigenevolution region implying that short-lived
{\it forbidden orbital parameters} should be observable in stellar
clusters. Some of these are indicated as open circles.  Eigenevolution
(i.e. classical tidal circularisation) on the main sequence with
$\lambda_{\rm ms}=24.7,\chi_{\rm ms}=8$ applied to the data in the
lower panel depopulates the eigenevolution region and circularises all
orbits with period less than about $12\,$d. The dashed lines are
constant peri-astron distances (eq.~\ref{eq_pk:rperi}) for $r_{\rm
peri} = \lambda\,R_\odot$ and $m_{\rm sys} = 2.2, 0.64, 0.2\,M_\odot$
(in increasing thickness). Note that horizontal and vertical cuts
through this diagramme produce eccentricity and period distribution
functions (as well as mass-ratio distributions) for short and
long-period systems in agreement with the observational
constraints. The initial orbits with $P>10^{8.5}\,$d come about due to
crowding.  From \cite{K95d}.}}
\label{fig_pk:e_p}
\end{center}
\end{figure}

\begin{figure}
\begin{center}
\rotatebox{0}{\resizebox{0.95 \textwidth}{!}{\includegraphics{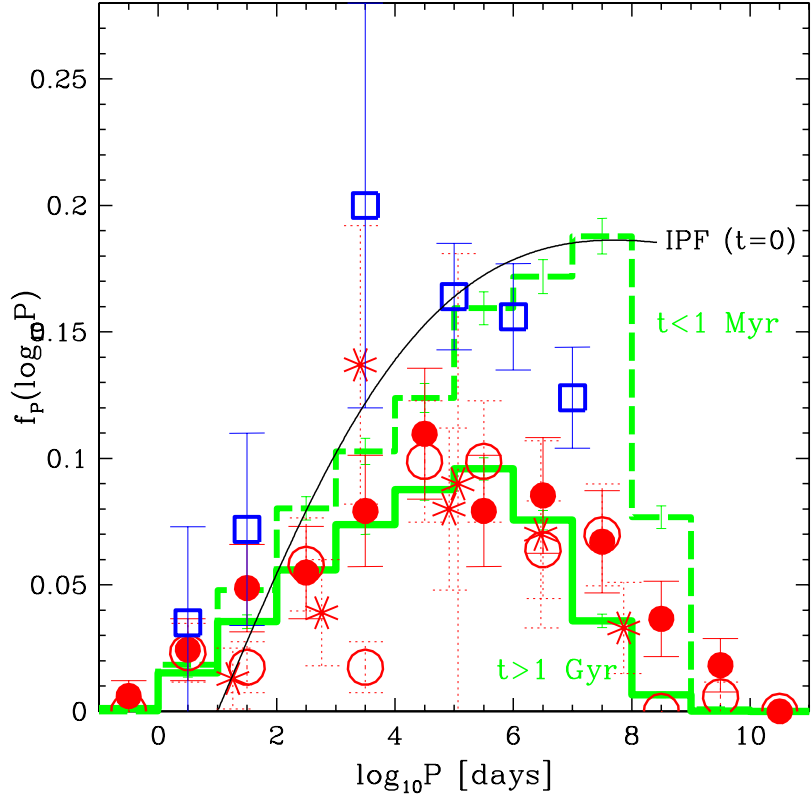}}}
\vskip -25mm
\caption{\small{The period distribution functions (IPF:
eq.~\ref{eq_pk:fpbirth} with eq.~\ref{eq_pk:birth_fp_param} and for
stellar masses $0.1\le m_i/M_\odot \le 1.1$).  The dashed histogramme
is derived from IPF using the eigenevolution and feeding algorithms,
and represents the binary-population at an age of about $10^5\,$yr,
while the solid histogramme follows from the dashed one after passing
through a cluster with initially $N=200$~binaries and
$r_{0.5}=0.8\,$pc. The agreement of the dashed histogramme with the
observational pre-main sequence data (as in fig.~\ref{fig_pk:fp}), and
of the solid histogramme with the observational main-sequence
(Galactic-field) data (also as in fig.~\ref{fig_pk:fp}) is good,
noting that the longest-period TTauri binary population is expected to
show some disruption. A full model of the Galactic field late-type
binary population has therewith been arrived at which unifies all
available but apparently discordant observational data (see also
figs~\ref{fig_pk:ftot}, \ref{fig_pk:e_p} and~\ref{fig_pk:qfin}).}}
\label{fig_pk:pfin}\index{period distribution}
\end{center}
\end{figure}

\begin{figure}
\begin{center}
\rotatebox{-90}{\resizebox{0.7 \textwidth}{!}{\includegraphics{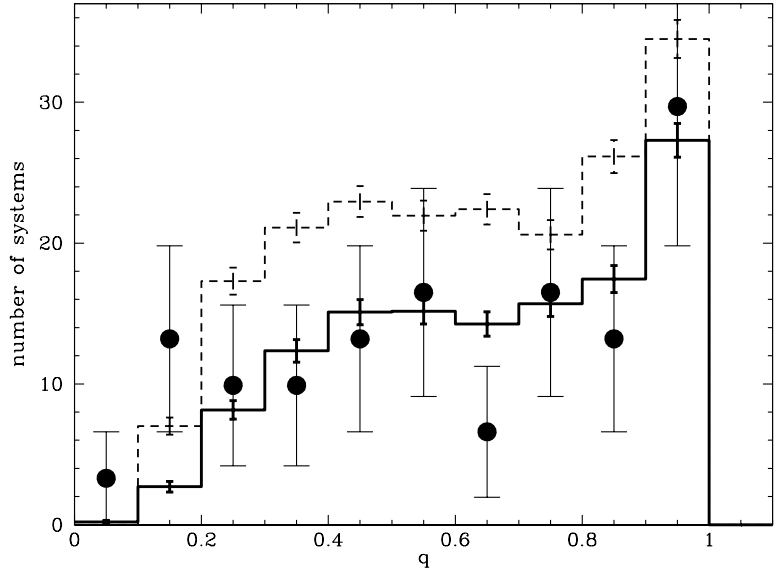}}}
\vskip -5mm
\caption{\small{The mass-ratio distribution for stars with $0.1\le
m/M_\odot\le 1.1$ is the solid histogramme, whereas the {\it initial}
mass-ratio distribution (random pairing from the canonical IMF; after
eigenevolution and feeding, at an age $\approx 10^5\,$yr) is shown as
the dashed histogramme. The solid histogramme follows from the dashed
one after passing through a cluster with initially $N=200$~binaries
and $r_{0.5}=0.8\,$pc.  The observational data (solid dots,
\citealt{ReidGizis97}) have been obtained after removing WD companions
and scaling to the model. This solar-neighbourhood 8~pc sample is not
complete and may be biased towards $q=1$ systems
\citep{Henryetal97}. Nevertheless, the agreement between model (solid
histogramme) and the data is striking.  A full model of the Galactic
field binary population has therewith been arrived at which unifies
all available but apparently discordant observational data (see also
figs~\ref{fig_pk:ftot}, \ref{fig_pk:e_p} and~\ref{fig_pk:pfin}).}}
\label{fig_pk:qfin}\index{mass ratio distribution}
\end{center}
\end{figure}

Fig.~\ref{fig_pk:e_p} shows an example of the overall model in terms
of the eccentricity--period diagramme. Fig.~\ref{fig_pk:pfin} and
fig~\ref{fig_pk:qfin} demonstrate that it nicely accounts for the
period and mass-ratio distribution data,
respectively.\index{eccentricity-period distribution}

Note that {\it initial} distributions are derived from {\it birth}
distributions. \index{birth distribution of binary stars} This is to
be understood in terms of these initial distributions being the
initialisation of $N-$body experiments, while the birth distributions
are more related to the theoretical distribution of orbital parameters
before dissipational and accretion processes have a major effect on
them. The birth distributions are, however, mostly an algorithmic
concept. Once the $N-$body integration is finished, e.g. when the
cluster is dissolved, the remaining binaries can be evolved to the
main-sequence distributions by applying the same eigenevolution
algorithm above, but with parameters
\begin{equation}
\lambda_{\rm ms}=24.7\, , \; \chi_{\rm ms}=8.
\end{equation}
The need for $\lambda_{\rm ms} < \lambda$ and $\chi_{\rm ms}>\chi$ to
ensure for example the tidal-circularisation period of $12\,$days for
G~dwarfs \citep{DM91}, is nicely consistent qualitatively with the
shrinking of pre-main-sequence stars and the emergence of radiative
cores that essentially reduce the coupling between the stellar
surface, where the dissipational forces are most effective, and the
center of the star.

The reader is also directed to \cite{MardlingAarseth01} where a model
of tidal circularisation is introduced into the $N-$body code.

Finally, the above work and the application to the ONC \index{Orion
Nebula Cluster} and Pleiades \citep{KAH} \index{Pleiades} suggests the
following hypothesis:

\vspace{2mm}

\centerline{ \fbox{\parbox{12cm}{{\sc Initial Binary Universality
Hypothesis}: The initial period (eq.~\ref{eq_pk:birth_fp_param}),
eccentricity (eq.~\ref{eq_pk:thermal}), and mass-ratio (random pairing
from canonical IMF) distributions constitute the parent distribution
of all late-type stellar
populations. \label{hyp_pk:bin_univ}}}}\index{Initial Binary
Universality Hypothesis}

\vspace{2mm}

\noindent Can this hypothesis be disproven?

\subsection{The initial binary population -- massive stars}
\label{sec_pk:initialpop_massive}\index{Binaries: massive stars}

The above semi-empirical distribution functions have been formulated
for late-type stars (primary mass $m_1\simless 1\,M_\odot$) as
it is for these that we have the best observational constraints.  It
is not clear yet if they are also applicable to massive binaries. 

An approach taken by \cite{ClarkePringle1992} is to constrain the
binary properties of OB stars by assuming that runaway OB stars are
ejected from star-forming regions. About 10--25~per cent of all
O~stars are runaway stars, while about 2~per cent of B~stars are
runaways.  This approach leads to the result that massive stars must
form in small$-N$ groups of binaries that are biased towards unit mass
ratio. This is a potentially powerful approach, but it can only
constrain the properties of OB binaries when they are ejected which
occurs after substantial dynamical encounters in the cluster core
which typically lead to the mass-ratio evolving towards unity as the
involved binaries harden. The true birth properties of massive
binaries therefore remain obscure, and one needs to resort to $N-$body
experiments to test various hypothesis given the observational
constrains. One such hypothesis could be, for example, to assume
massive stars form in binaries with birth pairing properties as for
low-mass stars (section~\ref{sec_pk:initialpop_lowmass}), i.e. most
massive primaries would have a low-mass companion, and to investigate
if this hypothesis leads to the observed number of runaway massive
stars through dynamical mass segregation to the cluster core and
partner exchanges through dynamical encounters there between the
massive stars.

Apart from the fraction of runaway stars, \index{runaway stars}direct
surveys have lead to some insights as to the binary properties of the
observed massive stars.  Thus, for example, \cite{Bainesetal06} report
a very high ($f\approx 0.7\pm 0.1$) binary fraction among Herbig Ae/Be
stars with the binary fraction increasing with increasing primary
mass. Furthermore, they find that the circum-binary disks and the
companions appear to be co-planar thereby supporting a fragmentation
origin rather than collisions or capture as the origin of massive
binaries.  Most O~stars are believed to exist as short-period binaries
with $q\approx 1$ \citep{GarciaMermilliod01}, at least in rich
clusters. On the other hand, small-$q$ appear to be favoured in less
substantial clusters such as the ONC, being consistent there with
random pairing \citep{Preibischetal99}.  \cite{Kouwenhovenetal05}
report the A and late-type B binaries in the Scorpius OB2 association
to have a mass-ratio distribution {\it not} consistent with random
pairing. The lower limit on the binary fraction is~0.52, while
\cite{Kouwenhovenetal07} update this to a binary fraction of 72~per
cent. They also find that the semi-major axis distribution contains
too many close pairs compared to a \cite{DM91} log-normal
distribution.  These are important constraints, but again, they are
derived for binaries in an OB association, which is an expanded
version of a dense star cluster (section~\ref{sec_pk:assoc}) and
therefore hosts a dynamically evolved population.

Given the above results, perhaps the massive binaries in the ONC
\index{Orion Nebula Cluster} represent the primordial population,
whereas in rich clusters and in OB associations \index{OB
associations} the population has already dynamically evolved through
hardening and companion exchanges to that observed there ($f_q$ rising
towards $q=1$). This possibility needs to be investigated using
high-precision $N$-body computations of young star clusters. The first,
simplest hypothesis to test would be to extend the pairing rules of
section~\ref{sec_pk:initialpop_lowmass} to all stellar masses, perform
{\it many} (because of the small number of massive stars) $N-$body
renditions of the same basic pre-gas expulsion cluster, and to
quantify the properties of the emerging stellar population at various
dynamical times \citep{K01}.

Another approach would be to constrain $a$ and $m_2$ for a given
$m_1\simgreat 5\,M_\odot$ such that 
\begin{equation}
E_{\rm bin} \approx E_{\rm k}
\end{equation}
(eq.~\ref{eq_pk:Ebin_Ek}). Or one can test the initial massive-star population
given by 
\begin{equation}
a < {r_c \over N_{\rm OB}}^{1\over 3}
\end{equation}
which follows from stating that the density of a massive binary,
$2\times3/(a^3\,4\,\pi)$, be larger than the cluster-core density,
$N_{\rm OB}\,3/(r_c^3\,4\,\pi)$.

So far, none of these possibilities have been tested, apart from
extending the {\sc Initial Binary Universality Hypothesis}
(p.~\pageref{hyp_pk:bin_univ}) to massive stars \citep{K01}.

\newpage

\section{Summary}
\label{sec_pk:summary}

The above material gives an outline of how to set up an initial, birth
or primordial stellar population such that it resembles observed
stellar populations. In section~\ref{sec_pk:initialpop_lowmass} a
subtle differentiation was performed between {\it initial} and {\it
birth} populations, in the sense that an initial population is derived
from a birth population through initial processes that act too
rapidly to be treated by an $N-$body integration.

An $N-$body stellar system is generated for numerical experiments by
specifying its 3D structure and velocity field
(section~\ref{sec_pk:initcond}), the mass distribution of its population
(section~\ref{sec_pk:IMF}) and the properties of its binary population
(section~\ref{sec_pk:binpop}).

Given the distribution function discussed here, and the existing
numerical results based on these, it is surprising how universal the
stellar and binary population turns out to be at birth. A dependence
of the IMF or the birth binary properties on the physical properties
of star-forming clouds cannot be detected conclusively. In fact, the
theoretical proposition that there should be a dependency is
falsified, except possibly (i) in the extreme-tidal field environment
at the Galactic centre, or (ii) in the extreme proto-stellar density
environment of ultra-compact dwarf galaxies, or (iii) for extreme
physical environments
(pp.~\pageref{sec_pk:gal_bulge}--\pageref{sec_pk:popIII}).

The unified picture that has emerged concerning the origin of stellar
populations is that stars form according to a universal IMF and mostly
in binary systems and in very dense clusters that expel their residual
gas and rapidly evolve to T- or OB-associations. If the latter are
massive enough, the dense embedded clusters evolve to populous OB
associations that may be expanding rapidly and that contain cluster
remnants which may reach globular cluster masses and beyond in intense
star-bursts. This unified picture naturally explains the high infant
weight loss and infant mortality of clusters, the binary properties of
field stars, possibly thick disks of galaxies and the existence of
population~II stellar halos around galaxies that have old globular
cluster systems.

Some open questions that remain are quite obviously, why the
star-formation product is so universal (within current constraints),
and how massive stars are distributed in binaries and if they form at
the centres of their clusters, why the cluster mass of $\approx
10^6\,M_\odot$ is special, and which star-cluster population is a full
solution to the inverse dynamical population synthesis problem
\index{dynamical population synthesis}
(p.~\pageref{page_pk:inv_dyn_pop_synth}). Naturally, many more
observations are required not only of topical high-redshift star-burst
systems, but also of the more mundane low-redshift and preferably
local star-forming objects and globular and open star clusters, to
further refine the above broad picture.

\subsubsection{Acknowledgements}
\label{sec_pk:acknowl}
It is a pleasure to thank Sverre Aarseth for organising a splendid and
much to be remembered Cambridge Nbody school in the Summer of~2006,
and also Christopher Tout for editing and proof-reading this work. I
am also indebted to Jan Pflamm-Altenburg who read parts of this
manuscript carefully, to Andreas K\"upper for producing the Plummer vs
King model comparisons and for carefully reading this whole text, and
to Joerg Dabringhausen, who supplied me with figures from his work.

%
%
%
%

{}
%


\printindex
\end{document}